\documentclass[11pt,a4paper]{article}

\usepackage{macros}
\usepackage[symbol]{footmisc}

\preprint{ }
\title{Class $\mathcal{S}$ on $S^2$}

\author[a]{Satoshi Nawata,}
\author[b]{Yiwen Pan \footnote{Corresponding author.}}
\author[a]{and Jiahao Zheng}

\affiliation[a]{Department of Physics and Center for Field Theory \& Particle Physics, Fudan University, 20005, Songhu Road, 200438 Shanghai, China}

\affiliation[b]{School of Physics, Sun Yat-Sen University, Guangzhou, Guangdong, China}

\emailAdd{snawata@gmail.com}
\emailAdd{panyw5@mail.sysu.edu.cn}
\emailAdd{azjh1997@gmail.com}
\abstract{We study  2d $\cN=(0,2)$ and $\cN=(0,4)$ theories derived from compactifying class $\cS$ theories on $S^2$ with a topological twist. We present concise expressions for the elliptic genera of both classes of theories, revealing the TQFT structure on Riemann surfaces $C_{g,n}$. Furthermore, our study highlights the relationship between the left-moving sector of the (0,2) theory and the chiral algebra of the 4d $\mathcal{N}=2$ theory. Notably, we propose that the (0,2) elliptic genus of a theory of this class can be expressed as a linear combination of characters of the corresponding chiral algebra.}

\begin{document}
\allowdisplaybreaks

\maketitle

\section{Introduction}

The study of quantum field theory (QFT) and string theory over the years has continually revealed deeper structures and interconnected webs of relationships. At the heart of this intricate web lies the 6d $\mathcal{N} = (2,0)$ superconformal field theory (SCFT), which describes the low-energy dynamics on the worldvolume of M5-branes in M-theory.
The 6d $\mathcal{N} = (2,0)$ SCFT stands out due to its maximal supersymmetry and the highest spacetime dimension that hosts a superconformal algebra. While directly handling the dynamics of 6d $\mathcal{N} = (2,0)$ SCFT is challenging due to the lack of a Lagrangian description, it serves as a central hub from which a plethora of lower-dimensional theories can be derived through compactifications on various manifolds. When M5-branes wrap a certain manifold $M$ with a suitable topological twist, it effectively gives rise to a lower-dimensional QFT $\cT[M]$ associated to the manifold, leading to a rich interplay between geometry and QFT. Starting from Gaiotto's construction \cite{Gaiotto:2009we}, subsequent development along this direction elucidated how various QFTs can be geometrically engineered from M5-branes, revealing a profound geometric structure underlying the space of QFTs.

A large family of 4d $\mathcal{N} = 2$ SCFTs $\cT[C]$ is constructed in \cite{Gaiotto:2009we} by considering M5-branes wrapping Riemann surfaces $C$ with punctures. These theories are collectively known as \emph{theories of class $\mathcal{S}$}. The complex moduli of the surfaces encode the gauge couplings of the SCFTs, and the punctures on the Riemann surface prescribing boundary conditions for the M5-brane determine the flavor symmetry and operator spectrum in the SCFTs. The class $\cS$ construction furthers our understanding of M5-branes by offering a concrete, lower-dimensional perspective on the dynamics of M5-branes, and at the same time offers a geometric viewpoint on the resulting 4d $\mathcal{N} = 2$ theories.

Exact supersymmetric partition functions play an essential role in enhancing this geometric viewpoint. In particular, superconformal indices stand out as simple, yet powerful observables that count BPS states (states that preserve a portion of supersymmetry). Since they are invariant under exactly marginal deformations, they encode crucial information about the Hilbert space even in the strong coupling regime. A series of outstanding works \cite{Gadde:2009kb,Gadde:2010te,Gadde:2011ik,Gadde:2011uv,Buican:2015ina} unveiled a topological quantum field theory (TQFT) structure underlying the superconformal indices of $\cT[C]$ by identifying the indices with correlation functions on the Riemann surfaces $C$. The TQFT description maps various physical manipulations on the 4d theory $\cT[C]$, such as gauging, Higgsing and insertion of non-local operators, to geometrical operations and objects on the corresponding Riemann surface $C$. In this geometric viewpoint, theories of class $\cS$ and its indices can be built by gluing simple building blocks, and generalized S-duality becomes apparent.

A remarkable development in the study of 4d $\cN=2$ SCFTs is the deep connection to 2d chiral algebra/vertex operator algebra (VOA) \cite{Beem:2013sza,Beem:2014rza,Lemos:2014lua,Xie:2019vzr},  referred to as an \emph{SCFT/VOA correspondence}. In \cite{Beem:2013sza}, it is shown that any 4d $\cN=2$ SCFT $\mathcal{T}^{\mathrm{4d}}$ contains a protected subsector consisting of so-called Schur operators restricted on a two-dimensional plane, which furnishes a 2d chiral algebra $\chi(\mathcal{T}^{\mathrm{4d}})$. Two notable examples of Schur operators are the $\SU(2)_R$ current and the flavor moment map operators, which represent the stress-energy tensor and Kac-Moody generators in $\chi(\mathcal{T}^{\mathrm{4d}})$. Through the correspondence, rich 4d physics is reincarnated in 2d context, inspiring deeper understanding and new construction of chiral algebras. In turn, by leveraging the rigidity of chiral algebras, we gain novel perspectives on 4d SCFTs \cite{Lemos:2015orc,Song:2015wta,Buican:2016arp,Xie:2016evu,Beem:2017ooy,Fredrickson:2017yka,Choi:2017nur,Beem:2018duj,Bonetti:2018fqz,Arakawa:2018egx,Arakawa:2023cki}. In particular, the Schur index, which is a special limit of 4d $\mathcal{N} = 2$ superconformal index, gets mapped to the vacuum character of $\chi(\mathcal{T}^{\mathrm{4d}})$. As discussed in \cite{Beem:2017ooy}, the fact that the stress-energy tensor in $\chi(\mathcal{T}^{\mathrm{4d}})$ is not a Higgs branch operator implies the existence of a certain null state (or descendant of null) in the Verma module of $\chi(\cT^{\mathrm{4d}})$. Through Zhu's recursion formula \cite{zhu1996modular,Gaberdiel:2008pr}, such a state translates to a modular differential equation that the Schur index should solve. In fact, additional null states may exist which lead to a set of flavored modular differential equations that all module characters of $\chi(\cT^{\mathrm{4d}})$ must satisfy \cite{Zheng:2022zkm,Pan:2023jjw}, putting stringent constraints on both the chiral algebra and the 4d physics.

In this paper, we push forward this research direction by further compactifying class $\cS$ theories on $S^2$ with a topological twist. The 4d $\cN=2$ SCFTs have an inherent $\SU(2)_R \times \U(1)_r$ $R$-symmetry. A topological twist on $S^2$ using $\U(1)_R \times \U(1)_r$ results in 2d $\cN=(0,2)$ theories \cite{Gadde:2015wta,Cecotti:2015lab}.  On the other hand, a twist with $\U(1)_r$ gives rise to 2d $\cN=(0,4)$ theories \cite{Putrov:2015jpa}. In this work, we provide remarkably simple closed-form expressions of the (0,2) theories analogous to Lagrangian class $\cS$ theories, and those of the (0,4) theories of all class $\cS$ theories of type $A$ with genus $g>0$. Schematically, the elliptic genera for both classes take the form 
\be 
\cI^{\textrm{2d}}_{g,n}= \cH^{g - 1} \prod_{i = 1}^n \mathcal{I}_{\lambda_i}(b_i)~.
\ee 
Here, $\cH$ denotes the contribution from a handle on a Riemann surface $C_{g,n}$, and $\mathcal{I}_{\lambda_i}$ represents the contribution from the $i$-th puncture. Surprisingly, each of these contributions, both $\cH$ and $\mathcal{I}_{\lambda_i}$, can be expressed as a product of theta functions. This structure, as we will demonstrate, naturally unveils the explicit TQFT construction of elliptic genera on Riemann surfaces $C_{g,n}$ for both classes of 2d theories.

Moreover, in the infrared, the left-moving sector of a (0,2) theory from compactifying $\mathcal{T}^{\textrm{4d}}$ exhibits a connection to the associated chiral algebra $\chi(\mathcal{T}^{\textrm{4d}})$. Concretely, we propose that the (0,2) elliptic genus is a linear combination of characters of the chiral algebra $\chi(\mathcal{T}^{\textrm{4d}})$, which can be verified using flavored modular linear differential equations. This relation suggests that the $(0,2)$ theory is endowed with the VOA $\chi(\mathcal{T}^{\textrm{4d}})$ as the infrared symmetry. Techniques for studying VOAs can be applied to gain insights into the Hilbert space and correlation functions of 2d $\cN=(0,2)$ theories at the infrared fixed point.

\bigskip 
This paper is structured as follows. In \S\ref{sec:02}, we explore 2d $\cN=(0,2)$ quiver gauge theory analogous to Lagrangian class $\cS$ theories. Our primary objective is to unveil the duality between these theories and Landau-Ginzburg (LG) models while also exploring their connection to VOAs. In \S\ref{sec:4d2drelation}, the focus is on establishing the relation between 2d $(0,2)$ theories and 4d $\mathcal{N}=2$ SCFTs. We consider a twisted compactification of 4d $\mathcal{N}=2$ SCFTs on $S^2$, which leads to 2d (0,2) quiver gauge theories. We further discuss the connection between the 4d $\mathcal{N}=2$ Schur index and the (0,2) elliptic genera.  Additionally, we put forth a conjecture that the (0,2) elliptic genus can be expressed as a linear combination of characters of the corresponding chiral algebra. 
In \S\ref{sec:02-SU2}, we examine 2D (0,2) $\mathrm{SU}(2) \times \mathrm{U}(1)$ quiver gauge theories. Here, we investigate our proposal on a case-by-case basis, demonstrating that these theories have LG duals. In specific cases, we identify the elliptic genus as a linear combination of characters of the associated VOA. When explicit characters are not available, we check that the elliptic genus solves the modular linear differential equations that constrain the VOA characters. In the end, we demonstrate that the (0,2) elliptic genera exhibit a TQFT structure on Riemann surfaces with a minimal number of U(1) gauge groups. In \S\ref{sec:02-SUN}, we extend the computation to $\mathrm{SU}(N) \times \mathrm{U}(1)$ gauge theories. Lastly, \S\ref{sec:02-nonLag} collects a few remarks on non-Lagrangian theories, providing a perspective on this particular area of study. 

In \S\ref{sec:04}, we study 2d $\mathcal{N} = (0,4)$ theories from another twisted compactification of $A$-type class $\mathcal{S}$ theories on $S^2$. Since theories in this class generally lack Lagrangian descriptions, we make use of the elliptic inversion formula to compute their elliptic genera in \S\ref{sec:04-A2} and \S\ref{sec:04-A3}.  Given that such a $(0,4)$ theory is characterized by a Riemann surface with punctures decorated by embedding $\SU(2) \hookrightarrow \SU(N)$, in \S\ref{sec:04-AN-1}, we propose a Higgsing procedure to derive the contributions to the (0,4) elliptic genus from the puncture data. We will show that the elliptic genera of all these theories can be reorganized as simple products of theta functions, and they exhibit a TQFT structure under the cut-and-join operations on Riemann surfaces. We end this section by commenting on future directions and open problems.

Appendix \ref{app:notation} consolidates the notations and conventions, and introduces definitions of special functions and modular forms used throughout this paper. In Appendix \ref{app:JK}, we revisit the definitions of Jeffrey-Kirwan (JK) residues, given their intricate nature and frequent reference in this paper.
For readers interested in in-depth calculations, Appendix \ref{app:02} offers detailed JK residue computations for (0,2) elliptic genera while Appendix \ref{app:04} provides those for (0,4) elliptic genera.

\section{\texorpdfstring{$\cN=(0,2)$}{N=(0,2)} gauge/Landau-Ginzburg duality and VOAs}\label{sec:02}

In this section, we study 2d $\cN=(0,2)$ quiver gauge theories analogous to 4d $\cN=2$ theories of class $\cS$ \cite{Gaiotto:2009we} with Lagrangian descriptions. We construct these 
(0,2) quiver gauge theories by gauging a basic building block consisting (0,2) chiral multiplets corresponding to a sphere with two maximal punctures and one minimal puncture. This family of 2d $\cN=(0,2)$ theories includes a class of theories obtained a particular twisted compactification on $S^2$, called \cite[Schur-like reductions]{Gadde:2015wta}, of class $\cS$ theories with Lagrangian descriptions. For this subclass, we study the relation between the elliptic genus of a (0,2) theory and characters of the chiral algebra of the corresponding class $\cS$ theory.  Additionally, we demonstrate that, under certain conditions, the (0,2) quiver gauge theories are dual to Landau-Ginzburg models.

\subsection{Relation between 2d (0,2) theories and 4d \texorpdfstring{$\cN=2$}{N=2} SCFTs}\label{sec:4d2drelation}

\subsubsection{4d \texorpdfstring{$\cN=2$}{N=2} SCFT and Schur index}\label{sec:Schur}
A 4d $\mathcal{N}=2$ superconformal theory has the symmetry algebra $\SU(2,2|2)$ is generated by supercharges $(Q_{\alpha}^I,\tilde Q_{\dot\alpha}^I)$, their superconformal partners $(S^{\alpha}_I,\tilde S^{\dot\alpha}_I)$ and other bosonic symmetry generators. 
The 4d $\mathcal{N}=2$ superconformal index counts the 1/8-BPS states that are annihilated by one supercharge and its conformal partner, say $\tilde{Q}_{\dot{-}}^1$ and $\tilde{S}^{\dot{-}}_1$. In other words, it provides a measure of the $\tilde{Q}_{\dot{-}}^1$-cohomology, which consists of states that saturate the bound\footnote{Compared with the notation in \cite{Beem:2013sza}, the supercharges $Q^1_-, \tilde Q^1_{\dot -}$ in this paper are identified with $Q^1_{-}, \tilde Q_{2\dot -}$ there. }
$$
\widetilde\delta_{1\dot{-}}:=\{\tilde{S}^{\dot{-}}_1,\tilde{Q}_{\dot{-}}^1\}=E-2 j_2-2 R+r~.
$$
Here, we use the Cartan generators $\left(E, j_1, j_2, R, r\right)$ of $\SU(2,2 | 2)$. Note that $j_{1,2}$ represents the angular momentum of $\mathrm{SO}(4) \simeq \mathrm{SU}(2)_1 \times \mathrm{SU}(2)_2$, and $(R, r)$ are quantum numbers associated with the $\cN=2$ superconformal $R$-symmetry $\mathrm{SU}(2)_R\times \mathrm{U}(1)_r$. We refer to Table \ref{tab:top-twist} for charges of supercharges under these symmetry groups.
Then, the  4d $\mathcal{N}=2$ superconformal index, denoted as $\cI^{\textrm{4d}}(p, q, t)$, is defined as follows:
\begin{align}
\cI^{\textrm{4d}}(p, q, t) = \Tr(-1)^F\,
e^{-\beta\,\tilde \delta_{{1}\dot{-}}}\,
p^{-j_1+j_2-r}\,
q^{j_1+j_2-r} \,
t^{R+r}\, \prod_{a}z_a^{f_a}\ , 
\end{align}
 The variables $z_a$ correspond to flavor fugacities, and $f_a$ represents flavor charges. Evaluating the 4d $\mathcal{N}=2$ superconformal index can be done using single-letter indices \cite{Gadde:2009kb, Gadde:2010te, Gadde:2011ik, Gadde:2011uv}.

The contribution of the half-hypermultiplet with representation $\lambda$ to the multi-particle index yields the elliptic gamma function \eqref{EGF}, given by
\be
\cI_{\frac12H}^{\textrm{4d}}(z;p,q,t)=\prod_{w\in \lambda}\prod_{i,j=0}^{\infty}\frac{1-z^{-w} p^{i+1} q^{j+1}/\sqrt{t}}{1-z^{w}\sqrt{t} p^i q^j}= \prod_{w\in \lambda}\Gamma(z^w\sqrt{t})~,
\ee
where $w$ runs over the weights of the representation $\lambda$.
The 4d $\mathcal{N}=2$ vector multiplet contributes as follows:
\be
\cI_{\textrm{vec}}^{\textrm{4d}}(z;p,q,t)=\frac{\kappa^{\textrm{rk}G} \Gamma(\tfrac{pq}{t})^{\textrm{rk}G} }{\left|W_G\right|} \prod_{\a \in \Delta} \frac{\Gamma(z^\alpha \frac{p q}{t}) }{\Gamma\left(z^\alpha\right)}~,\qquad\qquad \kappa=(p;p)(q;q)
\ee
where $\Delta$ represents the set of roots associated with the gauge group $G$, and $|W_G|$ is the order of the Weyl group of $G$.
Then, the 4d $\cN=2$ superconformal index of a quiver gauge theory can be schematically expressed as the contour integral 
\begin{equation}\label{SCI-int}
\cI^{\textrm{4d}}(\boldsymbol{a};q)=\oint_{|z|=1} \prod_{\textrm{gauge}} \frac{d z}{2 \pi i z} \cI_{\mathrm{vec}}^{\textrm{4d}}(z;p,q,t) \prod_{\textrm{matter}} \cI_{\frac12H}^{\textrm{4d}}(z;p,q,t)~ .
\end{equation}

The 4d $\cN=2$ superconformal index has various specializations \cite{Gadde:2011ik,Gadde:2011uv}. Among them, the Schur index can be obtained at the specialization of $t=q$, and it counts 1/4 BPS operators consisting of Higgs branch operators annihilated by $(Q_{-}^1, S^{-}_1)$ and $(\tilde{Q}_{\dot{-}}^1,\tilde{S}^{\dot{-}}_1)$.
In the Schur limit, the hypermultiplet contribution is reduced to
\be\label{Schur-limit-hyper}
\cI_{H}^{\textrm{4d}}=\Gamma(z^{\pm} \sqrt{t}):=\Gamma(z\sqrt{t})\Gamma(z^{-1}\sqrt{t})\quad \xrightarrow{t\to q} \quad   \cI_{H}^{\textrm{Schur}}=\frac{\eta(q)}{\vartheta_4(z)}
\ee
where $\eta(q)$ is the Dedekind eta function \eqref{eta}, 
and the vector multiplet contribution is reduced to
\be\label{Schur-limit-vec}
\cI_{\textrm{vec}}^{\textrm{4d}}=\frac{\kappa^{\textrm{rk}G} \Gamma(\tfrac{pq}{t})^{\textrm{rk}G} }{\left|W_G\right|}\prod_{\alpha \in \Delta} \frac{\Gamma\left(z^{\alpha} \frac{pq}{t}\right)}{\Gamma\left(z^\alpha\right)}\quad \xrightarrow{t\to q} \quad \cI_{\textrm{vec}}^{\textrm{Schur}}= \frac{\eta(q)^{2 \,\textrm{rk}G} }{\left|W_G\right|}\prod_{\alpha \in \Delta}  i\frac{\vartheta_1(z^\alpha)}{\eta(q)}~.
\ee

Leveraging the state/operator correspondence, a Schur state can be obtained from a Schur operator $\mathcal{O}(0)$ applied to the vacuum. The Schur operator at the origin (anti-)commutes with the set of the four supercharges. While shifting the operator from this point generally disrupts the BPS condition, one can change the position of the operator across the $\mathbb{R}_{34}^2=\mathbb{C}_{z, \bar{z}}$ plane using the twisted translation proposed in \cite{Beem:2013sza}:
$$
\mathcal{O}(z, \bar{z}):=e^{-z L_{-1}-\bar{z} \widehat{L}_{-1}} \mathcal{O}(0) e^{+z L_{-1}+\bar{z} \widehat{L}_{-1}}a
$$
where
$$
L_{-1}=P_{++}, \quad \widehat{L}_{-1}=P_{--}+R_1^2.
$$
The twisted translated Schur operator $\mathcal{O}(z, \bar{z})$ is also annihilated by the two supercharges
\be\mathbb{Q}_1:=Q_{-}^1+\tilde{S}^{\dot-}_1~, \qquad \mathbb{Q}_2:=\tilde{Q}_{\dot-}^1-S_1^{-}~.\ee 
Moreover, the $\bar{z}$-dependence of $\mathcal{O}(z, \bar{z})$ turns out to be $\mathbb{Q}_{1,2}$-exact. Consequently, at the level of $\mathbb{Q}_{1,2}$-cohomology, the cohomology class $\mathcal{O}(z):=[\mathcal{O}(z, \bar{z})]$ depends on the location holomorphically in $z$. Furthermore, the OPE coefficients of these Schur operators (as cohomology classes) are also holomorphic, forming a 2d vertex operator algebra (VOA)/chiral algebra on the plane $\mathbb{C}_{z, \bar{z}}$ as discussed in \cite{Beem:2013sza}.

The space of Schur operators defines the space of states of the associated VOA, and thus the Schur limit of the superconformal index, which counts the Schur operators with signs, equals the vacuum character of the associated VOA. The associated chiral algebra and the vacuum character are interesting and powerful invariants of 4d $\mathcal{N} = 2$ SCFTs, which capture various aspects of four-dimensional physics. Non-perturbative dynamics in four dimensions can be probed by surface defects \cite{Gukov:2006jk}. To study the relation of the chiral algebra, one can introduce half-BPS surface operators in $\mathcal{T}$ preserving the same nilpotent supercharges $\mathbb{Q}_{1,2}$ where the support of the surface defect transversely intersecting with the chiral algebra plane $\mathbb{C}_{z, \bar z}$ at the origin. From the perspective of the chiral algebra, a such surface defect introduces a non-trivial boundary condition at the origin, and the defect operators are acted on by the Schur operators in the 2d bulk through bulk-defect OPE. Therefore, it is believed that such surface defects introduce non-vacuum modules of the associated chiral algebra.

Similar to the original Schur index, one can count defect operators in the $\mathbb{Q}_{1,2}$ cohomology to obtain the defect Schur index. In general, it is difficult to compute graded dimensions of such operators from the first principle. However, a superconformal index in the presence of a surface defect can be evaluated with suitable manipulations \cite{Gaiotto:2012xa,Gadde:2013dda}. A notable example of the manipulations involves vortex defects, which can be derived using the Higgsing procedure on a 4d $\mathcal{N} = 2$ SCFT \cite{Gaiotto:2012xa}. The vortex defect index can be computed by an appropriate residue computation on the superconformal index of the theory $\mathcal{T}^\text{UV}$. For example, a vortex defect with vorticity $k$ in an $A_1$-type class $\mathcal{S}$ theory $\mathcal{T}_2[C_{g,n}]$ can be computed by (up to some factors $q$)
\begin{equation}
    \mathcal{I}^\text{def}_{g, n}(b_1, \ldots, b_n) = \operatorname{Res}_{b_{n + 1} \to q^{\frac{k + 1}{2}}} \frac{\eta(\tau)^2}{b_{n + 1}} \mathcal{I}_{g, n + 1}(b_1, \ldots, b_{n + 1}) \ .
\end{equation}
Here $b_i$ denote the SU(2) flavor fugacities that are manifest in the class $\mathcal{S}$ construction. 

In particular, the techniques to evaluate the Schur index with a surface defect have been developed to study the relation to the chiral algebra \cite{Pan:2021ulr,Pan:2021mrw,Zheng:2022zkm}.
The original Schur index of $\mathcal{T}$ can be viewed as a supersymmetric partition function on $S^3_{\varphi, \chi, \theta} \times S^1_t$ (with suitable background fields turned on)\footnote{Here the $S^3$ is viewed as a $T^2_{\varphi, \chi}$ fibering over an interval $[0,\pi/2]_\theta$. The points with $\theta = 0$ and $\theta = \pi/2$ form two special tori.}, and it localizes to a multivariate contour integral of an elliptic integrand $\mathcal{Z}(\mathsf{a}_i)$, where the integration variables $\mathsf{a}_i$ capture the holonomy of the dynamical gauge field along the temporal $S^1$ \cite{Pan:2019bor,Dedushenko:2019yiw}. To define a surface defect, one can also specify a BPS singular boundary condition of the gauge fields at the defect plane $\mathbb{R}^2\subset \bR^4$, or equivalently, at a particular $T^2$ in $S^3 \times S^1$. Such a singular background shifts the corresponding integration variables $\mathsf{a}_i \to \mathsf{a}_i + \lambda_i\tau$ where the $\lambda_i$ reflect the singular boundary condition. As the values of $\lambda_i$ vary,  the shifted integration variables eventually cross the integral contour. Consequently, the Schur index with a surface defect is given by integration around a different contour, instead of the unit-circle integral like \eqref{SCI-int}. The resulting defect Schur index is expected to be a linear combination of non-trivial characters of the associated chiral algebra.

\subsubsection{2d (0,2) theory and elliptic genus}\label{sec:EG}
2d $\cN=(0,2)$ supersymmetric field theories have attracted considerable attention due to their importance in theoretical physics and mathematical physics. In a 2d $\cN=(0,2)$ gauge theory, the matter content generically consists of chiral multiplets, Fermi multiplets and vector multiplets. A chiral multiplet $(\psi_{+}, \phi)$ contains a complex scalar field and a right-moving Weyl fermion $\psi_{+}$ while a vector multiplet $(A_\mu, \lambda_{-})$ contains a gauge field $A_\mu$ and gauginos $\lambda_{-},\overline \lambda_{-}$. A Fermi multiplet $(\psi_{-},E(\phi))$ consists of a left-moving Weyl fermion $(\psi_{-})$ and an $E$-term which is a holomorphic function of some chiral multiplets.
For a comprehensive explanation, we refer the reader to \cite{Witten:1993yc}.

In the analysis of 2d supersymmetric theories, a fundamental tool is the elliptic genus \cite{Witten:1986bf}, which counts BPS states protected under RG flow.  Conceptually, the elliptic genus can be understood as a partition function defined on a torus with a complex structure parameter $\tau$, where fermions exhibit periodic boundary conditions along the temporal circle. Moreover, the spatial circle allows for two distinct types of boundary conditions, Ramond and Neveu-Schwarz (NS), both applicable to the left- and right-moving sectors. For the sake of simplicity, we will focus on the Ramond-Ramond and NS-NS sectors, referring to them as the Ramond and NS sectors, respectively.

In the context of  $\cN=(0,2)$ gauge theory, the elliptic genus in the Ramond and NS sectors can be defined respectively by
\begin{align}
\cI^{(0,2)_\textrm{R}}(q,z)=&\operatorname{Tr}_{\mathrm{R}}(-1)^F q^{H_L} \overline{q}^{ H_R} \prod_a z_a^{f_a} \cr 
\cI^{(0,2)_\textrm{NS}}(q,z)=&\operatorname{Tr}_{\mathrm{NS}}(-1)^F q^{H_L}\overline{q}^{({H}_R-\frac{R}{2})} \prod_a z_a^{f_a} 
\end{align}
where the left- and right-moving Hamiltonians are $2 H_L=H+i P$ and $2H_R=H-i P$, respectively, in the Euclidean signature and $R$ represents the $\U(1)_R$ $R$-charge. In a superconformal theory, these operators correspond to the zero-mode generators $L_0$, $\overline{L}_0$, and $\overline J_0$ of the superconformal algebra.

Due to supersymmetry, only right-moving ground states ($H_R=0$) contribute to the elliptic genus in the Ramond sector, while right-moving chiral primary states ($H_R=\frac{R}{2}$) in the right-moving sector contribute to the elliptic genus in the NS sector. Consequently, the elliptic genus in both sector are holomorphic functions of $q$.

For $\cN=(0,2)$ theories described by a Lagrangian, the computation of the elliptic genus depends on the specific details of the gauge theory and its matter content, as outlined in \cite{Benini:2013nda,Benini:2013xpa}. Let us consider the contributions from different types of multiplets:
\begin{description}
    \item [Chiral Multiplet:]
The contribution of an $\cN=(0,2)$ chiral multiplet in a representation $\lambda$ of the gauge and flavor group is
\be \label{chiral}
\cI_{\textrm{chi}}^{(0,2)_\textrm{R}}(q,z)=\prod_{w \in \lambda} i \frac{\eta(q)}{\vartheta_1(z^w)}~,\qquad \cI_{\textrm{chi}}^{(0,2)_\textrm{NS}}(\tau, u)=\prod_{w \in \lambda}  \frac{\eta(q)}{\vartheta_4(q^{\frac{r-1}{2}} z^w)} .
\ee
\item[Fermi Multiplet:]
The contribution of an $\cN=(0,2)$ Fermi multiplet in a representation $\lambda$ of the gauge and flavor group is given by
\be \label{Fermi}
\cI_{\textrm{fer}}^{(0,2)_\textrm{R}}(q,z)=\prod_{w \in \lambda} i \frac{\vartheta_1(z^w)}{\eta(q)} ~,\qquad \cI_{\textrm{fer}}^{(0,2)_\textrm{NS}}(\tau, u)=\prod_{w \in \lambda}  \frac{\vartheta_4(q^{\frac{r}{2}} z^w)}{\eta(q)} .
\ee
\item[Vector Multiplet:] The contribution of an $\cN=(0,2)$ vector multiplet with gauge group $G$ is 
\be \label{Vector}
\cI_{\textrm{vec}}^{(0,2)_{\textrm{R} \mid \textrm{NS}}}(q,z)= \frac{\eta(q)^{2 \,\textrm{rk}G} }{\left|W_G\right|}\prod_{\alpha \in \Delta} i \frac{\vartheta_1(z^\alpha)}{\eta(q)}~.
\ee
\end{description}
Note that the elliptic genera in the NS sector for chiral and Fermi multiplet depend on the $R$-charge $r$ of the multiplet.

Then, the elliptic genus of a quiver gauge theory can be schematically expressed as the Jeffrey-Kirwan (JK) residue integral \cite{jeffrey1995localization,brion1999arrangement,szenes2003toric,Benini:2013xpa}
\begin{equation}\label{EG}
\cI^{(0,2)_{\textrm{R} \mid \textrm{NS}}}=\int_{\textrm{JK}} \prod_{\textrm{gauge}} \frac{d z}{2 \pi i z} \cI_{\mathrm{vec}}^{(0,2)_{\textrm{R} \mid \textrm{NS}}}(q,z) \prod_{\textrm{matter}} \cI_{\textrm{chi}}^{(0,2)_{\textrm{R} \mid \textrm{NS}}}(q,z) \cI_{\textrm{fer}}^{(0,2)_{\textrm{R} \mid \textrm{NS}}}(q,z) .
\end{equation}
In this section, our computation focuses on NS elliptic genera to compare with characters of the associated VOA. Nevertheless, Ramond elliptic genera can be obtained from the NS ones simply by replacing $\vartheta_4$ by $\vartheta_1$.

Since (0,2) theories are chiral theories, one must pay attention to anomalies. 
 Let us consider a (0,2) theory with a global symmetry $F$ described by a simple Lie algebra. The 't Hooft  anomaly coefficient $k_F$ associated with this symmetry can be determined by 
\begin{equation}\label{anomaly1}
\Tr\gamma^3f^af^b = k_F\delta^{ab}~,
\end{equation}
Here, $f^a$ represents the generators of $F$, $\gamma^3$ denotes the gamma matrix that quantifies chirality, and the trace is taken over Weyl fermions in the theory. For a global anomaly, the computation involves evaluating the difference between the sums over the sets of (0,2) chiral and Fermi multiplets:
\begin{equation}\label{anomaly}
k_F = \sum_{\Phi \in \text{(0,2) chiral}} T(R^\Phi_{F}) - \sum_{\Psi \in \text{(0,2) Fermi}} T(R^\Psi_{F}),
\end{equation}
where $T(R_F)$ represents the index of the representation $R_F$ of $F$. Note that the (0,2) supermultiplet of the gauge invariant field strength can be treated as a Fermi multiplet, and the gaugino contributes to the anomaly if it is charged under $F$. In this paper, we focus solely on $\SU(N)$ groups as instances of non-Abelian symmetries. For these groups, the indices are given by $T(\square) = \frac{1}{2}$ and $T(\mathbf{adj}) = N$.
 In cases where the theory possesses two $\mathrm{U}(1)$ symmetries, $\mathrm{U}(1)_1$ and $\mathrm{U}(1)_2$, with charges $f_1$ and $f_2$ respectively, a mixed 't Hooft anomaly can emerge:
\begin{equation}
k_{12} = \Tr(\gamma^3 f_1 f_2)~.
\end{equation}
Certainly, any gauge anomaly must vanish for the theory to be well-defined.

In particular, the anomaly associated with the $\U(1)_R$-symmetry is related to the right-moving central charge $c_R$ as
\begin{equation}\label{cR}
c_R = 3\Tr(\gamma^3R^2)~.
\end{equation}
To determine $\U(1)_R$-charges of various fields, the $c$-extremization is performed \cite{Benini:2012cz,Benini:2013cda} if the theory meets the following two assumptions:
\begin{enumerate}[nosep]
    \item the theory is bounded, and the energy spectrum is bounded from below, 
    \item the classical vacuum moduli space is normalizable.
\end{enumerate}
Once $c_R$ is determined, the left-moving central charge can be obtained from the gravitational anomaly, which is the difference between the number of chiral and Fermi multiplets
\begin{equation}\label{grav}
c_R - c_L = \Tr(\gamma^3)~.
\end{equation}
 
The presence of anomalies can be observed directly at the level of (0,2) elliptic genera \cite{Putrov:2015jpa}. Let us focus on an $\SU(N)$ global symmetry whose fugacities are denoted by $b_{1,\ldots,N}$ with $\prod_{i=1}^N b_i=1$ in an elliptic genus. Then, the corresponding 't Hooft anomaly can be seen in the shift of the elliptic genus as 
\begin{equation}\label{thooft-anomaly}
\cI^{(0,2)}(\boldsymbol{b})\to\left(q b_i / b_j\right)^{2 k_F} \cI^{(0,2)}(\boldsymbol{b}) \qquad \textrm{as } \quad b_i \rightarrow q b_i, b_j \rightarrow b_j / q~.
\end{equation}
For a U(1) 't Hooft anomaly, the elliptic genus behaves as
\be 
\cI^{(0,2)}(c)\to (-q^{1/2}c)^{k_F} \cI^{(0,2)}(c) \qquad \textrm{as } \quad c\to qc~,
\ee 
where $c$ is the corresponding U(1) fugacity.
In particular, for a theory to be gauge anomaly-free, the integrand of its elliptic genus \eqref{EG} must be invariant under shifts of the gauge fugacities $z_i \rightarrow q z_i$.

\subsubsection{Twisted compactifications of 4d \texorpdfstring{$\cN=2$}{N=2} SCFTs on \texorpdfstring{$S^2$}{S2}}\label{sec:twist}
A 2d $\cN=(0,2)$ theory can be obtained from a 4d $\mathcal{N} = 1$ gauge theory, and in particular, a 4d $\mathcal{N} = 2$ Lagrangian SCFT, by a compactification on $S^2$ with a certain topological twist. Such a reduction is referred to as Schur-like reduction in \cite{Gadde:2015wta}. (See also \cite[\S5]{Cecotti:2015lab}.)

The first explicit supersymmetric localization of 4d $\cN=1$ theories on $T^2\times S^2$ was carried out in \cite{Closset:2013sxa}. Subsequently, the exploration of its relationship with the 2d $\cN=(0,2)$ elliptic genera was undertaken in the vicinity of the same period  \cite{Benini:2015noa,Honda:2015yha,Putrov:2015jpa,Gadde:2015wta,Cecotti:2015lab}. In this context, let us summarize the key aspects involved in the twisted compactification of 4d $\cN=1$ theories on $S^2$.

\begin{table}[ht]\centering
$$\begin{array}{c|ccccc|cc|cc}
 & \SU(2)_1 & \SU(2)_2 & \SU(2)_R & \U(1)_r  & \U(1)_f & \U(1)_{T^2} & \U(1)_{S^2} & \U(1)^{(0,2)} & \U(1)^{(0,4)}  \\
\hline \hline 
Q_{-}^1 & -\frac{1}{2} & 0 & \frac{1}{2}& 1&0& -1& -1& 0 & 0 \\
Q_{+}^1 & \frac{1}{2} & 0 & \frac{1}{2}& 1&0& 1& 1& 2 & 2 \\
Q_{-}^2 & -\frac{1}{2} & 0 & -\frac{1}{2}& 1&0& -1& -1& -1 & 0  \\
Q_{+}^2 & \frac{1}{2} & 0 & -\frac{1}{2}& 1&0& 1& 1& 1 & 2  \\
\tilde{Q}_{\dot-}^1 & 0 & -\frac{1}{2} & \frac{1}{2}& -1&0& -1& 1& 1 & 0  \\
\tilde{Q}_{\dot+}^1 & 0 & \frac{1}{2} & \frac{1}{2}& -1&0& 1& -1& -1 & -2 \\
\tilde{Q}_{\dot-}^2 & 0 & -\frac{1}{2} & -\frac{1}{2}& -1&0& -1& 1& 0 & 0 \\
\tilde{Q}_{\dot+}^2 & 0 & \frac{1}{2} & -\frac{1}{2}& -1&0& 1& -1& -2 & -2 \\
\hline 
q  & 0  & 0 & \frac{1}{2} &0&1&0&0&0&0\\ 
\tilde q  & 0  & 0 &\frac{1}{2}&0&-1&0&0&1&0\\ 
\Phi  & 0  & 0 &0&2&0&0&0&1&2\\ 
\end{array}$$
\caption{Symmetries of 4d $\cN=2$ supercharges and fields. The 4d $\cN=1$ chirals $(q,\tilde q)$ constitute an $\cN=2$ hypermultiplets and $\Phi$ represents the $\cN=1$ adjoint chiral in an $\cN=2$ vector multiplet.  The fifth column denotes the $\U(1)_f$ flavor symmetry that distinguishes $q$ and $\tilde q$. The topological twist of $\U(1)_{S^2}$ with $\U(1)_{R+\frac12(r-f)}$ results in the $\cN=(0,2)$ supersymmetry wheareas the twist with $\U(1)_r$ yields the $\cN=(0,4)$ supersymmetry.}
\label{tab:top-twist}
\end{table}

First of all,  the generic holonomy group $\Spin(4)=\SU(2)_1\times \SU(2)_2$ reduces to $\U(1)_{T^2}\times \U(1)_{S^2}$ on the background $T^2\times S^2$ where $\U(1)_{T^2}$ (resp. $\U(1)_{S^2}$) is the U(1) subgroup of the diagonal (resp. anti-diagonal) subgroup of $\SU(2)_1\times \SU(2)_2$. In order to define covariant constant supercharges on $S^2$, we perform a topological twist of  $\U(1)_{S^2}$ along with a global $\U(1)_\frakR$-symmetry. Additionally, in a Lagrangian theory, the $\U(1)_\frakR$ charge of a 4d $\cN=1$ chiral multiplet must be an integer to ensure a well-defined compactification on $S^2$.

Under this compactification, a 4d $\cN=1$ chiral multiplet with  $\U(1)_\frakR$-charge $\mathfrak{r}$ decomposes into $(1-\mathfrak{r})$ (0,2) chiral multiplets if $\mathfrak{r}<1$, or $(\mathfrak{r}-1)$ (0,2) Fermi multiplets if $\mathfrak{r}>1$. When $\mathfrak{r}=1$, the 4d chiral multiplet does not contribute to the 2d theory.

In general, the $T^2\times S^2$ partition function involves the summation of magnetic fluxes of gauge fields. Nevertheless, if the $\U(1)_\frakR$-charges of all chiral multiplets are non-negative, then the contributions from all non-zero flux sectors vanish, and only the zero flux contribution remains.

The $R$-symmetry of 4d $\cN=2$ superconformal theory is $\SU(2)_R\times \U(1)_r$. To perform the twisted compactification above, we treat the theory as a $\mathcal{N} = 1$ theory by selecting a $\U(1)_\mathfrak{R} \subset \SU(2)_R\times \U(1)_r$. 
Let us consider the choice $\frakR=R+\frac{r}{2}$. As in Table \ref{tab:top-twist}, only the supercharges $Q_{-}^1,\tilde{Q}_{\dot-}^2$ are neutral under $\mathfrak{R} + 2(j_1 - j_2)$ and therefore survive under this twist. These two supercharges share the same $\U(1)_{T^2}$ charge $2(j_1 + j_2) = -1$, hence this twist leads to an ${\cal N}=(0,2)$ supersymmetry. The symmetry $\frakR=R+\frac{r}{2}$ is referred to as the 2d (0,2) $\U(1)_R$-symmetry.

With the choice of $\mathfrak{R}$, the adjoint chiral $\Phi$ in an $\cN=2$ vector multiplet has $\mathfrak{r}=1$, which does not contribute to the 2d theory. Consequently, an ${\cal N}=2$ vector multiplet simply reduces to a (0,2) vector multiplet. On the other hand, the hypermultiplet ($q,\tilde q$) is assigned a fractional charge $\frakr=\frac{1}{2}$. In order for their $\frakR$-charges to be an integer, we further twist with the flavor symmetry $\U(1)_f$, which acts on the two chirals ($q,\tilde q$) with opposite charges. Under the resulting  $\frakR=R+\frac{1}{2}(r-f)$, the scalars $q$ and $\tilde{q}$ acquire charges $\frakr=0$ and $\frakr=1$, respectively. Hence, an $\cN=2$ hypermultiplet reduces to a (0,2) chiral multiplet in the representation of $q$. (See Table \ref{tab:top-twist}.) In this paper, the above process of compactification on $S^2$ with the topological twist is referred to as the \emph{(0,2) reduction}.

For a Lagrangian theory, an interesting observation emerges when comparing the Schur limit \eqref{Schur-limit-hyper} and \eqref{Schur-limit-vec} with the contributions of (0,2) chirals \eqref{chiral} and vector multiplet \eqref{Vector} to the elliptic genus. The integrand of the elliptic genus for the 2d (0,2) theory after the reduction coincides with that of the Schur index of the original 4d $\mathcal{N}=2$ SCFT, upon suitable shifts of U(1) flavor fugacities. However, it is important to note that the computation of the (0,2) elliptic genus involves the JK residue integral whereas the Schur index is evaluated by a contour integral along the unit circles.  We will provide a more detailed account of this comparison in the subsequent discussion.

\subsubsection{Relation to VOAs}

In the following analysis, we explore the 2d (0,2) quiver gauge theories obtained by gauging the free 2d (0,2) theories associated with the three-punctured spheres. This includes, as a subclass, the theories from the (0,2) reduction of Lagrangian class $\mathcal{S}$ theories.  We will demonstrate that under a certain condition, surprisingly, these quiver theories including the reduced class $\cS$ theories are dual to Landau-Ginzburg (LG) models.

Let us discuss the central charges of the (0,2) reduction of Lagrangian class $\cS$ theories. As illustrated below \eqref{cR}, if the two assumptions are satisfied, the central charges are determined by the $c$-extremization. However, the vacuum moduli space of the (0,2) reduction of a class $
\cS$ theory is non-compact, so the second assumption of the $c$-extremization is violated. Specifically, in such situations, a flavor current related to a non-compact direction is non-holomorphic so that it does not mix with the $R$-symmetry current \cite{Benini:2013cda,Sacchi:2020pet}. Consequently, the naive application of the $c$-extremization to a (0,2) theory of this class can yield negative central charges. Nonetheless, the left-moving central charge turns out to coincide with that of the VOA of the parent class $\cS$ theory \cite[\S5.2]{Cecotti:2015lab}
\be 
c_L^{\textrm{naive}}=-12 c_{4d}~,
\ee 
if we assign the ``wrong'' $\U(1)_R$-charges to chiral multiplets as a result of the naive application of the $c$-extremization
\be \label{wrong}
r_\Phi=1~.
\ee 
To get the physical central charge, we must enforce no mixing with any non-holomorphic flavor current arising from non-compact directions in the moduli space. At a practical level, we assign zero $\U(1)_R$-charges 
\be\label{correct} r_\Phi=0 \ee
to the chiral multiplets which parameterize the non-compact directions.  This procedure yields the correct positive central charges where the left-moving $c_L$ is shifted from $c_L^\textrm{naive}$ of the VOA associated to the 4d theory \cite{Cecotti:2015lab,Eager:2019zrc} by
\be \label{c-shift}
c_L=c_L^{\textrm{naive}}+3n_h=c_L^{\textrm{naive}}+12(5 c^{4 d}-4 a^{4 d})~,
\ee 
where $n_h$ represents the number of hypermultiplets and $ a^{4 d},c^{4 d}$ are the anomaly coefficients in the corresponding class $\cS$ theory. This
suggests that the (0,2) theory flows to an SCFT fixed point despite the non-compact moduli space. 

Exploring 2d (0,2) theories at the IR fixed point through the viewpoint of associated VOA is of significant importance because the BPS operators in the (0,2) IR CFT constitute a VOA. As elucidated in \cite{Cecotti:2015lab,Dedushenko:2017osi}, the VOA associated with a (0,2) theory of this type arises from a BRST reduction of the $bc\beta\gamma$ system at zero gauge coupling, where the $\beta \gamma$ systems in the corresponding gauge representations come from the free $(0,2)$ chiral multiplets, and small $b c$ ghosts in the adjoint of the gauge group from the free vector multiplets. The conformal weights of an involved $\beta \gamma$ system are $\left(h_\beta, h_\gamma\right) = (1-\lambda, \lambda)$ where $\lambda$ is related to the correct $R$-charge assignment of the IR CFT. The state space of the resulting VOA is independent of the parameter $\lambda$, but the parameter affects the stress-energy tensor $T$ of the (0,2) IR CFT. In fact, the shift \eqref{c-shift} in central charge can be traced back to the difference between the stress-energy tensor $T$ and $T^\chi$ of the VOA associated to the 4d theory by the derivative of the $\U(1)$ flavor current $J$ of the $\beta \gamma$ system, schematically \cite{Dedushenko:2017osi,Eager:2019zrc}:
\begin{equation}
T=T^{\chi}+\left(\frac12-\lambda\right)\partial J~.
\end{equation}
Therefore, the deviation from $\lambda=\frac12$ reflects the non-compact nature of the vacuum moduli space, which invalidates the naive application of the $c$-extremization.

From our previous discussions, we have observed that the integrand of the (0,2) elliptic genus of the reduction of 4d $\cN=2$ Lagrangian SCFT agrees with the integrand of the Schur index (upon appropriate redefinitions of U(1) flavor fugacities), albeit with distinct contour choices. As discussed at the end of \S\ref{sec:Schur}, a different contour of the integrand is expected to provide the Shcur index with a surface defect, which is intrinsically tied to non-vacuum characters of the corresponding VOA $\chi(\mathcal{T}^\text{4d})$. Hence, we investigate the relationship between the (0,2) reduction of the class $\mathcal{S}$ theory and the corresponding VOA from this perspective.

There are two primary tools for us to investigate this relation: the elliptic genus \cite{Benini:2013nda,Benini:2013xpa} and modular (linear) differential equations \cite{Mathur:1988gt,Mathur:1988na,Pan:2023jjw}. Combining these tools, we present an intriguing conjecture proposing that the NS elliptic genus of the (0,2) reduction of a Lagrangian class $\mathcal{S}$ can be expressed as a linear combination of characters $\ch^{\chi(\cT_N[C_{g,n}])}_{\lambda_i}$ of the corresponding VOA:
\be \label{sum-characters}
\cI^{(0,2),N}_{g,n}=\sum_i a_i \ch^{\chi(\cT_N[C_{g,n}])}_{\lambda_i}
\ee 
where $a_i\in \bQ$, and $\lambda_i$ represent the highest weights of representations of $\chi(\cT_N[C_{g,n}])$. We remark that the U(1) flavor fugacities in the elliptic genus need to be appropriately redefined to precisely align with the characters of the VOA. Moreover, since the theory is dual to an LG model, the elliptic genus can be simply expressed as a product of theta and eta functions which can be viewed as free field characters of suitable 2d $bc \beta \gamma$ systems. Consequently, it forms a Jacobi form with its index determined by the 't Hooft anomaly of the global symmetry.

\subsection{SU(2)\texorpdfstring{$\times$}{x}U(1) gauge theories, LG duals and VOAs}\label{sec:02-SU2}

\begin{figure}[ht]
    \centering
    \includegraphics[width=4cm]{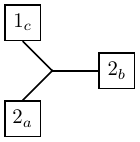}
    \caption{Basic building block $U_2$ for SU(2) theory}
    \label{fig:SU2-trinion}
\end{figure}

In this subsection, we consider 2d $\mathcal{N} = (0,2)$ gauge theories with $\SU(2)$ and $\U(1)$ gauge groups. Analogous to the 4d $\mathcal{N} = 2$ superconformal theories of class $\mathcal{S}$, the 2d theories we consider will have the building blocks depicted in Figure \ref{fig:SU2-trinion}.

The basic building block in class $\cS$ is the theory corresponding to a three-punctured sphere $C_{0,3}$.  In the case of type $A_1$, the 4d $\mathcal{N}=2$ theory $T_2$ is a free theory of 8 half-hypermultiplets with the flavor symmetry $\SU(2)_a\times \SU(2)_b\times \SU(2)_c$. (See the left of Figure \ref{fig:04-block}.)  We use $\U(1)_c\subset \SU(2)_c$ for the topological twist on $S^2$, and the (0,2) reduction of $T_2$ using this flavor symmetry leads to the theory of free (0,2) chiral multiplets in the representation $(\boldsymbol{2}, \boldsymbol{2},1)$ of $\SU(2)_a\times \SU(2)_b\times \U(1)_c$ flavor symmetry. This (0,2) theory, akin to $T_2$, serves as the fundamental building block and is denoted as $U_2$. In quiver notation, we will depict this theory as a vertex with three external legs corresponding to $\SU(2)_a\times \SU(2)_b\times \U(1)_c$  flavor groups (see Figure \ref{fig:SU2-trinion}).

It follows from \eqref{chiral} that  the NS elliptic genus of the $U_2$ theory is 
\be \label{U2}
    \mathcal{I}_{U_2}^{(0,2)}(a, b;c) =  \frac{\eta(q)}{\vartheta_4(q^{-\frac12} a^{ \pm 1} b^{ \pm 1}\tilde c)} =  \frac{\eta(q)}{\vartheta_4( a^{ \pm 1} b^{ \pm 1}c)}~.
\ee
As emphasized in \eqref{correct}, the $\U(1)_R$-charge of the (0,2) chiral multiplet is zero $r=0$, but we shift the $\U(1)_c$ flavor fugacity by $c=q^{-\frac12}\tilde c$. Note that the factors with repeated sign $\pm$ in the arguments are all multiplied. (See \eqref{signs}.) On the other hand, 
as seen in \eqref{Schur-limit-hyper}, the Schur limit of the superconformal index of the $T_2$ theory is given by \cite{Gadde:2009kb}
\be 
\cI_{T_2}^{\textrm{4d}}=\Gamma(\sqrt{t}  a^{ \pm 1} b^{ \pm 1} c^{ \pm 1})\quad \xrightarrow{t\to q}\quad   \cI_{T_2}^{\textrm{Schur}}= \frac{\eta(q)}{\vartheta_4( a^{ \pm 1} b^{ \pm 1}c)}
\ee 
Therefore, the redefinition of the $\U(1)_c$ flavor fugacity in \eqref{U2} ensures that the elliptic genus of the $U_2$ theory coincides with the Schur index of the $T_2$ theory. In this way, the elliptic genus can be decomposed into characters of the corresponding VOA.

Moreover, the contribution of a vector multiplet is the same in both the Schur index \eqref{Schur-limit-vec} and the elliptic genus \eqref{Vector}. The $\SU(2)$ vector multiplet contribution is
\be \label{SU2-vector}
\mathcal{I}_{\textrm{vec}}^{(0,2)} (a)=-\frac{\vartheta_1(a^{\pm2})}{2}~,
\ee 
and the $\SU(2)$ gauging leads to no gauge anomaly.

As a Riemann surface $C_{g,n}$ can be constructed by gluing pants, the corresponding 4d $\cN=2$ theory $\mathcal{T}_2[C_{g,n}]$ can be obtained by gauging the $T_2$ theories. However, as observed, the distinction between the 4d $T_2$ theory and 2d $U_2$ theory lies in the global symmetry. To construct (0,2) quiver gauge theories using the $U_2$ building block, we will outline the U(1) gauging method.

\begin{figure}[ht]
    \centering
    \includegraphics[width=13cm]{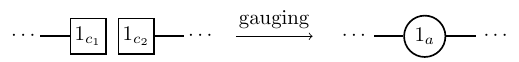}
    \caption{U(1) gauging}
    \label{fig:U1-gauging}
\end{figure}

Given a $\U(1)_{c_1}\times \U(1)_{c_2}$ global symmetry, we gauge the anti-diagonal part $\sfa=(\sfc_1-\sfc_2)/2$ with keeping the diagonal $\sfd=(\sfc_1+\sfc_2)/2$  as a global symmetry. To cancel gauge anomaly during the gauging process, we include Fermi multiplets with $\U(1)_a$ gauge charges $\pm2$. The $\U(1)_R$-charge for these Fermi multiplets is taken to be $r=0$, a value determined by the $c$-extremization.  At the level of the elliptic genus, the gauging procedure is given by:
\bea \label{U1-gauging}
\frac{\eta(q)}{\vartheta_4(c_1\cdots )} \frac{\eta(q)}{\vartheta_4(c_2\cdots )} \quad \to \quad &  \eta(q)^2\int_{\textrm{JK}}\frac{d a}{2 \pi i a} \frac{\eta(q)}{\vartheta_4( da\cdots )} \frac{\eta(q)}{\vartheta_4( d a^{-1}\cdots )}\frac{\vartheta_4(a^{\pm2})}{\eta(q)^2}\cr
\eea 
The details of U(1) gauging will be seen in  examples in \S\ref{sec:SU2-genus1-puncture2}, \S\ref{sec:SU2-genus2} and below.

Through U(1) gauging, one can construct a family of (0,2) quiver gauge theories. However, when a U(1) gauge group is involved, a (0,2) theory is no longer the (0,2) reduction of a class $\cS$ theory in general.  Nonetheless, as we will demonstrate, (0,2) theories of genus $g>0$ with $(g-1)$ $\U(1)$ gauge groups are dual to each other, making them frame-independent. We will further propose that the (0,2) reduction of the class $\cS$ theory $\cT_2[C_{g>0,n}]$ of type $A_1$ on $S^2$ is closely related to the corresponding (0,2) quiver theory of genus $g$ with $(g-1)$ $\U(1)$ gauge groups. Notably, if one replaces $\vartheta_4(a^{\pm2})$ by $\vartheta_1(a^{\pm2})$ in the U(1) gauging, then the integrand of the elliptic genus is the same as that of the corresponding Schur index (up to a factor of $2^{g-1}$), though their integration contours differ; the elliptic genus uses the JK prescription while the Schur index uses the maximal tori of the gauge groups. Additionally, the (0,2) theory of this class turns out to be dual to an LG model.

Furthermore, we explore generalized quiver gauge theories by gluing the $U_2$ theories, extending the (0,2) reduction of the class $\mathcal{S}$ theories.  Specifically, we demonstrate that a quiver gauge theory with $g$ loops and $g$ U(1) gauge nodes is dual to an LG model.

\subsubsection{SU(2) SQCD}

\begin{figure}[ht]
    \centering
    \includegraphics[width=4.5cm]{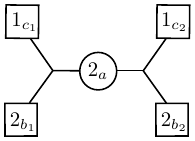}
    \caption{Quiver diagram for SU(2) SQCD with four fundamental chirals. In the diagram, a circle node represents a gauge group, and square nodes denote flavor groups. The number $N$ within a node means $\mathrm{SU}(N)$ for $N>1$, and $\mathrm{U}(1)$ for $N=1$.}
    \label{fig:SU2-SQCD}
\end{figure}

First, let us consider the SU(2) gauge theory with four fundamental chiral multiplets, whose quiver diagram is presented in Figure \ref{fig:SU2-SQCD}. As described in \cite{Sacchi:2020pet}, the naive application of the $c$-extremization leads to the ``wrong'' central charges of the theory 
\be \label{cc-SU2-SQCD-wrong}
c_L^{\textrm{naive}}=-14~, \qquad c_R^{\textrm{naive}}=-9~,
\ee 
while the enforcement of the zero $R$-charge to the chiral multiplets results in the ``correct'' positive central charges
\be \label{cc-SU2-SQCD}
c_L=10~, \qquad c_R=15~.
\ee 
Nevertheless, the ``wrong'' left-moving central charge $c_L^{\textrm{naive}}$ is equal to the central charge of the corresponding VOA $\frakso(8)_{-2}$

This theory was first studied in \cite{Putrov:2015jpa} and its elliptic genus is evaluated there as 
\bea\label{EG-SU2-SQCD}
\cI^{(0,2),2}_{0,4}=&\int_{\JK}\frac{da}{2\pi ia}    \mathcal{I}_{U_2}^{(0,2)}(b_1,a;c_1)  \mathcal{I}_{\textrm{vec}}^{(0,2)} (a) \mathcal{I}_{U_2}^{(0,2)}(b_2,a;c_2)~\cr 
=&\frac{\eta(q)^5\vartheta_1(c_1^2c_2^2)}{\vartheta_1(c_1^2)\vartheta_1(c_2^2) \vartheta_1(c_1c_2b_1^\pm b_2^\pm)}~.
\eea 
As pointed out in \cite{Putrov:2015jpa,Gadde:2015wta,Dedushenko:2017osi,Sacchi:2020pet}, the theory exhibits an LG dual description.  This dual description involves two chiral multiplets $\Phi_{1,2}$, and one chiral meson multiplet $\wt\Phi_{i,j=1,2}$ with $\U(1)_R$-charge of 0, as well as a Fermi multiplet $\Psi$ with $\U(1)_R$-charge of 1, and they form a $J$-type superpotential
$$
W=\Psi (\Phi_1\Phi_2+\operatorname{det} \wt\Phi )~.
$$
The six chiral multiplets $\Phi$ can be regarded as the $\wedge^2 \boldsymbol{4}=\boldsymbol{6}$ representation of $\SU(4)$ so that the superpotential can be expressed as $W=\Psi \operatorname{Pf}(\Phi)$.
Then, the expression \eqref{EG-SU2-SQCD} can be naturally understood as the elliptic genus of the LG model.

More remarkably, it has been revealed in \cite{Dedushenko:2017osi,Eager:2019zrc} that the space of BPS states in the SU(2) SQCD has a relation to the VOA $\frakso(8)_{-2}$, which is the VOA $\chi(\cT_2[C_{0,4}])$ of the corresponding 4d $\cN=2$ theory \cite{Beem:2013sza}. 
Concretely, it was found \cite{Eager:2019zrc} that the elliptic genus \eqref{EG-SU2-SQCD} can be written as a linear combination of characters of $\frakso(8)_{-2}$
\begin{equation}
\cI^{(0,2),2}_{0,4}(q,b,c)=\ch_0^{\frakso(8)_{-2}}(q,b,c)-\ch_{-2 \omega_4}^{\frakso(8)_{-2}}(q,b,c)~.
\end{equation}
where $\ch_0$ is the vacuum character and $\ch_{-2 \omega_4}$ is one of the  three non-vacuum characters of $\frakso(8)_{-2}$. Therefore, the space of BPS states furnishes a non-trivial representation of the associated chiral algebra $\chi(\mathcal{T}_2[C_{0,4}])$.

The VOA $\mathfrak{so}(8)_{-2}$ contains many null states in its vacuum module. For example, the Sugawara condition $T - T_\text{Sug} = 0$ is a trivial null state, simply stating that the stress-energy tensor $T$ is given by the Sugawara stress-energy tensor. The Sugawara condition is part of the so-called Joseph relations
\begin{equation}
    (J^A J^B)|_\mathfrak{R} = 0, \qquad
    (J^A J^B)_{\boldsymbol{1}} \sim T \ ,
\end{equation}
which correspond to more null states or descendants of null. Using Zhu's recursion relations \cite{zhu1996modular}, the null states (or their descendants) that are uncharged under the Cartan of $\mathfrak{so}(8)$ may lead to flavored modular differential equations that any $\mathfrak{so}(8)_{-2}$-character must satisfy. Concretely, there are 10 equations of weight-two, 4 equations of weight-three and 1 equation of weight-four that together constrain the characters of $\mathfrak{so}(8)_{-2}$ \cite{Zheng:2022zkm,Pan:2023jjw}. In particular, the above elliptic genus $\cI^{(0,2),2}_{0,4}(q,b,c)$ is a linear combination of characters, and therefore a solution to the set of equations. Reversing the logic, the fact that $\cI^{(0,2),2}_{0,4}(q,b,c)$ solves the set of modular differential equations predicts that it must be some linear combination of VOA characters.

\subsubsection{SU(2) linear quivers}

\begin{figure}[ht]
    \centering
    \includegraphics[width=0.95\textwidth]{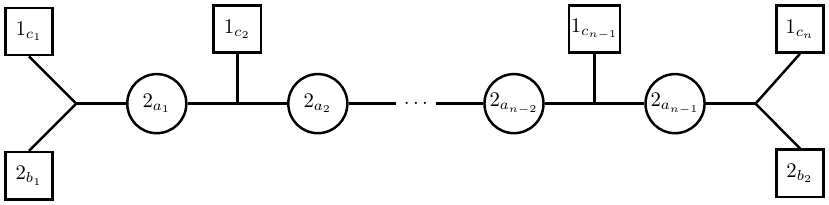}
    \caption{SU(2) linear quiver}
    \label{fig:SU2-linear-quiver}
\end{figure}

Consider the SU(2) linear quiver theory with $(n-1)$ SU(2) gauge nodes as in Figure \ref{fig:SU2-linear-quiver}. The theory has manifest flavor symmetry $\SU(2)^2 \times \U(1)^n$. Its central charge is given by 
\be 
c_L=2(n+3) ~,\qquad  c_R=3(n+3) ~.
\ee 
As explained in \cite[Appendix D]{Putrov:2015jpa}, \eqref{EG-SU2-SQCD} can be interpreted as the elliptic inversion formula:
\be \label{SU2-EI}
\frac{\eta(q)^5\vartheta_1(c_1^2c_2^2)}{\vartheta_1(c_1^2)\vartheta_1(c_2^2) \vartheta_1(c_1c_2b_1^\pm b_2^\pm)}=\int_{\textrm{JK}} \frac{da}{2\pi i a}\frac{\eta(q)^8\vartheta_1(a^{\pm2})}{2\vartheta_4(c_1b_1^\pm a^\pm)\vartheta_4(c_2b_2^\pm a^\pm)}~.
\ee 
Therefore, starting from a collection of $U_2$ theories we can repeatedly gauge the $\SU(2)$ flavor symmetries to construct a linear quiver theory, and the elliptic genus takes the following simple form:
\bea
\label{EG-SU2-linear}
&\cI^{(0,2),2}_{0,n+2}\cr 
=&\int_{\JK}\frac{d\boldsymbol{a}}{2\pi i\boldsymbol{a}}    \mathcal{I}_{U_2}^{(0,2)}(b_1,a_1;c_1) \mathcal{I}_{\textrm{vec}}^{(0,2)} (a_{1}) \mathcal{I}_{U_2}^{(0,2)}(b_2,a_{n-1};c_n)\prod_{i=1}^{n-2} \mathcal{I}_{U_2}^{(0,2)}(a_i,a_{i+1};c_{i+1})  \mathcal{I}_{\textrm{vec}}^{(0,2)} (a_{i+1}) ~\cr 
=&\frac{\eta(q)^{n+3}\vartheta_1(\prod_{i=1}^n c_i^2)}{\vartheta_\alpha(b_1^\pm b_2^\pm\prod_{i=1}^n c_i)\prod_{i=1}^n \vartheta_1(c_i^2) }~, \qquad \begin{cases}\alpha=1 & n \textrm{ even} \\ \alpha=4 & n \textrm{ odd} \end{cases}.
\eea 
As also found earlier in \cite[\S4.3]{Sacchi:2020pet}, the above linear quiver theory has an LG description which consists of $n$ chiral multiplets $\Phi_{k=1,\ldots,n}$ and one chiral meson multiplet $\wt\Phi_{i, j=1,2}$  with $\U(1)_R$-charge $r=0$, and one Fermi multiplet $\Psi$ with $\U(1)_R$-charge $r=1$, forming a $J$-type superpotential
$$
W = 
    \Psi (\prod_{i=1}^n\Phi_i+\operatorname{det} \wt\Phi) ~.
$$
It is worth mentioning that with the shift of the $\U(1)_{c_i}$ flavor fugacities in \eqref{U2}, one must take this shift into account to correctly read off the $\U(1)_R$-charges of the superfields in the LG model from \eqref{EG-SU2-linear}.

\subsubsection{Genus one with one puncture and SU(2) with adjoint chiral}\label{sec:SU2-adjoint}

\begin{figure}[ht]
    \centering
    \includegraphics[width=4cm]{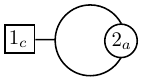}
    \caption{Genus one with one puncture for SU(2)}
    \label{fig:SU2-genus1-1}
\end{figure}

Let us consider the theory corresponding to the Riemann surface of genus one with one $\U(1)$-puncture. Because of $\boldsymbol{2}\otimes\boldsymbol{2}=\boldsymbol{3}\oplus \boldsymbol{1}$, the theory is a product of an $\SU(2)$ gauge theory with one adjoint chiral and a free chiral. 
The elliptic genus can be evaluated as
\bea\label{EG-genus1-puncture1}
\cI^{(0,2),2}_{1,1}=&\int_{\JK}\frac{da}{2\pi ia}    \mathcal{I}_{U_2}^{(0,2)}(a,a^{-1};c_1)  \mathcal{I}_{\textrm{vec}}^{(0,2)} (a)~\cr 
=&\frac{\eta(q)}{\vartheta_1(c_1^2 )}=\frac{\eta(q)}{\vartheta_4(c_1)}\cdot \frac{\vartheta_4(c_1)}{\vartheta_1(c_1^2)}~.
\eea 
The first factor is the contribution of the free chiral, which is the vacuum character of the $\beta\gamma$-ghost, while the second factor is the elliptic genus of the $\SU(2)$ theory with one adjoint chiral. Following this computation, the SU(2) gauge theory with one adjoint chiral has an LG dual given by one chiral and one Fermi multiple. From the viewpoint of the original gauge theory, the first factor is the contributions from $\Tr(\Phi)$ while the second one is from $\Tr(\Phi^2)$ and its superpartner for $\cN=(2,2)$ supersymmetry, where $\Phi$ is the (0,2) chiral multiplet charged with $\boldsymbol{2}\otimes \boldsymbol{2}$ under SU(2) gauge group.

The naive application of the $c$-extremization leads to the ``wrong'' central charges 
\be \label{cc-SU2-genus1-puncture1-naive}
c_L^{\textrm{naive}}=-9-1~, \qquad c_R^{\textrm{naive}}=-9~.
\ee  
A miscalculation persists even if zero U(1)$_R$-charge of the chiral multiplet is assumed, incorrectly yielding the computed right-moving central charge to be three. To rectify this, we need to read the dimension of the moduli space. In fact, the moduli space of the chiral multiplet is $\mathbb{C}^2$ spanned by $\Tr(\Phi)$ and $\Tr(\Phi^2)$, which implies that the maximal torus U(1) of the gauge group remains unbroken in the infrared. Consequently, the accurate values of the central charges are
\be \label{cc-SU2-genus1-puncture1}
c_L=5~, \qquad c_R=6~.
\ee

We note that the elliptic genus enjoys the symmetry $c_1\leftrightarrow c_1^{-1}$, which is the SU(2) Weyl group. Moreover, the elliptic genus admits the expansion with SU(2) characters
\be \label{su2-characterexp}
\cI^{(0,2),2}_{1,1}=\frac{i q^{-\frac{1}{12}}}{c_1-c_1^{-1}}\bigg(1+ \ch_{\frac12}^{\SU(2)}(c_1^2)q+ \ch_{2}^{\SU(2)}(c_1^2)q^2+\Big[1+\ch_{\frac12}^{\SU(2)}(c_1^2)+\ch_{\frac32}^{\SU(2)}(c_1^2)\Big]q^3+\cdots \bigg)
\ee 
where $\ch_{j}^{\SU(2)}$ is the spin-$j$ character of $\SU(2)$. Hence, one can observe that the $\U(1)_{c_1}$ flavor symmetry gets enhanced to $\SU(2)_{c_1}$ in the infrared.

The 4d $\mathcal{N}=4$ theory with $\SU(2)$ gauge group has the small $\mathcal{N}=4$ superconformal algebra as its associated VOA  \cite{Beem:2013sza}. First of all, the central charge of the VOA is $c_{2d}=-9$ which agrees with the naive left-moving central charge of the SU(2) adjoint chiral \eqref{cc-SU2-genus1-puncture1-naive}. Moreover, the elliptic genus of the SU(2) adjoint chiral can be viewed as a special $bc \beta \gamma$ system \cite{Bonetti:2018fqz}\footnote{The conformal weights of $b, c, \beta, \gamma$ are $3/2, -1/2, 1, 0$.}, and is also a linear combination of the characters of the associated VOA
\be \label{N=4}
  \frac{i\vartheta_4(c_1)}{\vartheta_1(c_1^2)} = \ch_0^{\cN=4}(q,c_1)+\ch_{1}^{\cN=4}(q,c_1)~.
\ee 
Here $\ch_0^{\cN=4}$ is the vacuum character and $\ch_{1}^{\cN=4}$ is the character of the non-vacuum irreducible module of the VOA \cite{Adamovic:2014lra}, and both characters are shown to satisfy three common flavored modular differential equations from null states in the VOA \cite{Pan:2021ulr}. Note that the flavor symmetry enhancement to SU(2) is supported from the viewpoint of the VOA since both the small $\mathcal{N}=4$ superconformal algebra and the $bc\beta\gamma$-ghost VOA are endowed with an SU(2) flavor symmetry\footnote{The $\SU(2)$ current of this particular $bc \beta \gamma$ system is given by $J^+ = \beta$, $ J^0 = bc + 2\beta \gamma$, $J^- = \beta \gamma \gamma + \gamma b c - 3/2 \partial \gamma$ \cite{Bonetti:2018fqz}.}.

An $\mathcal{N}=(0,2)$ vector multiplet and an $\mathcal{N}=(0,2)$ adjoint chiral multiplet form an $\mathcal{N}=(2,2)$ vector multiplet. Consequently, there exists no distinction between the left- and right-moving sectors in the $\mathrm{SU}(2)$ adjoint chiral. This is supported by the equality of central charges for both sectors, which are both $c_L=c_R=3$. Moreover, the elliptic genus of this theory can be expressed using characters of the small $\mathcal{N}=4$ superconformal algebra as in \eqref{N=4}. This suggests that the infrared limit of the $\mathcal{N}=(2,2)$ $\mathrm{SU}(2)$ vector multiplet is equipped with the small $\mathcal{N}=4$ superconformal algebra in both the left- and right-moving sectors, suggesting the \emph{supersymmetry enhancement}. 
It deserves further investigation to determine the exact infrared Hilbert space on a torus by modular invariance as demonstrated in \cite{Gadde:2014ppa, Gadde:2016khg, Guo:2019rrc}.

\subsubsection{Genus one with two punctures}\label{sec:SU2-genus1-puncture2}

Now consider the (0,2) reduction of the class $\cS$ theory $\cT_2[C_{1,2}]$, which is an $\SU(2)^2$ gauge theory coupled to two bi-fundamentals. As we saw in the previous example, we conjecture that the diagonal U(1) gauge group is unbroken in the infrared for a genus one theory. As a result, the central charges are given by
\be\label{cc-g1p2} c_L =7~,\quad  c_R=9~.\ee
The quiver diagram is given in Figure \ref{fig:SU2-(1,2)-quiver-1}. 

\begin{figure}[ht]
    \centering
 \includegraphics[width=6cm]{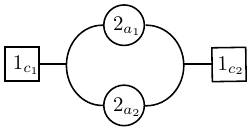}
    \caption{The (0,2) reduction of the class $\cS$ theory $\cT_2[C_{1,2}]$.}
    \label{fig:SU2-(1,2)-quiver-1}
\end{figure}

The elliptic genus of this theory is computed by the JK residue computation of the integral
\begin{align}\label{EG-genus1-puncture2}
    \mathcal{I}^{(0,2),2}_{1,2}(c_1,c_2)
    =& \int_\text{JK}
    \frac{d\boldsymbol{a}}{2\pi i \boldsymbol{a}}
   \mathcal{I}_{U_2}^{(0,2)}(a_1,a_2;c_1)  \mathcal{I}_{U_2}^{(0,2)}(a_1^{-1},a_2^{-1};c_2)  \mathcal{I}_{\textrm{vec}}^{(0,2)} (a_1)\mathcal{I}_{\textrm{vec}}^{(0,2)} (a_2)\cr 
    =& - \frac{\eta(q)^2}{\prod_{i = 1}^2 \vartheta_1({c}_i^2)} \ .
\end{align}
The elliptic genus implies the presence of an LG dual for the theory. This LG dual might comprise two free chiral multiplets. Potentially, there could be equal numbers of chiral and Fermi multiplets with identical charges, in addition to them. As seen in \eqref{su2-characterexp}, both the $\U(1)_{c_i}$ flavor symmetries get enhanced to $\SU(2)_{c_i}$ in the infrared. 

 Although the module structure of the corresponding VOA  $\chi(\cT_2[C_{1,2}])$ is not explicitly known, we argue that the elliptic genus is a linear combination of the module characters. One crucial piece of evidence is that the elliptic genus is a solution to the same flavored modular differential equations \cite{Zheng:2022zkm} that the Schur index of $\mathcal{T}_2[C_{1,2}]$ satisfies. For instance, the elliptic genus satisfies a weight-two differential equation,
\begin{align}
  0 = \Bigg[
  D_q^{(1)}
  - \frac{1}{4} \sum_{i = 1,2} D_{c_i}^2
  -\frac{1}{4}& \ \sum_{\alpha_i = \pm} E_1 \begin{bmatrix}
    1 \\ c_1^{\alpha_1}c_2^{\alpha_2}
  \end{bmatrix}
  \sum_{i = 1,2}\alpha_i D_{c_i}
  - \sum_{i = 1,2} E_1 \begin{bmatrix}
    1 \\ c_i^2
  \end{bmatrix}D_{c_i} \\
  & \ + 2 \bigg(
  E_2 + \frac{1}{2} \sum_{\alpha_i = \pm}E_2 \begin{bmatrix}
    1 \\ c_1^{\alpha_1}c_2^{\alpha_2}
  \end{bmatrix}
  + \sum_{i = 1,2} E_2 \begin{bmatrix}
    1 \\ c_i^2
  \end{bmatrix}
  \bigg) \Bigg] \mathcal{I}^{(0,2),2}_{1,2} \ . \nonumber
\end{align}
The definition of the twisted Eisenstein series is given in Appendix \ref{app:Eisenstein}.

\begin{figure}[ht]
    \centering
 \includegraphics[width=6cm]{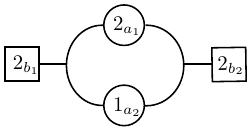}\qquad \qquad\includegraphics[width=6cm]{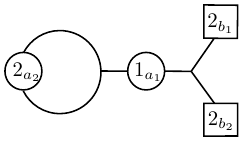}
    \caption{$\SU(2) \times \U(1)$ gauge theories corresponding to the genus-one Riemann surface with two punctures.}
    \label{fig:U1-(1,2)-quiver-1}
\end{figure}

Furthermore, by gauging the anti-diagonal of the $\U(1)$ flavor symmetry of the two $U_2$ theories, one can construct two other quiver theories as $\SU(2) \times \U(1)$ gauge theories, corresponding to the genus-one Riemann surface with two punctures, as in Figure \ref{fig:U1-(1,2)-quiver-1}. Notably, these theories do not originate from class $\cS$. Nonetheless, the central charges of these theories are computed as
\be c_L =13~,\quad  c_R=15~,\ee 
assuming that the diagonal U(1) gauge group is unbroken.

Let us first consider the left theory in Figure \ref{fig:U1-(1,2)-quiver-1}. The elliptic genus of this theory is
\begin{equation}
    \mathcal{I}_{1,2}'(b_1,b_2;d_1) = \eta(q)^2\int_\text{JK}
    \frac{d\boldsymbol{a}}{2\pi i \boldsymbol{a}}
  \mathcal{I}_{U_2}^{(0,2)}(a_1,b_1;d_1a_2)  \mathcal{I}_{U_2}^{(0,2)}(a_1^{-1},b_2;d_1a_2^{-1})  \mathcal{I}_{\textrm{vec}}^{(0,2)} (a_1) \frac{\vartheta_4(a_2^{ \pm 2})}{\eta(q)^2} \ . \nonumber
\end{equation}
Here, we introduce the $\mathrm{U}(1)_d$ flavor symmetry, which rotates the two chiral multiplets $U_2^{(0,2)}$ with the same phase and has a fugacity $d_1$. The detailed calculations of the JK residues can be found in Appendix \ref{app:genus1-puncture2-2}, and the final result is neatly summarized in a more simplified form
\begin{equation}\label{Iprime-g1p2}
       \mathcal{I}_{1,2}'(b_1,b_2;d_1) =-2\frac{\eta(q)^2\vartheta_4({d}_1^2)^2}{
    \vartheta_1({d}_1^2 {b}_1^\pm {b}_2^\pm)}=2\frac{\vartheta_4({d}_1^2)^2}{\eta(q)\vartheta_1({d}_1^4)} \cdot \frac{\eta(q)^3\vartheta_1({d}_1^{-4})}{
    \vartheta_1({d}_1^2 {b}_1^\pm {b}_2^\pm)} \ .
\end{equation}
The first factor always appears when the quiver gauge theory has a U(1) gauge node. This can be understood as the contribution of two Fermi fields $\Gamma_{1,2}$ with $\U(1)_{d_1}$ flavor charge 2 and one chiral field $\Phi$ with $\U(1)_{d_1}$ flavor charge 4. 
The second factor of the elliptic genus can be interpreted as the contribution from one Fermi field $\Psi$ and one chiral meson field $\wt\Phi$ with $\SU(2)_{b_1}\times \SU(2)_{b_2}$ flavor symmetry. The field content and their charges are summarized as follows:
\be\label{LG-g1p2}
\begin{array}{c|cc|cc}
     & \Gamma_{1,2}&\Psi &\Phi &\wt\Phi \\
     \hline
 \U(1)_{d_1}    & 2&-4 &4&2 \\
  \SU(2)_{b_i}    & \emptyset & \emptyset& \emptyset& \boldsymbol{2} \\
\end{array}
\ee 
Therefore, a generic $J$-type superpotential is 
$$
W=\Psi(\Phi +\operatorname{det} \wt\Phi) ~.
$$
The coefficient 2 in \eqref{Iprime-g1p2} means that the theory is dual to two (decoupled) copies of this LG model.

 The elliptic genus of the right theory in Figure \ref{fig:U1-(1,2)-quiver-1} can be computed as the JK residue of the integrand
\begin{equation}
    \mathcal{I}_{1,2}''(b_1,b_2;d_1) = \eta(q)^2\int_\text{JK}
    \frac{d\boldsymbol{a}}{2\pi i \boldsymbol{a}}
  \mathcal{I}_{U_2}^{(0,2)}(a_2,a_2^{-1};da_1)  \mathcal{I}_{U_2}^{(0,2)}(b_1,b_2;da_1^{-1})  \mathcal{I}_{\textrm{vec}}^{(0,2)} (a_2) \frac{\vartheta_4(a_1^{ \pm 2})}{\eta(q)^2} \ . \nonumber
\end{equation}
The detailed computations of the JK residues are provided in Appendix \ref{app:genus1-puncture2-3}, while the total residue is presented in a compact form
\begin{equation}\label{Ipp-g1p2}
    \mathcal{I}_{1,2}''(b_1,b_2;d_1)
    = 
    \frac{\vartheta_4({d}_1^2)^2}{\eta(q)\vartheta_1({d}_1^4)}\cdot\frac{\eta(q)^3\vartheta_1({d}_1^8)}{\vartheta_1({d}_1^{-4}  {b}_1^{\pm2} {b}_2^{\pm2})}\cdot
    \prod_{i = 1}^2 \frac{\vartheta_1({b}_i^4)}{\vartheta_1({b}_i^{-2})}
      \ .
\end{equation}
The elliptic genus is different from \eqref{Iprime-g1p2} so that the two theories in Figure \ref{fig:U1-(1,2)-quiver-1} are not dual to each other. When the number of U(1) gauge nodes is equal to the genus, the theory is contingent on the quiver diagram, unlike class $\cS$ theory. 

Since \eqref{Ipp-g1p2} is expressed as a product of theta functions, the theory is also dual to the following LG model. The Fermi multiplets $\Gamma_{1,2}$ and $\Psi$ and the chiral multiplets  $\Phi$ and $\wt\Phi_{\pm\pm}$ are similar to the aforementioned LG theory.  A key distinction arises from the inclusion of two additional Fermi multiplets, $\Xi_{1,2}$, and two chiral multiplets, $\Sigma_{1,2}$. Because of these additions, the manifest flavor symmetry is $\U(1)_{b_1}\times \U(1)_{b_2}$ at UV. The charges of these fields are summarized as follows:
$$
\begin{array}{c|ccc|ccc}
            & \Gamma_{1,2}&\Psi &\Xi_{j}&\Phi&\wt\Phi_{\e_1\e_2} &\Sigma_j \\ \hline
\U(1)_{d_1} & 2 &8 &0&4&-4&0 \\
\U(1)_{b_i} & 0 & 0&4\delta_{ij}& 0& \e_i2 &-2\delta_{ij}
\end{array}
$$
where $\e_i=\pm$. Then, a generic $J$-type superpotential is 
$$
W=\Psi\operatorname{det} \wt\Phi+\Xi_1  \Phi \wt\Phi_{-+} \Sigma_1\Sigma_2+\Xi_2  \Phi \wt\Phi_{+-} \Sigma_1\Sigma_2 ~.
$$
The expansion of the elliptic genus \eqref{Ipp-g1p2} in terms of $q$ reveals the fugacities $b_{1,2}$ arrange themselves as characters of $\SU(2)_{b_1}\times \SU(2)_{b_2}$, implying the symmetry enhancement from  $\U(1)_{b_1}\times \U(1)_{b_2}\to \SU(2)_{b_1}\times \SU(2)_{b_2}$ at IR.

For the quiver theories in Figure \ref{fig:U1-(1,2)-quiver-1}, the corresponding VOA for the IR CFT is yet to be identified.  Given their duality to the LG models, the technique in \cite{Dedushenko:2015opz} offers a potential method for identifying the VOA. This remains an area for further study.

\subsubsection{Genus one with three punctures}\label{sec:SU2-genus1-puncture3}

\begin{figure}[ht]
    \centering
 \includegraphics[width=6cm]{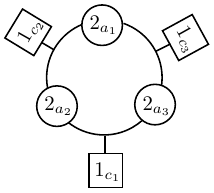}
    \caption{The (0,2) reduction of the class $\cS$ theory $\cT_2[C_{1,3}]$.}
    \label{fig:SU2-(1,3)-quiver-1}
\end{figure}

Now let us consider quiver theories of genus one with three punctures. The (0,2) reduction of the class $\cS$ theory $\cT_2[C_{1,3}]$ is the $\SU(2) \times \SU(2) \times \SU(2)$ gauge theory in Figure \ref{fig:SU2-(1,3)-quiver-1}. Considering that the diagonal U(1) gauge group remains unbroken, the central charges of the theory are given by
\be 
c_L=9~, \quad c_R=12~.
\ee 
The elliptic genus is the JK residue of the integrand
\begin{align}\label{EG-genus1-puncture3}
    \mathcal{I}^{(0,2),2}_{1,3}
    =& \int_\text{JK}
    \frac{d\boldsymbol{a}}{2\pi i \boldsymbol{a}}
   \mathcal{I}_{U_2}^{(0,2)}(a_1^{-1},a_2;c_2)  \mathcal{I}_{U_2}^{(0,2)}(a_2^{-1},a_3;c_1)  \mathcal{I}_{U_2}^{(0,2)}(a_3^{-1},a_1;c_3)  \prod_{i=1}^3\mathcal{I}_{\textrm{vec}}^{(0,2)} (a_i)\cr 
    =&  \frac{\eta(q)^3}{\prod_{i = 1}^3 \vartheta_1({c}_i^2)}
\end{align}
which signals the existence of an LG dual which includes three free chiral multiplets, associated to the three punctures. In addition to them, there could be equal numbers of chiral and Fermi multiplets with the same charges. All the U(1) flavor symmetries get enhanced to SU(2) in the infrared, as illustrated in \eqref{su2-characterexp}.

\begin{figure}[ht]
    \centering
 \includegraphics[width=5cm]{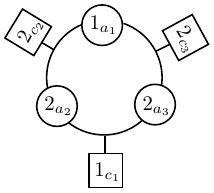}\qquad \qquad \raisebox{0.7cm}{\includegraphics[width=7.5cm]{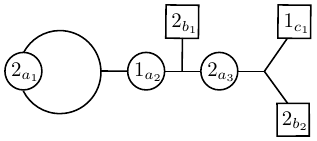}}
    \caption{$\SU(2) \times \SU(2) \times \U(1)$ gauge theories corresponding to the genus-one Riemann surface with three punctures.}
    \label{fig:U1-(1,3)-quiver-1}
\end{figure}
Let us now consider $\SU(2) \times \SU(2) \times \U(1)$ gauge theories for genus one with three punctures as illustrated in Figure \ref{fig:U1-(1,3)-quiver-1}. The central charges of the theory are given by
\be 
c_L=15~, \quad c_R=18~.
\ee 
The elliptic genus of the left quiver theory of Figure \ref{fig:U1-(1,3)-quiver-1} is 
\begin{align}\label{EG-genus1-puncture3-2}
    &\mathcal{I}^{\prime}_{1,3}\cr
    =& \eta(q)^2\int_\text{JK}
    \frac{d\boldsymbol{a}}{2\pi i \boldsymbol{a}}
   \mathcal{I}_{U_2}^{(0,2)}(a_2,c_2^{-1};d_1a_1)  \mathcal{I}_{U_2}^{(0,2)}(a_2^{-1},a_3;c_1)  \mathcal{I}_{U_2}^{(0,2)}(a_3^{-1},c_3;d_1a_1^{-1})   \frac{\vartheta_4(a_1^{ \pm 2})}{\eta(q)^2}\prod_{i=2}^3\mathcal{I}_{\textrm{vec}}^{(0,2)} (a_i)\cr 
    =& 2\frac{\vartheta_4({d}_1^2)^2}{\eta(q)\vartheta_1({d}_1^4)} \cdot \frac{\eta(q)^4\vartheta_1(c_1^{2}{d}_1^{4})}{
    \vartheta_1(c_1^2)\vartheta_4(c_1{d}_1^2 {b}_1^\pm {b}_2^\pm)}~.
\end{align}
The computational details are given in Appendix \ref{app:SU2-g1p3-2}. The form of the elliptic genus signals an LG description. The LG model is similar to \eqref{LG-g1p2}, but there is additional chiral multiplet $\Phi_2$ and $\U(1)_{c_1}$ flavor symmetry:
$$
\begin{array}{c|cc|ccc}
     & \Gamma_{1,2}  &\Psi &\Phi_1&\Phi_2 &\wt\Phi \\
     \hline
 \U(1)_{d_1}    & 2 &4&-4&0&-2 \\
  \U(1)_{c_1}    & 0 &2&0&-2&-1 \\
  \SU(2)_{b_i}    & \emptyset & \emptyset& \emptyset& \emptyset& \boldsymbol{2} \\
\end{array}
$$
Therefore, a generic $J$-type superpotential is 
$$
W=\Psi(\Phi_1\Phi_2 +\operatorname{det} \wt\Phi) ~.
$$
The coefficient 2 in \eqref{EG-genus1-puncture3-2} means that the theory is dual to two (decoupled) copies of this LG model.

There is yet another quiver gauge theory with gauge group $\SU(2) \times \SU(2) \times \U(1)$ as in the right of Figure \ref{fig:U1-(1,3)-quiver-1} whose elliptic genus is
\begin{align}\label{EG-genus1-puncture3-3}
    &\mathcal{I}^{\prime\prime}_{1,3}\cr
    =& \eta(q)^2\int_\text{JK}
    \frac{d\boldsymbol{a}}{2\pi i \boldsymbol{a}}
   \mathcal{I}_{U_2}^{(0,2)}(a_1,a_1^{-1};d_1a_2)  \mathcal{I}_{U_2}^{(0,2)}(b_1,a_3;d_1a_1^{-1})  \mathcal{I}_{U_2}^{(0,2)}(a_3^{-1},b_2;c_2)   \frac{\vartheta_4(a_1^{ \pm 2})}{\eta(q)^2}\prod_{i=2}^3\mathcal{I}_{\textrm{vec}}^{(0,2)} (a_i)\cr 
    =&\frac{\vartheta_4({d}_1^2)^2}{\eta(q)\vartheta_1({d}_1^4)} \cdot \frac{\eta(q)^4\vartheta_1(c_1^{4}{d}_1^{8})}{
    \vartheta_1(c_1^2)\vartheta_1(c_1^{-2}{d}_1^{-4} {b}_1^{\pm2} {b}_2^{\pm2})}\cdot
    \prod_{i = 1}^2 \frac{\vartheta_1({b}_i^4)}{\vartheta_1({b}_i^{-2})}
      \ ,
\end{align}
suggesting an LG description of the theory.

Since \eqref{EG-genus1-puncture3-3} is expressed as a product of theta functions, the theory is also dual to the following LG model. The Fermi multiplets $\Gamma_{1,2}$ and $\Psi$ and the chiral multiplets  $\Phi_{1,2}$ and $\wt\Phi_{\pm\pm}$ are similar to the aforementioned LG theory.  A key distinction arises from the inclusion of two additional Fermi multiplets, $\Xi_{1,2}$, and two chiral multiplets, $\Sigma_{1,2}$. Because of these additions, the manifest flavor symmetry is $\U(1)_{b_1}\times \U(1)_{b_2}$ at UV. The charges of these fields are summarized as follows:
$$
\begin{array}{c|ccc|cccc}
            & \Gamma_{1,2}&\Psi &\Xi_{j}&\Phi_1&\Phi_2&\wt\Phi_{\e_1\e_2} &\Sigma_j \\ \hline
\U(1)_{d_1} & 2 &8 &0&4&0&-4&0 \\
  \U(1)_{c_1}    & 0 &4&0&0&2&-2&0 \\
\U(1)_{b_i} & 0 & 0&4\delta_{ij}& 0&0& \e_i2 &-2\delta_{ij}
\end{array}
$$
where $\e_i=\pm$. Then, a generic $J$-type superpotential is 
$$
W=\Psi\operatorname{det} \wt\Phi+\Xi_1  \Phi_1\Phi_2 \wt\Phi_{-+} \Sigma_1\Sigma_2+\Xi_2  \Phi_1\Phi_2 \wt\Phi_{+-} \Sigma_1\Sigma_2 ~.
$$
Like \eqref{Ipp-g1p2}, the expansion of the elliptic genus \eqref{EG-genus1-puncture3-3} shows that a flavor symmetry is enhanced to $\SU(2)_{b_1}\times \SU(2)_{b_2}$ at IR.

\subsubsection{Genus two}\label{sec:SU2-genus2}

\begin{figure}[ht]
    \centering
    {\raisebox{.7cm}{\includegraphics[width=7cm]{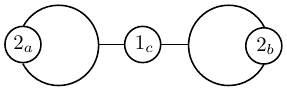}}} \qquad \includegraphics[width=3.8cm]{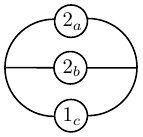}
    \caption{Genus two for SU(2)}
    \label{fig:SU2-genus2}
\end{figure}
For the genus-two case (with no puncture), we have two different quiver descriptions as depicted in Figure \ref{fig:SU2-genus2}. As seen in \S\ref{sec:SU2-adjoint}, when a diagonal SU(2) subgroup of $\SU(2)^2$ in the $U_2$ theory is gauged, the Cartan subgroup U(1) remains unbroken. Therefore, for the theory of genus two, the $\U(1)^2$ gauge group is unbroken. Considering this fact, the central charges of this theory are computed as
\be 
c_L=10~, \quad c_R=9~. 
\ee
It is straightforward to show that the two different descriptions are dual to each other. Certainly, the Lagrangian descriptions provide the different expressions of the elliptic genera 
\bea
\cI^{(0,2),2}_{\includegraphics[width=1cm]{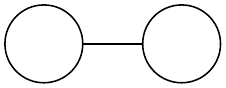}}=&\int_{\JK}\frac{da}{2\pi ia}\frac{db}{2\pi ib}\frac{dc}{2\pi ic}    \mathcal{I}_{U_2}^{(0,2)}(a,a^{-1};dc)     \mathcal{I}_{U_2}^{(0,2)}(b,b^{-1};dc^{-1})  \mathcal{I}_{\textrm{vec}}^{(0,2)} (a)\mathcal{I}_{\textrm{vec}}^{(0,2)} (b)\vartheta_4(c^{\pm2})~\cr 
\cI^{(0,2),2}_{\includegraphics[width=.5cm]{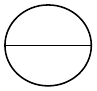}}=& \int_{\JK}\frac{da}{2\pi ia}\frac{db}{2\pi ib}\frac{dc}{2\pi ic}     \mathcal{I}_{U_2}^{(0,2)}(a,b;dc) \mathcal{I}_{U_2}^{(0,2)}(a^{-1},b^{-1};dc^{-1})   \mathcal{I}_{\textrm{vec}}^{(0,2)} (a)\mathcal{I}_{\textrm{vec}}^{(0,2)} (b) \vartheta_4(c^{\pm2})
\eea 
where the last factors $\vartheta_4(c^{\pm2})$ correspond to the contributions from the U(1) vector multiplet and the two Fermi multiplets. Both JK residues can be straightforwardly evaluated, and the results agree, suggesting that the two theories are dual to each other. 

Alternatively, the two elliptic genera can be computed from other building blocks. The former theory can be obtained by gluing two copies of the theory of genus one with one puncture whose elliptic genus is given in \eqref{EG-genus1-puncture1}. Similarly, the latter can be obtained by gluing the two punctures in the theory of genus one with two punctures whose elliptic genus is given in \eqref{EG-genus1-puncture2}. In this approach, it becomes more evident that they are identical, and moreover the JK residue provides a remarkably simple result
\bea \label{EG-genus2}
\cI^{(0,2),2}_{2,0}=&\eta(q)^2\int_{\JK}\frac{dc}{2\pi ic}\frac{\vartheta_4(c^{\pm2})}{\vartheta_1(d^{2}c^{\pm2})}=\frac{2\vartheta_4(d^{2})^2}{\eta(q)\vartheta_1(d^{4})}~.
\eea 

This can be compared to the S-duality in class $\cS$ theory. While the process of U(1) gauging may not naturally fit in the context of the (0,2) reduction of the class $\cS$ theory as explained below \eqref{U1-gauging}, the gluing procedure explained above is analogous to the class $\cS$ construction. In fact, since the flavor symmetries get enhanced to SU(2) for genus one theories, we can gauge the anti-diagonal SU(2) to obtain the (0,2) reduction of the class $\cS$ theory $\cT_2[C_{2,0}]$ of genus two whose elliptic genus is
\bea  \label{genus-2-classS}
&-\int_{\JK}\frac{dc}{2\pi ic}\cI^{(0,2),2}_{1,1}(cd) \frac{\vartheta_1(c^{\pm2})}{2}\cI^{(0,2),2}_{1,1} (c^{-1}d) =-\int_{\JK}\frac{dc}{2\pi ic}\cI^{(0,2),2}_{1,2}(cd,c^{-1}d) \frac{\vartheta_1(c^{\pm2})}{2} ~\cr 
 =&-\frac{\eta(q)^2}{2}\int_{\JK}\frac{dc}{2\pi ic}\frac{\vartheta_1(c^{\pm2})}{\vartheta_1(d^{2}c^{\pm2})}=\frac{\vartheta_1(d^{2})^2}{\eta(q)\vartheta_1(d^{4})}~.
\eea 
Recalling that the Jacobi theta functions $\vartheta_1$ and $\vartheta_4$ are related by \eqref{theta14}, the result differs from \eqref{EG-genus2} merely by the shift $d\to q^{1/4}d$ of the $\U(1)_d$ flavor fugacity, up to a factor. Indeed, comparing the SU(2) vector multiplet contribution \eqref{SU2-vector} and U(1) gauging \eqref{U1-gauging} at the level of elliptic genera, the difference appears only in $\vartheta_1(a^{\pm2})$ and $\vartheta_4(a^{\pm2})$, up to a factor of 2 which is the order of the Weyl group of SU(2). Hence, the duality between the two (0,2) theories in Figure \ref{fig:SU2-genus2} is analogous to the S-duality in the class $\cS$ theory $\cT_2[C_{2,0}]$.

As a result, one can expect the relation between the elliptic genera (\ref{EG-genus2},\ref{genus-2-classS}) and characters of the VOA $\chi(\cT_2[C_{2,0}])$.
In 4d, the associated VOA of the genus-two $A_1$-theory $\mathcal{T}_2[C_{2,0}]$ of class $\mathcal{S}$ is studied in \cite{Kiyoshige:2020uqz,Beem:2021jnm}. In particular, null states are present at level-four and six, which are expected to give rise to a weight-four and a weight-six flavor differential equation \cite{Zheng:2022zkm}. The above elliptic genus \eqref{genus-2-classS} is indeed a solution to these two equations once the $\U(1)$ fugacity $d$ is rescaled $d \to d^{1/2}$. Therefore, it is natural to argue that the elliptic genus is some linear combination of module characters of $\chi(\mathcal{T}_2[C_{2,0}])$
\be 
\frac{\vartheta_1(d)^2}{\eta(q)\vartheta_1(d^{2})}=\sum_i a_i \ch^{\chi(\cT_2[C_{2,0}])}_{\lambda_i}(d)~, \qquad a_i \in \bQ.
\ee 
Likewise, we can argue that, upon rescaling $d \to d^{1/2}$, the elliptic genus \eqref{EG-genus2} of the genus two theory constructed from the $U_2$ theories can be written in a similar way as
\be 
\frac{2\vartheta_4(d)^2}{\eta(q)\vartheta_1(d^{2})}=2q^{1/4}d \sum_i a_i \ch^{\chi(\cT_2[C_{2,0}])}_{\lambda_i}(q^{1/2}d)~.
\ee

\subsubsection{General Riemann surfaces and TQFT structure}\label{sec:generalSU2}

\begin{figure}[ht]
    \centering
    \includegraphics[width=\textwidth]{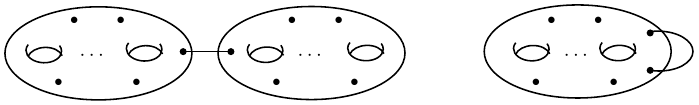}
    \caption{The gluing minimal punctures leads to a new Riemann surface, and (0,2) elliptic genera are consistent with the cut-and-join procedure on Riemann surfaces $C_{g,n}$.}
    \label{fig:02-TQFT}
\end{figure}

For a (0,2) quiver theory of genus $g>0$ constructed from the $U_2$ theory, the minimal number of U(1) gauge groups is $g-1$. Hence, based on the previous results, we can consider a (0,2) quiver theory analogous to the class $\cS$ theory $\cT_2[C_{g>0,n}]$ where the numbers of SU(2) and U(1) gauge groups are $2(g-1)+n$ and $g-1$, respectively. At a generic point of the moduli space of chiral multiplet, we conjecture that $\U(1)^g$ gauge group is unbroken. Consequently, the central charges of the theory are
\be \label{SU2-(g,n)-c}
c_L=7g-4+2n~, \qquad  c_R=3(2g-1+n)~.
\ee 
Regardless of their quiver descriptions (or frames), these theories all flow to the same infrared (IR) theory.
By introducing U(1) flavor fugacities $c_i$ for the external punctures and $d_i$ for the U(1) gauging at UV, the elliptic genus of the theory can be expressed in a simple form
\be \label{SU2-02-EG}
\cI^{(0,2),2}_{g>0,n}(c_1,\ldots,c_n)=\prod_{j=1}^{g-1}\frac{2\vartheta_4(d_j^2)^2 }{\eta(q)\vartheta_1(d_{j}^4)}\prod_{i=1}^n \frac{\eta(q)}{\vartheta_1(c_i^2)}~.
\ee 
Therefore, the flavor symmetry associated to each puncture gets enhanced to SU(2) at IR as seen in \eqref{su2-characterexp}.
Remarkably, the integral formula \eqref{EG-genus2} guarantees that the above form of the (0,2) elliptic genera is consistent with the TQFT structure as in Figure \ref{fig:02-TQFT}
\begin{align}
\mathcal{I}_{g=g_1+g_2, n_1+n_2-2}^{(0,2),2}=&\int_{\mathrm{JK}} \frac{d a}{2 \pi i a} \mathcal{I}^{(0,2),2}_{g_1, n_1}(\ldots,d_{g-1}a) \mathcal{I}^{(0,2),2}_{g_2, n_2}(d_{g-1}a^{-1},\ldots) \vartheta_1(a^{\pm2})~,\cr 
    \mathcal{I}_{g + 1, n - 2}^{(0,2),2} = &\int_\text{JK} \frac{da}{2\pi i a} 
    \mathcal{I}_{g, n}^{(0,2),2}(\ldots,d_{g} a, d_{g}a^{-1}) \vartheta_1(a^{\pm2})~.
\end{align}

As we recall, the elliptic genus of the (0,2) reduction of the class $\cS$ theory of genus one with $n$ punctures is 
\be 
\cI^{(0,2),2}_{1,n}(c_1,\ldots,c_n)=\prod_{i=1}^n \frac{\eta(q)}{\vartheta_1(c_i^2)}~,
\ee 
which exhibit the enhancement to SU(2) for each puncture in the infrared. 
To construct the (0,2) reduction of the class $\cS$ theory $\mathcal{T}_2[C_{g>0,n}]$ of type $A_1$, we can repeatedly gauge anti-diagonal SU(2) of genus one theories with multiple punctures. Upon the rescaling $d\to q^{1/4}d$, the resulting elliptic genus differs from \eqref{SU2-02-EG} by merely a factor of $(2q^{1/2}d^2)^{g-1}$. In this sense, the TQFT structure of the elliptic genus can be attributed to the class $\cS$ construction.

From the perspective of the chiral algebra, we observe a connection to class $\cS$ theory. The VOA corresponding to any Lagrangian $\mathcal{N} = 2$ SCFT can be constructed using the gauging method described in \cite{Beem:2013sza}. Yet, putting this method into practice is intricate. Detailing the VOA structure and its representation theory is notably challenging, even for class $\cS$ theories of type $A_1$.
Nevertheless, if we change $\vartheta_4(a^{\pm2})$ to $\vartheta_1(a^{\pm2})$ for the $\U(1)$ gauging \eqref{U1-gauging}, the JK integrand of the elliptic genus, when derived from the UV Lagrangian, coincides with that of the Schur index of $\cT_2[C_{g>0,n}]$ up to a factor, and the JK residue gives
\be \label{genus-gn-change}
\prod_{j=1}^{g-1}\frac{2\vartheta_1(d_j^2)^2 }{\eta(q)\vartheta_1(d_{j}^4)}\prod_{i=1}^n \frac{\eta(q)}{\vartheta_1(c_i^2)}~.
\ee 
 Building upon the results in  \cite{Pan:2021ulr,Pan:2021mrw,Zheng:2022zkm}, certain residues of the integrand correspond to the Schur index with surface defects of Gukov-Witten type\footnote{Up to some prefactors of $q$ to account for the different stress-energy tensors involved \cite{Bianchi:2019sxz}.}. Given its role as a surface defect index, one expects \eqref{genus-gn-change} to satisfy all the flavored modular differential equations associated to the VOA $\chi(\mathcal{T}_2[C_{g,n}])$. 
In light of these observations, we propose that the elliptic genus \eqref{SU2-02-EG} can be written in terms of a linear combination of characters of the VOA $\chi(\cT_2[C_{g>0,n}])$ upon an appropriate redefinition of the U(1) fugacities and a certain overall factor.

When the number of U(1) gauge groups is $g$ for a theory of genus $g$, it admits an LG dual, but it depends on the quiver description as we saw in the examples of \S\ref{sec:SU2-genus1-puncture2} and \S\ref{sec:SU2-genus1-puncture3}. It would be interesting to study the VOA structures for theories of this type.

\subsection{SU(\texorpdfstring{$N$}{N})\texorpdfstring{$\times$}{x}U(1) gauge theories, LG duals and VOAs}\label{sec:02-SUN}

\begin{figure}[ht]
    \centering
    \includegraphics[width=4cm]{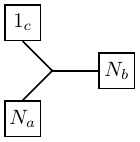}
    \caption{ A building block $U_N$, representing $N^2$ (0,2) chiral multiplets with $\SU(N)_a\times\SU(N)_b\times\U(1)_c$ flavor symmetry.}
    \label{fig:SUN-trinion}
\end{figure}

Now let us move on to cases of higher rank. For class $\cS$ of type $A_{N-1}$, punctures are classified by a partition of $N$, and theories do not admit a Lagrangian description in general. As seen in \S\ref{sec:twist}, we perform the (0,2) reduction for 4d $\cN=2$ Lagrangian theories of class $\cS$. Consequently, the basic building block is a sphere with two maximal punctures and one minimal puncture, corresponding to $N^2$ hypermultiplets. (See the right of Figure \ref{fig:04-block}.) Its (0,2) reduction yields (0,2) $N^2$ chiral multiplets, with the flavor symmetry represented as $\U(N^2)$ and includes the subgroup $\SU(N)_a \times \SU(N)_b \times \U(1)_x$. This particular (0,2) theory, labeled as $U_N$, serves as our fundamental building block, with its quiver illustrated in Figure \ref{fig:SUN-trinion}. As highlighted in \eqref{correct},  the $c$-extremization is invalid for theories of this class, and the $\U(1)_R$-charge of the (0,2) chiral multiplet is $r=0$. The NS elliptic genus of $U_N$ is given by
\begin{equation}\label{EG-UN}
    \mathcal{I}_{U_N}^{(0,2)}(a, b, c) = \prod_{i,j = 1}^N \frac{\eta(q)}{\vartheta_4(q^{-\frac12}\tilde c a_i b_j)}  = \prod_{i,j = 1}^N \frac{\eta(q)}{\vartheta_4(c a_i b_j)} 
\end{equation}
where we redefine the $\U(1)_c$ flavor fugacity by $c=q^{-\frac12}\tilde c$.
Note that we impose the condition $\prod_{i=1}^Na_i=1=\prod_{j=1}^Nb_j$ on the $\SU(N)$ fugacities.
On a related note, as seen in \eqref{Schur-limit-hyper}, the Schur limit of the superconformal index for a sphere with two maximal punctures and one minimal puncture is given by 
\be 
\cI^{\textrm{4d}}=\prod_{i,j = 1}^N\Gamma(\sqrt{t} (c a_i b_j)^{ \pm 1})\quad \xrightarrow{t\to q}\quad   \cI^{\textrm{Schur}}=\prod_{i,j = 1}^N\frac{\eta(q)}{\vartheta_4(c a_i b_j)} ~.
\ee 
Consequently, by redefining the $\U(1)_c$ flavor fugacity as in \eqref{EG-UN}, the elliptic genus of the $U_N$ theory agrees with the Schur index above.

The gauging procedure of the $U_N$ theories is as usual. 
The contribution of a vector multiplet is the same in both the Schur index \eqref{Schur-limit-vec} and the elliptic genus \eqref{Vector}. The $\SU(N)$ vector multiplet contribution is
\be 
\mathcal{I}_{\textrm{vec}}^{(0,2)} (a)=\frac{\eta(q)^{2N}}{N!}\prod_{A\neq B} i \frac{\vartheta_1(a_A/a_B)}{\eta(q)}~.
\ee 
and the $\SU(N)$ gauging leads to no gauge anomaly. In this way, the integrand of the superconformal index and the (0,2) elliptic genus agree for a class $\cS$ Lagrangian theory.

However, in 2d (0,2) theories, one can also gauge the U(1) symmetry of the $U_N$ theory. This U(1) gauging is similar to the $U_2$ case, but the U(1) gauge charges of the two Fermi multiplets must be $\pm N$ to avoid gauge anomaly. Following the $c$-extremization, the $\U(1)_R$-charge for these Fermi multiplets is assigned to be $r=0$. Consequently, the gauging procedure is then applied to the elliptic genus as described:
\bea \label{U1-gauging-2}
\frac{\eta(q)}{\vartheta_4(q,c_1\cdots )} \frac{\eta(q)}{\vartheta_4(q,c_2\cdots )}  \quad \to \quad &  \eta(q)^2\int_{\textrm{JK}}\frac{d a}{2 \pi i a} \frac{\eta(q)}{\vartheta_4( da\cdots )} \frac{\eta(q)}{\vartheta_4( d a^{-1}\cdots )}\frac{\vartheta_4(a^{\pm N})}{\eta(q)^2}\cr
\eea 
Once a (0,2) quiver theory involves a U(1) gauging of the $U_N$ theories, the theory is no longer the (0,2) reduction of a class $\cS$ theory. Nonetheless, one can consider the elliptic genus of the theory in the infrared.

\subsubsection{Gauge/LG duality for linear quivers}
\begin{figure}[ht]
    \centering
    \includegraphics[width=0.95\textwidth]{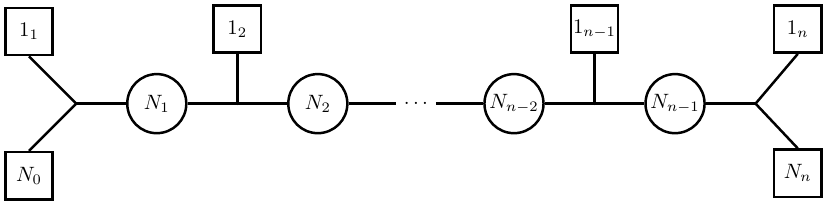}
    \caption{An SU($N$) linear quiver where the subscripts are added solely for node numbering purposes.}
    \label{fig:SUN-linear-quiver}
\end{figure}

The (0,2) reduction of class $\cS$ theory of higher rank is quite restricted because the class $\cS$ theory should have a Lagrangian description. Consequently, a theory sought to be built upon the $\SU(N)$ gauging of the $U_N$ theories. One such example is a linear quiver as in Figure \ref{fig:SUN-linear-quiver}. The central charges are given by
\be 
c_L=2(N^2+n-1) ~,\qquad c_R=3(N^2+n-1)~.
\ee 

As the simplest example, we can consider the (0,2) $\SU(N)$ SQCD with $N$ fundamentals and $N$ anti-fundamentals. 
As found in \cite[Eqn.(4.7)]{Putrov:2015jpa}, the computation of its elliptic genus is equivalent to the higher rank rendition of the elliptic inversion formula \eqref{SU2-EI} 
\begin{equation}\label{SUN-EI} 
\frac{
  \eta(q)^{N^2+1}
  \vartheta_1(c_1^Nc_2^N)
  }{
  \vartheta_\alpha(c_1^N)
  \vartheta_\alpha(c_2^N) 
  \prod_{A,B=1}^N\vartheta_1(c_1c_2b_{0,A} b_{2,B}^{-1})}
  =
  \frac{\eta(q)^{2N^2}}{N!}
  \int_{\textrm{JK}} 
  \frac{d\boldsymbol{a}}{2\pi i\boldsymbol{a}} 
  \prod_{A,B,i=1}^N
  \frac{
    \prod_{j\neq i}\vartheta_1(a_i/a_{j})
    }{
    \vartheta_4(c_1b_{0,A}a_i^{-1})
    \vartheta_4(c_2b_{2,B}^{-1} a_i)
    }~,
\end{equation}
where $\alpha =1$ for even $N$ and $\alpha =4$ for odd $N$. 
To evaluate the elliptic genus of a linear quiver, we repeatedly apply the elliptic inversion formula \eqref{SUN-EI}, and it therefore takes a simple form
\begin{equation}
    \mathcal{I}_{0,n,2}^{(0,2),N}
    =
    \frac{
      \eta(q)^{N^2+n-1}
      \vartheta_\alpha( \prod_{i=1}^nc_i^N)
      }{
      \prod_{i=1}^n 
      \vartheta_\beta\left( c_i^N\right)
      \cdot  
      \prod_{A,B=1}^N
      \vartheta_\gamma( b_{0,A}~b_{n,B}^{-1}\prod_{i=1}^nc_i) 
      } 
\end{equation}
where 
\be 
\alpha=\begin{cases}
1 & n\cdot N \textrm{ even} \\
4& n\cdot N \textrm{  odd}
\end{cases}~,\qquad 
\beta=\begin{cases}
1 & N \textrm{ even} \\
4&N \textrm{  odd}
\end{cases}~,\qquad 
\gamma=\begin{cases}
1 & n \textrm{ even} \\
4& n \textrm{  odd}
\end{cases}~.
\ee
Taking into account the shift of the $\U(1)_{c_i}$ fugacities in \eqref{U2}, the form of the elliptic genus tells us that the theory is dual to an LG model with one Fermi multiplet $\Psi$, $n$ chiral multiplets $\Phi$ and one chiral meson multiplet $\tilde\Phi_{i,j=1,\ldots,n}$, forming a $J$-type superpotential
$$
W=\Psi (\prod_{i=1}^n\Phi_i+\operatorname{det} \wt\Phi) ~.
$$

\subsubsection{Gauge/LG duality for circular quivers}

\begin{figure}[ht]
    \centering
    \includegraphics[width=0.6\textwidth]{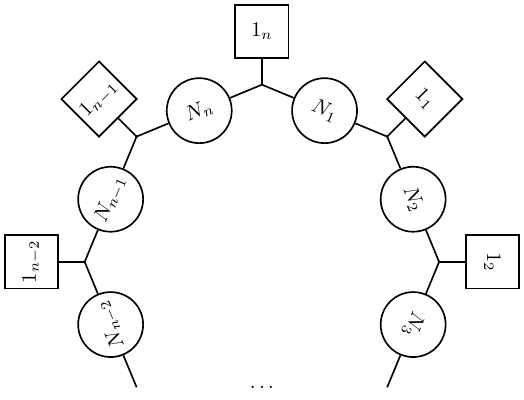}
    \caption{An SU($N$) circular quiver where the subscripts are added solely for node numbering purposes.}
    \label{fig:SUN-circular-quiver}
\end{figure}

The other class of the (0,2) reduction of class $\cS$ theory is a circular quiver as in Figure \ref{fig:SUN-circular-quiver}. Again, for a genus-one theory, the diagonal Cartan subgroup $\U(1)^{N-1}$ is unbroken. As a result, the central charges for genus $1$ with $n$ punctures are given by
\be 
c_L=2n+3(N-1)~,\qquad c_R=3(n+N-1)~.
\ee 
Some of the explicit JK residue computations of elliptic genera can be found in Appendix \ref{app:SU3-g1p1} and \ref{app:SU3-g1p2}.

As the simplest example, let us consider the theory at genus-one with one puncture. In other words, it is an $\SU(N)$ gauge theory with one adjoint chiral, which is the (0,2) reduction of the 4d $\mathcal{N} = 4$ $\SU(N)$ theory, and an additional free chiral multiplet. The evaluation of its elliptic genus can be understood as  
an elliptic inversion formula of another kind 
\bea \label{inversion2}
    \mathcal{I}_{1,1}^{(0,2),N}=&\frac{\eta(q)^{N}}{N!\vartheta_4(c)^N}\int_{\textrm{JK}} \frac{d\boldsymbol{a}}{2\pi i\boldsymbol{a}}\prod_{j\neq i} \frac{\vartheta_1(a_i/a_{j})}{\vartheta_4(c a_i/a_{j})}\cr 
    =&  \frac{\eta(q)}{\vartheta_\alpha (c^N)}~,
\eea  
where  $\alpha =1$ for even $N$ and $\alpha =4$ for odd $N$. Taking into account the shift of the flavor fugacity in \eqref{EG-UN}, the elliptic genus can indeed be written as
\be 
\mathcal{I}_{1,1}^{(0,2),N}= \frac{\eta(q)}{\vartheta_1 (\tilde c)}\cdot \frac{\vartheta_1 (\tilde c)}{\vartheta_1 (\tilde c^2)} \cdot \frac{\vartheta_1 (\tilde c^2)}{\vartheta_1 (\tilde c^3)} \cdots\frac{\vartheta_1 (\tilde c^{N-1})}{\vartheta_1 (\tilde c^N)}~
\ee 
which are contributions from $\operatorname{Tr}(\Phi), \operatorname{Tr}(\Phi^2), \ldots \operatorname{Tr}(\Phi^N)$ as well as their superpartners where $\Phi$ is the $\boldsymbol{N}\otimes \overline{\boldsymbol{N}}$ chiral multiplet in the UV quiver theory. Thus, the moduli space of the chiral multiplet is $\bC^N$ spanned by $\operatorname{Tr}(\Phi^i)$ ($i=1,\ldots,N$), consistent with the right-moving central charge $c_R=3N$.

We can remove a free hypermultiplet factor $\eta(\tau)/\vartheta_4(c)$ from the above expression and obtain
\begin{equation}
    \mathcal{I}^{(2,2), N}_{\mathcal{N} = 4} = \frac{\vartheta_4(c)}{\vartheta_\alpha(c^N)} \ .
\end{equation}
We note that this expression is precisely the vacuum character of $N-1$ copies of $bc \beta \gamma$ systems labeled by $i = 1, \ldots, N-1$, with the following conformal weights $h$ and $\U(1)$ charges $m$.
\begin{center}
\begin{tabular}{c|c|c}
& $h$ & $m$ \\
\hline
$b_i$ & $\frac{1}{2}(d_i + 1)$ & $\frac{1}{2}(d_i - 1)$ \\
$c_i$ & $- \frac{1}{2}(d_i - 1)$ & $ - \frac{1}{2}(d_i - 1)$ \\
$\beta_i$ & $\frac{1}{2}d_i$ & $\frac{1}{2}d_i$ \\
$\gamma_i$ & $1 - \frac{1}{2} d_i$ & $- \frac{1}{2}d_i$\\
\end{tabular}
\end{center}
Here $d_i = i + 1$ denotes the degree of the $i$-th invariant of $\SU(N)$. The vacuum character reads (up to a factor of $i$)
\begin{equation}
    q^{\frac{1}{8}(N^2 - 1)} \prod_{i = 1}^{N - 1} \frac{
      (c^{d_i - 1}q^{\frac{d_i + 1}{2}};q)
      (c^{- d_i + 1}q^{\frac{1 - d_i}{2}};q)
    }{
      (c^{d_i}q^{\frac{d_i}{2}};q)
      (c^{-d_i} q^{1 - \frac{d_i}{2}};q)
    } = \frac{\vartheta_4(c)}{\vartheta_\alpha(c^N)} \ .
\end{equation}
The $bc \beta \gamma$ system serves as a free field realization of the chiral algebra $\chi^{\mathcal{N} = 4, N}$ of the 4d $\mathcal{N} = 4$ $\SU(N)$ SYM \cite{Bonetti:2018fqz}, and therefore is a reducible module of $\chi^{\mathcal{N} = 4, N}$. Hence, the above vacuum character is naturally a reducible module character of $\chi^{\mathcal{N} = 4, N}$.

As discussed at the end of \S\ref{sec:SU2-adjoint}, the combination of a (0,2) vector multiplet and an adjoint chiral multiplet forms a (2,2) vector multiplet. Consequently, there is no distinction between the left- and right-moving sectors. The analysis above suggests that the infrared theory is described by the chiral algebra $\chi^{\mathcal{N} = 4, N}$.

As a generalization of this case, the circular quiver in Figure \ref{fig:SUN-circular-quiver} can be obtained by SU($N$) gauging of the ends of the linear quiver in Figure \ref{fig:SUN-linear-quiver}. The elliptic genus is given by
\be 
\cI^{(0,2),N}_{1,n}=\prod_{i=1}^n \frac{\eta(q)}{\vartheta_\alpha(c_i^N)}~.
\ee 
where again $\alpha =1$ for even $N$ and $\alpha =4$ for odd $N$. Extrapolating from the $\mathcal{N} = 4$ discussion, we conjecture that the elliptic genus continues to be a module character of the chiral algebra of the 4d $\mathcal{N} = 2$ circular quiver theory.

\subsubsection{General Riemann surfaces}

Other quiver theories cannot be obtained from the (0,2) reduction of class $\cS$ theory because they involve a U(1) gauge group. Nonetheless, one can consider $\SU(N)\times \U(1)$ quiver gauge theory of genus $g>0$ with $n$ punctures where the numbers of SU($N$) and U(1) gauge groups are $2(g-1)+n$ and $g-1$, respectively. We assume that $\U(1)^{g(N-1)}$ gauge group is unbroken so that the complex dimension of the moduli space of chiral multiples is $(Ng-1+n)$. Thus, the central charges are given by
\be 
c_L=(3N+1)g-4+2n~, \qquad  c_R=3(Ng-1+n)~.
\ee

By introducing U(1) flavor fugacities $c_i$ for the external punctures and $d_i$ for the U(1) gauging, the elliptic genus of the theory can be expressed in a simple form (up to sign)
\be \label{02-general-EG}
\cI^{(0,2),N}_{g>0,n}=\prod_{j=1}^{g-1}\frac{(-1)^\b N\vartheta_\beta(d_j^N)^2 }{\eta(q)\vartheta_1(d_{j}^{2N})}\prod_{i=1}^n \frac{\eta(q)}{\vartheta_\alpha(c_i^N)}~,
\ee 
where  $\alpha =1,\beta=4$ for even $N$ and $\alpha =4,\beta=1$ for odd $N$. 
This form is independent of quiver descriptions. Therefore, regardless of the quiver descriptions, we claim these theories all flow to the same infrared (IR) theory. Applying the formula 
\be  \eta(q)^2\int_{\JK}\frac{dc}{2\pi ic}\frac{\vartheta_4(c^{\pm N})}{\vartheta_\a(d^{N}c^{\pm N})}=\frac{(-1)^\b N\vartheta_\b(d^{N})^2}{\eta(q)\vartheta_1(d^{2N})}~,\ee 
one can convince oneself that \eqref{02-general-EG} is consistent with the TQFT structure as in Figure \ref{fig:02-TQFT}.

The distinctions between SU(2) and SU($N$) become evident in theories of genus $g$ that have $g$ U(1) gauge nodes. As demonstrated in Appendix \ref{app:SU3U1-genus1-puncture2}, the elliptic genus evaluation for the left quiver theory depicted in Figure \ref{fig:SU3U1-(1,2)-quiver-1} reveals that it is dual to an LG model. Contrarily,  the explicit evaluation shows that the elliptic genus of the right quiver theory in Figure \ref{fig:SU3U1-(1,2)-quiver-1} does not factorize into theta functions.  This observation implies that the right quiver theory does not possess an LG dual description.

\begin{figure}[ht]
    \centering
 \includegraphics[width=6cm]{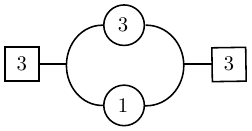}\qquad \qquad\includegraphics[width=6cm]{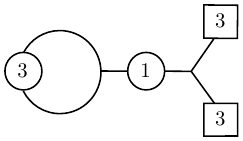}
    \caption{$\SU(3) \times \U(1)$ gauge theories corresponding to the genus-one Riemann surface with two punctures.}
    \label{fig:SU3U1-(1,2)-quiver-1}
\end{figure}

\subsection{Comments on non-Lagrangian cases}\label{sec:02-nonLag}

The (0,2) reduction described in \S\ref{sec:twist} can potentially be applied to non-Lagrangian class $\mathcal{S}$ theories, including the trinion theories $T_N$ and the Argyres-Douglas theories. However, to perform a consistent reduction, the $(0,2)$ $\U(1)$ $\frakR$-charges, represented as $\frakR=R + \frac{r - f}{2}$, must take integer values. This imposes stringent conditions on which class $\mathcal{S}$ theories can undergo consistent reduction. Upon examining the Higgs and Coulomb branch operators, the reduction seems possible for simple theories such as the $T_3$ theory and the $(A_1, D_4)$ Argyres-Douglas theory. On the other hand, the $(A_1, D_{2n + 1})$ theories do not admit consistent reduction since the dimensions of their Coulomb branch operators are fractional, resulting in a non-integral value for $R + \frac{r - f}{2}$. In the following, we will discuss potential candidates for the $(0,2)$ elliptic genus in these non-Lagrangian cases.

Recall that from \S\ref{sec:EG}, (0,2) elliptic genus in the Ramond sector\footnote{In the previous subsections, we consider (0,2) elliptic genus in the NS sector. Nonetheless, it is straightforward to transform it to the Ramond sector simply by replacing $\vartheta_4$ by $\vartheta_1$.} is expected to be a Jacobi form of weight-0 with a non-zero index which captures the 't Hooft anomaly of the flavor symmetry. Additionally, we further conjecture in \eqref{sum-characters} that the (0,2) elliptic genus for a class $\mathcal{S}$ theory $\mathcal{T}[C_{g,n}]$ should be some linear combination of the module characters of the associated VOA $\chi(\mathcal{T}[C_{g,n}])$. 
Indeed, both of the expected properties place strong constraints on the form of an elliptic genus.

The $T_3$ trinion theory of type $A_2$ is endowed with $E_6$ flavor symmetry \cite{Minahan:1996cj,Argyres:2007cn}. As uncovered in \cite[Figure 19]{Gaiotto:2009we}, the SU(2) gauging of the $T_3$ trinion theory leads to the infinite coupling limit of SU(3) $N_f=6$ superconformal theory. 
The corresponding VOA is the affine Lie algebra $(\widehat\frake_6)_{-3}$ with level $-3$ \cite{Beem:2013sza}. The algebra $\left(\widehat{\mathfrak{e}}_6\right)_{-3}$ is endowed with irreducible representations with the following highest weights \cite{arakawa2018joseph}:
\be \label{E6HW}
0, \ \ -3 \omega_1, \ \ -3 \omega_6, \ \ \omega_1-2 \omega_3,\ \ \omega_6-2 \omega_5,\ \ -2 \omega_2,\ \ -\omega_4\ .
\ee
Using the pure spinor formalism \cite{Aisaka:2008vw,Eager:2019zrc}, the following combination of the $\left(\widehat{\mathfrak{e}}_6\right)_{-3}$ characters is considered
\be \label{E6comb}
\mathcal{I}^{\mathfrak{e}_6} (\boldsymbol{m},q)= \ch_0^{\left(\hat{\mathfrak{e}}_6\right)_{-3}}(\boldsymbol{m},q) - \ch_{-3 \omega_1}^{\left(\hat{\mathfrak{e}}_6\right)_{-3}}(\boldsymbol{m},q).
\ee
This partition function can be expressed by a combination of theta functions as follows:
\begin{equation}\label{E6EG}
\mathcal{I}^{\mathfrak{e}_6}(\boldsymbol{m},q)=\frac{
    \eta(q)^{10}\big(\Theta^{\mathfrak{e}_6}_{\omega_{1}}(\widetilde{\boldsymbol{m}},q)-\Theta^{\mathfrak{e}_6}_{\omega_{6}}(\widetilde{\boldsymbol{m}},q)\big)
    }{
    \prod_{w\in\mathbf{S}}\vartheta_{1}(\boldsymbol{m}_{\mathfrak{d}_5}^{w})
        }~,
\end{equation}
where the two theta functions $\Theta^{\mathfrak{e}_6}_{\omega_{1},\omega_{6}}$ are defined  \cite{Sakai:2017ihc,10.3792/pjaa.69.247} as
\begin{align*}
    \Theta^{\mathfrak{e}_6}_{\omega_{1}}(\boldsymbol{m},q)&=
    \frac{q^{1/6}}{2}
    \sum_{k=1}^{4}\sigma_k m_0
    \vartheta_k(m_0^3q)
    \prod_{j=1}^5\vartheta_k(m_j)~,
    \\
    \Theta^{\mathfrak{e}_6}_{\omega_{6}}(\boldsymbol{m},q)&=
    \frac{q^{1/6}}{2}
    \sum_{k=1}^{4}\sigma_k m_0^{-1}
    \vartheta_k(m_0^3q^{-1})
    \prod_{j=1}^5\vartheta_k(m_j)~,
    \end{align*}
with $-\sigma_{1}=\sigma_{2}=\sigma_{3}=-\sigma_{4}=1$.
Here, $\boldsymbol{m}=(m_{0},\boldsymbol{m}_{\mathfrak{d}_5})=(m_0,m_1,\ldots ,m_5)$ are the fugacities for $\mathfrak{e}_6$ ($m_{i>0}$ are also  fugacities for the subalgebra $\mathfrak{d}_5$) in the orthogonal basis.\footnote{In contrary, the fugacities in \cite{Eager:2019zrc} are expressed in the alpha basis \cite{Feger:2012bs,Feger:2019tvk}.} In the numerator of \eqref{E6EG}, we use $\widetilde{\boldsymbol{m}}=(m_0^2, m_1,\ldots ,m_5)$, and in the denominator, $\mathbf{S}=[0,0,0,0,1]$ is the spin representation of $\mathfrak{d}_5$.

Using the branching rules, one can establish the relationships between the $\mathfrak{e}_6$ fugacities $\boldsymbol{m}$ and the $\mathfrak{a}_3\oplus\mathfrak{a}_3\oplus\mathfrak{a}_3$ fugacities $x_{i},y_{j},z_k$ in the fundamental weight basis (or the omega basis in \cite{Feger:2012bs,Feger:2019tvk}):
\bea 
&m_0=x_{1}^{\frac{1}{2}}z_{2}^{-\frac{1}{2}},&m_{3}=x_{1}^{\frac{1}{2}}y_{1}^{-1}z_{2}^{\frac{1}{2}},~\cr
&m_{1}=x_{1}^{-\frac{1}{2}}x_{2}^{-1}z_{1}z_{2}^{\frac{1}{2}},~&m_{4}=x_{1}^{\frac{1}{2}}y_{1}y_{2}^{-1},~\cr 
&m_{2}=x_{1}^{\frac{1}{2}}x_{2}z_{1}z_{2}^{\frac{1}{2}},~
&m_{5}=x_{1}^{\frac{1}{2}}y_{2}z_{2}^{\frac{1}{2}}~.
\eea
Recall from (\ref{thooft-anomaly}) that the 't Hooft anomaly can be read off from the shift property of \eqref{E6EG}. In particular, we focus on the three simple $\SU(3)$ flavor subgroups with the fugacities $x_{i},y_{j}, z_k$:
\begin{align}
x_1\rightarrow x_1q \,,\ x_2\rightarrow x_2/q,   \qquad& \mathcal{I}^{\mathfrak{e}_6}\rightarrow \frac{q^2x_1z_2^2}{x_2^3}~\mathcal{I}^{\mathfrak{e}_6} \nonumber 
\\ 
y_1\rightarrow y_1q \,,\ y_2\rightarrow y_2/q,   \qquad& \mathcal{I}^{\mathfrak{e}_6}\rightarrow (qy_1/y_2)^9~\mathcal{I}^{\mathfrak{e}_6} \nonumber  
\\
z_1\rightarrow z_1q \,,\ z_2\rightarrow z_2/q,    \qquad& \mathcal{I}^{\mathfrak{e}_6}\rightarrow \frac{q^2z_1^3}{x_1z_2}~\mathcal{I}^{\mathfrak{e}_6} ~.
\end{align}

In four dimensions, the $E_6$ theory is related to the $\SU(3)$ SQCD through the Argyres-Seiberg duality, where two of the $\SU(3)$ flavor symmetries of the former are identified with the two $\SU(3)$ flavor symmetries of the latter. If the above $\mathcal{I}^{\mathfrak{e}_6}$ truly represents the elliptic genus of the reduced $E_6$ theory, then its $\SU(3)^2$ 't Hooft anomaly should match with that of the reduced $\SU(3)$ SQCD. The elliptic genus $\mathcal{I}^{(0,2),3}_{0,2,2}$ of 2d (0,2) $\SU(3)$ SQCD with $3$ fundamentals and $3$ anti-fundamentals is the $N=3$ specialization of \eqref{SUN-EI}, and it has the  shift properties
\begin{align}
c_i\rightarrow c_iq,\ \ \qquad & \mathcal{I}^{(0,2), 3}_{0,2,2}\rightarrow q^\frac{9}{2}c_i^9~\mathcal{I}^{(0,2), 3}_{0,2,2} \nonumber 
\\ 
b_{i,1}\rightarrow b_{i,1}q,\ b_{i,2}\rightarrow b_{i,2}/q,   \qquad &\mathcal{I}^{(0,2), 3}_{0,2,2}\rightarrow (qb_{i,1}/b_{i,2})^9~\mathcal{I}^{(0,2), 3}_{0,2,2} \nonumber  
\end{align}
where $c_{i}$ are the two $\U(1)$ fugacities, and $b_{i,j}$ are the two $\SU(3)$ fugacities. However, by comparison, the two shift behaviors do not match if we identify the $SU(3)$ fugacities as $y_1 = b_{0,1}, y_2 = b_{0,2}, z_1 = b_{2, 1}, z_2 = b_{2,2}$. Hence, when we perform SU(2) gauging on the expression given in \eqref{E6EG}, it appears that we do not arrive at the (0,2) elliptic genus for $\SU(3)$ SQCD with $N_f=6$. It could be worthwhile to explore alternative combinations of $(\widehat\frake_6)_{-3}$ characters, distinct from \eqref{E6comb}, in order to compare with the $\SU(3)$ SQCD with $N_f=6$. Indeed, by the logic of \cite{Eager:2019zrc}, a linear combination of the vacuum character and the character with the highest weight $-\omega_4$ is the most promising starting point.

\bigskip

Argyres-Douglas theories \cite{Argyres:1995jj,Argyres:1995xn} constitute another interesting class of non-Lagrangian theories, whose construction involves a higher-order pole of the Higgs field in the Hitchin system \cite{Xie:2012hs}.

Let us first consider the $(A_1, D_4)$ theory. The rank-one theory contains a Coulomb branch operator with conformal dimension $\Delta = -3/2$, and therefore an integral $r$-charge $r = 2\Delta = 3$, suggesting a possible $S^2$ reduction and a corresponding (0,2) elliptic genus. The associated VOA is given by the Kac-Moody algebra $\widehat{\fraksu}(3)_{-3/2}$ \cite{Buican:2015ina,Cordova:2015nma}. The level $k = -3/2$ with respect to the $\SU(3)$ flavor symmetry is called \emph{boundary admissible} in Mathematics literature. There are four irreducible admissible highest weight modules with affine weights
\begin{equation}
    - \frac{3}{2}\widehat{\omega}_0, \quad
    - \frac{3}{2}\widehat{\omega}_1, \quad
    - \frac{3}{2}\widehat{\omega}_2, \quad
    \widehat{\rho} = - \frac{1}{2}\sum_{i = 0}^2 \widehat{\omega}_i \ ,
\end{equation}
where $\widehat{\rho}$ is the affine Weyl vector. The characters are given by
\begin{align}\
    \operatorname{ch}_{-\frac{3}{2}\widehat{\omega}_0}
    = & \ \frac{\eta(\tau) \vartheta_1(\mathsf{b}_1 - 2 \mathsf{b}_2|2\tau)
    \vartheta_1(- \mathsf{b}_1 - \mathsf{b}_2|2\tau)
    \vartheta_1(- 2 \mathsf{b}_1 + \mathsf{b}_2|2\tau)
    }{
    \eta(2\tau)
    \vartheta_1(\mathsf{b}_1 - 2 \mathsf{b}_2|\tau)
    \vartheta_1(- \mathsf{b}_1 - \mathsf{b}_2|\tau)
    \vartheta_1(- 2 \mathsf{b}_1 + \mathsf{b}_2|\tau)
    } \ ,\cr
    \operatorname{ch}_{-\frac{3}{2}\widehat{\omega}_1}
    = & \ -\frac{\eta(\tau) \vartheta_4(\mathsf{b}_1 - 2 \mathsf{b}_2|2\tau)
    \vartheta_4(- \mathsf{b}_1 - \mathsf{b}_2|2\tau)
    \vartheta_1(- 2 \mathsf{b}_1 + \mathsf{b}_2|2\tau)
    }{
    \eta(2\tau)
    \vartheta_1(\mathsf{b}_1 - 2 \mathsf{b}_2|\tau)
    \vartheta_1(- \mathsf{b}_1 - \mathsf{b}_2|\tau)
    \vartheta_1(- 2 \mathsf{b}_1 + \mathsf{b}_2|\tau)
    } \ ,\cr
    \operatorname{ch}_{-\frac{3}{2}\widehat{\omega}_2}
    = & \ -\frac{\eta(\tau) \vartheta_1(\mathsf{b}_1 - 2 \mathsf{b}_2|2\tau)
    \vartheta_4(- \mathsf{b}_1 - \mathsf{b}_2|2\tau)
    \vartheta_4(- 2 \mathsf{b}_1 + \mathsf{b}_2|2\tau)
    }{
    \eta(2\tau)
    \vartheta_1(\mathsf{b}_1 - 2 \mathsf{b}_2|\tau)
    \vartheta_1(- \mathsf{b}_1 - \mathsf{b}_2|\tau)
    \vartheta_1(- 2 \mathsf{b}_1 + \mathsf{b}_2|\tau)
    } \ , \cr
    \operatorname{ch}_{-\frac{1}{2}\widehat{\rho}}
    = & \ - \frac{\eta(\tau) \vartheta_4(\mathsf{b}_1 - 2 \mathsf{b}_2|2\tau)
    \vartheta_1(- \mathsf{b}_1 - \mathsf{b}_2|2\tau)
    \vartheta_4(- 2 \mathsf{b}_1 + \mathsf{b}_2|2\tau)
    }{
    \eta(2\tau)
    \vartheta_1(\mathsf{b}_1 - 2 \mathsf{b}_2|\tau)
    \vartheta_1(- \mathsf{b}_1 - \mathsf{b}_2|\tau)
    \vartheta_1(- 2 \mathsf{b}_1 + \mathsf{b}_2|\tau)
    } \ ,\nonumber
\end{align}
where the first one is the vacuum character of $\widehat{\fraksu}(3)_{-3/2}$ as well as the Schur index of the $(A_1, D_4)$ theory. The modular $S$-matrix is given by
\begin{equation}
    S = - \frac{1}{2}\begin{pmatrix}
      1 & 1 & 1 & -1\\
      1 & 1 & -1 & 1\\
      1 & -1 & 1 & 1\\
      -1 & 1 & 1 & 1
    \end{pmatrix} \ .
\end{equation}
There are four eigenvectors of $S$, with eigenvalues respectively $(-1, -1, -1, 1)$,
\begin{equation}
    \operatorname{ch}_{- \frac{3}{2}\widehat{\omega}_0} -  \operatorname{ch}_{- \frac{3}{2}\widehat{\rho}}, \quad
    \operatorname{ch}_{- \frac{3}{2}\widehat{\omega}_0} +  \operatorname{ch}_{- \frac{3}{2}\widehat{\omega}_1}, \quad
    \operatorname{ch}_{- \frac{3}{2}\widehat{\omega}_0} +  \operatorname{ch}_{- \frac{3}{2}\widehat{\omega}_2}, \quad
\end{equation}
and finally
\begin{equation}
    \operatorname{ch}_{- \frac{3}{2}\widehat{\omega}_0} -  \operatorname{ch}_{- \frac{3}{2}\widehat{\omega}_1} -
    \operatorname{ch}_{- \frac{3}{2}\widehat{\omega}_2}
    + \operatorname{ch}_{- \frac{3}{2}\widehat{\rho}} \ .
\end{equation}
Unfortunately, neither the $+1$ eigenvector nor any linear combination of the $-1$ eigenvectors behaves consistently under the shift of both $\SU(3)$ flavor fugacities $b_1$ and $b_2$. Consequently, no linear combination of the $\widehat{\fraksu}(3)_{-3/2}$ characters satisfies the expected properties of a (0,2) elliptic genus.

Let us also consider the $(A_1, D_{2n + 1})$ Argyres-Douglas theories which enjoy $\SU(2)$ flavor symmetry. Since these theories contain Coulomb branch operators with fractional $r$-charge, a valid (0,2) reduction is \emph{not} anticipated. Below, we contend that from the VOA perspective, a (0,2) elliptic genus does not exist. The associated VOAs are given by $\widehat{\mathfrak{su}}(2)_{k = -\frac{4n}{2n + 1}}$ \cite{Buican:2015ina,Cordova:2015nma}. In this case, the level $k = - \frac{4n}{2n + 1} = -2 + \frac{2}{2n + 1}$ is also boundary admissible, and the VOA has admissible affine weights given by (where $u:= 2n + 1$) \cite{Kac:1988qc,2016arXiv161207423K}
\begin{equation}
    \widehat{\lambda}_{k, j} = (k + \frac{2j}{u}) \widehat{\omega}_0 - \frac{2j}{u} \widehat{\omega}_1 \ , \qquad
    j = 0, 1,2, \ldots, u - 1 = 2n \ .
\end{equation}
They are highest weights of irreducible highest weight modules $L(\widehat{\lambda}_{k, j})$ of $\widehat{\mathfrak{su}}(2)_{k = -\frac{4n}{2n + 1}}$, whose characters are given by a simple formula
\begin{equation}
    \operatorname{ch}L(\widehat{\lambda}_{k, j})
    = z^{- \frac{2j}{u}} q^{\frac{j^2}{2u}}
    \frac{\vartheta_1(2\mathsf{z} - j \tau | u \tau)}{\vartheta_1(2 \mathsf{z}|\tau)} \ .
\end{equation}
Here, $z$ is the flavor $\SU(2)$ fugacity. The modular $S$-matrix is given by
\begin{equation}\label{modularS}
    S_{j j'} = \sqrt{\frac{2}{u^2 (k + 2)}} e^{\pi i(j + j')}e^{\pi i (j j'(k + 2))}
    \sin \left(\frac{\pi}{k + 2} \right)\ .
\end{equation}

However, none of the eigenvectors of the $S$-matrix transforms itself (up to a factor) under the shift $\mathsf{z} \to \mathsf{z} + \tau $ since each character transforms in the following manner (the subscript follows a cyclic rule such that $j \sim j + 2n + 1 $):
\begin{equation}
    \operatorname{ch}(\widehat{\lambda}_{k, j}) \xrightarrow{\mathsf{z} \to \mathsf{z} + \tau} (b^2 q)^{-k} \operatorname{ch}(\widehat{\lambda}_{k, j - 2}) \ .
\end{equation}
This implies that a Jacobi form from $\operatorname{ch}(\widehat{\lambda}_{k,j})$ must take the form
\begin{equation}
\textrm{const} \cdot    \sum_{j = 0}^{2n} \operatorname{ch}(\widehat{\lambda}_{k,j}) \ ,
\end{equation}
which is never an eigenvector of the $S$-matrix \eqref{modularS}. Therefore, we conclude that no linear combination of the $\widehat{\mathfrak{su}}(2)_{k = -\frac{4n}{2n + 1}}$ characters satisfies the expected properties of a (0,2) elliptic genus.

\section{\texorpdfstring{$\cN=(0,4)$}{N=(0,4)} elliptic genera for class \texorpdfstring{$\mathcal{S}$}{S} theories on \texorpdfstring{$S^2$}{S2}}\label{sec:04}

In this section, we study $\mathcal{N} = (0,4)$ theories obtained by a distinct twisted compactification of class $\mathcal{S}$ theories of type $A$ on $S^2$. In these theories, we perform a topological twist on $\U(1)_{S^2}$ with 4d $\cN=2$ superconformal $R$-symmetry $\U(1)_r\subset \SU(2)_R\times \U(1)_r$ as discussed in \cite{Kapustin:2006hi,Putrov:2015jpa}. Referencing Table \ref{tab:top-twist}, the four supercharges $Q_{-}^I,\tilde{Q}_{\dot-}^I$ ($I=1,2$) survive under this twist, and they possess the identical $\U(1)_{T^2}$ charge. Therefore, this twist preserves 2d ${\cal N}=(0,4)$ supersymmetry and thus, we refer to this twisted compactification as \emph{the (0,4) reduction} of class $\cS$ theories. 
For 4d $\cN=2$ SCFT, $\U(1)_r$-charges of operators are integral, eliminating the need for an additional twist by a flavor symmetry. The 2d ${\cal N}=(0,4)$ supersymmetry has $\SO(4)_R\cong \SU(2)_R^{-} \times \SU(2)_R^{+}$ as the UV $R$-symmetry where 4d $\SU(2)_R$ is identified with 2d $\SU(2)_R^{-}\subset \SO(4)_R$.  This subgroup subsequently evolves into the affine $\widehat\fraksu(2)$ Lie algebra within the small $\cN=4$ superconformal algebra in the infrared.
Given the (0,4) reduction of a class $\cS$ theory, we consider its IR SCFT on the Higgs branch where $\SU(2)_R^{+}$ becomes the small $\cN = 4$ superconformal $R$-symmetry in the right-moving sector. For a detailed analysis of the symmetries within this context, readers are directed to \cite{Putrov:2015jpa}.

In the (0,4) reduction, a 4d $\cN=2$ hypermultiplet reduces to a 2d $\cN=(0,4)$ hypermultiplet (two (0,2) chirals with opposite charges). Likewise, a 4d $\cN=2$ vector multiplet reduces to a 2d $\cN=(0,4)$ vector multiplet ((0,2) vector $+$ (0,2) adjoint Fermi). Consequently, for a Lagrangian theory, the basic building blocks in 2d are as follows:
\begin{figure}[ht]
    \centering
    \includegraphics[width=4cm]{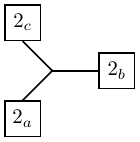} \qquad   \qquad  \includegraphics[width=4cm]{Figures/SUN.pdf}
    \caption{Left: A building block for type $A_1$, representing eight (0,2) chiral multiplets with $\SU(2)^3$ flavor symmetry. Right:  A building block for Lagrangian theories of type $A_{N-1}$, representing a (0,4) free hypermultiplet with $\SU(N)_a\times\SU(N)_b\times\U(1)_c$ flavor symmetry.}
    \label{fig:04-block}
\end{figure}

For type $A_1$, it corresponds to a sphere with three punctures. For type $A_{N-1}$, a sphere with one minimal puncture and two maximal punctures gives rise to this building block. For simplicity in notation (without distinguishing types of punctures), we denote its contribution to a (0,4) elliptic genus as $\mathcal{I}_{0,3}^{(0,4)}$. The explicit contributions of this and the vector multiplet are provided as follows:
\begin{align}
   \mathcal{I}_{0,3}^{(0,4)}(a,b,c) = &\prod_{\substack{i,j=1}}^N\frac{\eta(q)^2}{
    \vartheta_1(v(c a_i b_j)^\pm)
  } \ , \cr 
  \mathcal{I}_\text{vec}^{(0,4)}(a)= &  \frac{\big(\vartheta_1( v^2 )\eta(q) \big)^{N-1}}{N!}\prod_{\substack{A, B=1 \\ A\ne B}}^N \frac{\vartheta_1(v^2  a_A /{a}_B)
    \vartheta_1(a_A / a_B)}{\eta(q)^2}\ ,\label{04vec}
\end{align}
where the $\SU(N)$ fugacities condition is implicitly imposed 
\be 
\prod_{\substack{i=1}}^Na_i=1=\prod_{\substack{i=1}}^Nb_i~.
\ee 
The fugacity $v$ is for the Cartan subgroup of the anti-diagonal of $\SU(2)_R^{-} \times \SU(2)_R^{+}$ $R$-symmetry that commutes with the supercharges. In this physical setup, it is argued in \cite{Gadde:2013sca,Putrov:2015jpa} that the (0,4) elliptic genus is expected to be the Vafa-Witten partition function \cite{Vafa:1994tf} on $C_{g,n}\times S^2$.
In \cite{Putrov:2015jpa}, the S-duality of (0,4) theories with genus zero is confirmed by evaluating the elliptic genera. Notably, using the elliptic inversion formula \eqref{SU2-EI}, the elliptic genus of the (0,4) reduction of the non-Lagrangian $T_3$ trinion theory was obtained there.  The primary focus of this paper is to explore the (0,4) theories with a genus greater than zero, using elliptic genera.

In theories with a genus of zero, one can determine the right-moving central charge using the ${\mathrm{SU}(2)}_R^+$ anomaly
\begin{equation}\label{04-cR}
c_R = 6(2k_R)=6(n_h-n_v)~.
\end{equation}
Here, $k_R$ denotes the ${\mathrm{SU}(2)}_R^+$ anomaly coefficient, which can be computed as in \eqref{anomaly1} at UV, and $2k_R$ represents the level of the affine ${\mathrm{SU}(2)}_R^+$-symmetry. For a class $\cS$ theory $\cT_N[C_{g=0,n}]$ lacking Lagrangian description, $n_h$ and $n_v$ can be evaluated from partitions of $N$ assigned to punctures. The explicit treatment can be found in \cite{Chacaltana:2012zy,Putrov:2015jpa}, and we omit the details here. The left-moving central charge $c_L$ can be derived from the gravitational anomaly \eqref{grav}. Note that $2k_R$ represents the quaternionic dimension of the Higgs branch. Moreover, the $q\to 0$ limit of the elliptic genus agrees with the Hilbert series of the Higgs branch \cite{Hanany:2010qu} (the computational techniques are developed in \cite{Benvenuti:2006qr,Feng:2007ur,Gray:2008yu}).



On the other hand, the situation drastically changes for theories with a genus greater than zero.  In a theory with genus $g>0$, the $\U(1)^{(N-1)g}$ gauge symmetry remains unbroken on the Higgs branch \cite{Hanany:2010qu,Tachikawa:2015bga}.\footnote{The term ``Higgs branch'' refers to the moduli space of hypermultiplets where the gauge symmetry is ``maximally Higgsed''. Namely, it is the configuration in which the number of massless Abelian vector multiplets is minimized. This particular branch was previously referred to as the ``Kibble branch'' in \cite{Hanany:2010qu}.} Since a theory is expected to flow to a CFT on its Higgs branch, the right-moving central charge is equal to six times the quaternionic dimension of the Higgs branch. A general formula for the quaternionic dimension of the Higgs branch of an $A_{N - 1}$ type class $\mathcal{S}$ is discussed in \cite{Tachikawa:2015bga}. The gravitational anomaly is given by $c_R-c_L=2(n_h-n_v)$ where $n_h,n_v$ are the effective number of hypermultiplets and vector multiplets, respectively. Therefore, the central charges for the (0,4) theory of genus $g>0$ are given by
\bea \label{04-cc}
c_R=&6(n_h-n_v+g(N-1))= 6\Big[N-1+\frac{1}{2}\sum_{i = 1}^n (N^2-N - \dim_{\mathbb{C}} \mathcal{O}_i) \Big]~ ,\cr 
c_L=&4(n_h-n_v)+6g(N-1)=(4+2g) (N-1) + 2 \sum_{i = 1}^n (N^2-N - \dim_{\mathbb{C}} \mathcal{O}_i)
\eea
where $\mathcal{O}_i$ represents the nilpotent orbit corresponding to the $i$-th puncture \cite{Chacaltana:2012zy}. For a nilpotent orbit labelled by a partition $\lambda$ of $N$, the complex dimension is calculated as
$$
\operatorname{dim}_{\mathbb{C}} O_\lambda=N^2-\sum_i l_i^2
$$
where $\lambda^t=\left[l_1, l_2 \cdots\right]$ is the transpose of the Young diagram $\lambda$.
As we will see below, the $q\to 0$ limit of the elliptic genus is no longer equal to the Hilbert series of the Higgs branch. This is very similar to the relation between the Hall-Littlewood index and the Higgs branch Hilbert series \cite{Gadde:2011uv} in which the agreement can be seen for theories with genus zero but not higher.

In the following, we present closed-form expressions for the (0,4) elliptic genera of theories where the genus $g > 0$. If a theory has a Lagrangian description with a gauge group of adequately low total rank, one can straightforwardly compute the elliptic genus through the JK-residue method. To determine the elliptic genus of non-Lagrangian theories at higher genus, we exploit the inversion formula in \cite{Gadde:2015xta,Putrov:2015jpa,Agarwal:2018ejn}, namely performing additional gauging in Lagrangian theories. For detailed calculations, readers can refer to Appendix \ref{app:04}, which provides explicit JK residue computations of (0,4) elliptic genera.
The resulting closed-form expressions are remarkably simple, aligning well with the TQFT structure on punctured Riemann surfaces $C_{g,n}$.

\subsection{Type \texorpdfstring{$A_1$}{A1}}\label{sec:04-A1}
Class $\cS$ theories of type $A_1$ all have Lagrangian descriptions and are completely specified by the genus $g$ and the number of (regular) punctures $n$. We shall focus on theories at genus $g \ge 1$ with an arbitrary number of punctures. To compute elliptic genera, we can gauge the basic building block illustrated in the left of Figure \ref{fig:04-block} by using \eqref{04vec}. For $g = 1, n = 1$, the elliptic genus is computed by JK-residue where one only encounters non-degenerate poles,
\begin{align}\label{SU2-max}
    \mathcal{I}_{1,1}^{(0,4),2}(c)
    = & \int_{\textrm{JK}} \frac{da}{2\pi i a} \cI_{0,3}^{(0,4)}(a,a^{-1},c)\cI_{\textrm{vec}}^{(0,4)}(a) \cr
    = & \ \frac{\eta(q)^2 \vartheta_1(v^4)}{
      \vartheta_1(v^2)
      \vartheta_1(v^2 c^{\pm2})
    } \ .
\end{align}
The expression is a simple ratio of the theta functions, and the LG dual theory is described in \cite[\S2.2.3]{Putrov:2015jpa}. A similar computation can be performed for $g = 1, n \ge 1$, which yields
\begin{align}
    \mathcal{I}^{(0,4), 2}_{1, n}(c_1,\ldots,c_n)
    = & \ \prod_{i = 1}^n \frac{\eta(\tau)^2 \vartheta_1(v^{4})}{\vartheta_1(v^{2}) \vartheta_1(v^{2}  c_i^{\pm2})} \ .
\end{align}
While the Higgs (Kibble) branch Hilbert series was computed in \cite[\S4.2.2]{Hanany:2010qu} for $n=2$, the relation between the elliptic genus and the Hilbert series is unclear. Consequently, although the form of the elliptic genus suggests the existence of an LG dual theory,  its precise description remains unknown to us. 

It is straightforward to obtain the elliptic genus for higher genera. For example, the theory of genus two 
\bea 
    \mathcal{I}^{(0,4), 2}_{2,0}=& \int_{\textrm{JK}} \frac{da}{2\pi i a} \cI_{1,2}^{(0,4),2}(a,a^{-1})\cI_{\textrm{vec}}^{(0,4)}(a)= \int_{\textrm{JK}} \frac{da}{2\pi i a} (\cI_{1,1}^{(0,4),2}(a))^2\cI_{\textrm{vec}}^{(0,4)}(a) \cr
    = &\frac{\vartheta_{1}(v^{2})\vartheta_{1}(v^{4})}{\eta(q)^2}
\eea 
Moreover, we can increase the genus $g$ by gluing together any number of pairs of punctures, and 
the final result takes a simple form
\begin{equation}
    \mathcal{I}_{g,n}^{(0,4),2}(c_1,\ldots,c_n)=\bigg(\frac{\vartheta_{1}(v^{2})\vartheta_{1}(v^{4})}{\eta(q)^2}\bigg)^{g-1}
    \prod_{i=1}^n\frac{
      \eta(q)^{2} \vartheta_{1}(v^{4})
    }{
      \vartheta_{1}(v^{2})
      \vartheta_{1}(v^{2} c_i^{\pm2})
    } \ .
\end{equation}
This result is consistent with the cut-and-join TQFT structure on $C_{g,n}$ so that
\begin{align}
    \mathcal{I}^{(0,4),2}_{g + 1, n}(c_1,\ldots,c_n)
    = &\int_{\textrm{JK}} \frac{da}{2\pi i a} \mathcal{I}_{g, n + 2}^{(0,4), 2}(c_1, \cdots, c_n,a,a^{-1}) \mathcal{I}^{(0,4)}_\text{vec}(a) \ ,\cr
    \mathcal{I}^{(0,4),2}_{g_1+g_2, n_1+n_2}(c_1,\ldots,c_{n_1+n_2})=&\int_{\textrm{JK}}\frac{da}{2\pi i a}\mathcal{I}_{g_1, n_1+1}^{(0,4), 2}(c_1, \cdots,a)\mathcal{I}^{(0,4)}_\text{vec}(a)\mathcal{I}_{g_2, n_2+1}^{(0,4), 2}(c_{n_1+1}, \cdots,a^{-1})~.\nonumber
\end{align}

\subsection{Type \texorpdfstring{$A_2$}{A2}}\label{sec:04-A2}

For class $\cS$ theories of type $A_2$, there are two types of regular punctures the minimal punctures with flavor symmetry $\U(1)$ and the maximal punctures with flavor symmetry $\SU(3)$. We denote the number of these punctures as $ n_1 $ and $ n_3 $, respectively. To begin with, we can gauge the basic building block illustrated in the right of Figure \ref{fig:04-block} by using \eqref{04vec} to compute the elliptic genus for $g = 1$ with several minimal punctures. This can be swiftly computed using the JK residue from which we can postulate a general formula
\be 
  \mathcal{I}^{(0,4),3}_{g = 1; n_1, 0}(c_1,\ldots,c_{n_1})
  =  \prod_{i = 1}^{n_1}
  \frac{\eta(q)^2\vartheta_1(v^{6})}{
    \vartheta_1(v^{2})
    \vartheta_1(v^{3}c_i^{\pm3})
  }~. 
\ee

\begin{figure}[ht]
    \centering
    \includegraphics[width=13cm]{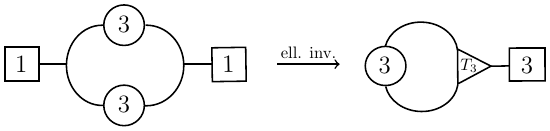}
    \caption{Application of the elliptic inversion formula leads to the elliptic genus for a maximal puncture.}
    \label{fig:elliptic-inversion1}
\end{figure}
To access the elliptic genus in the presence of maximal punctures, we apply the elliptic inversion formula in \cite{Putrov:2015jpa} that computes the (0,4) elliptic genus of the $T_3$ theory. Specifically, starting from $g=1, n_1=2$, we use the inversion formula \eqref{SU2-EI} to obtain $g=1,n_3=1$ (Figure \ref{fig:elliptic-inversion1})
\bea \label{SU3-max}
  \mathcal{I}^{(0,4),3}_{g = 1; 0, n_3=1}(b)=& \frac{\eta(q)^5}{2\vartheta_1(v^{2}z^{\pm2})}\int_{\textrm{JK}} \frac{ds}{2\pi i s} \frac{\vartheta_1(s^{ \pm 2}) \vartheta_1(v^{-2})}{\vartheta_1(v^{-1} s^{ \pm 1} z^{ \pm 1})} \mathcal{I}^{(0,4),3}_{g = 1; n_1, 0}(s^{\frac13}/r,s^{-\frac13}/r)\cr 
   = &\ \frac{\eta(q)^6\vartheta_1(v^{2})\vartheta_1(v^{4}) \vartheta_1(v^{6})}{
    \prod_{\substack{A, B = 1}}^3 \vartheta_1(v^{2} b_{A} /b_{B})}
\eea 
where the SU(3) fugacities for the maximal puncture is identified by $(b_1, b_2, b_3) =(r z, r / z, r^{-2})$.
The detailed computations of the elliptic genus for type $A_2$ theories are collected in Appendix \ref{app:04-SU3}. In summary, the elliptic genus for the (0,4) reduction of class $\cS$ theory $\cT_3[C_{g;n_1,n_3}]$ is given by the following simple form
\begin{align}
  \mathcal{I}^{(0,4),3}_{g; n_1, n_3}
  = \bigg(
  \frac{\vartheta_1(v^{2}) \vartheta_1(v^{4})^2 \vartheta_1(v^{6})}{
    \eta(q)^4
  }
  \bigg)^{g - 1}
  \mathcal{I}^{(0,4),3}_{1; n_1, 0}
  \mathcal{I}^{(0,4),3}_{1; 0, n_3} \ ,
\end{align}
where
\begin{align}
  \mathcal{I}^{(0,4),3}_{g = 1; 0, n_3}
  = \ \prod_{i = 1}^{n_3}   \frac{\eta(q)^6\vartheta_1(v^{2})\vartheta_1(v^{4}) \vartheta_1(v^{6})}{
    \prod_{\substack{A, B = 1}}^3 \vartheta_1(v^{2} b_{iA}^{} /b_{iB}^{})
  }\ .
\end{align}

\subsection{Type \texorpdfstring{$A_3$}{A3}}\label{sec:04-A3}

For type $A_3$ theories, there are four types of regular punctures whose partitions and flavor symmetries are given as follows:
\begin{itemize}[nosep]
    \item $[3,1]  \qquad \qquad    \U(1)$
    \item $[2,1,1]\qquad \quad \SU(2)\times \U(1)$
    \item $[2,2] \qquad \qquad \SU(2)$
        \item $[1,1,1,1]\qquad \, \SU(4)$
\end{itemize}
 We use the notations $ n_1, n_2, n_3, n_4 $ to denote the numbers of these punctures, respectively. 
For $g = 1$ only with minimal punctures ($n_2 = n_3 = n_4 = 0$), it is straightforward to compute the elliptic genus
\begin{equation}\label{SU4-min}
    \mathcal{I}^{(0,4), 4}_{1; n_1, 0,0,0}
    = \prod_{i = 1}^{n_1} \frac{\eta(q)^2 \vartheta_1(v^{8})}{
      \vartheta_1(v^{2})
      \vartheta_1(v^{4} c_i^{\pm4})
    } \ .
\end{equation}

\begin{figure}[ht]
    \centering
    \includegraphics[width=13cm]{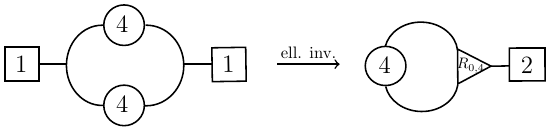}
    \caption{Application of the elliptic inversion formula leads to the elliptic genus for a [2,1,1] puncture.}
    \label{fig:elliptic-inversion2}
\end{figure}
Applying the elliptic inversion formula in \cite{Agarwal:2018ejn},  we can calculate the elliptic genus of genus-one theories that have [2,1,1], and [2,2] punctures. As a specific case, when there is solely one [2,1,1] puncture, the elliptic genus is as follows:
\begin{align}
    \mathcal{I}^{(0,4),4}_{g = 1; 0,1,0,0}=& \frac{\eta( q)^5}{2 \vartheta_1(v^2 z^{ \pm 2})} \int_{\mathrm{JK}} \frac{d s}{2 \pi i s} \frac{\vartheta_1(s^{ \pm 2}) \vartheta_1(v^{-2})}{\vartheta_1(v^{-1} s^{ \pm 1} z^{ \pm 1})}  \mathcal{I}^{(0,4),3}_{g = 1;2,0,0,0}(s^{\frac14}/r,s^{-\frac14}/r)\cr
    = &\frac{ \eta(q)^6 \vartheta_1(v^{6}) \vartheta_1(v^{8})}{
        \vartheta_1(v^{2})^2
        \vartheta_1(v^{2} z^{\pm2 })
        \vartheta_1(v^{3} z^{\pm}  r^{\pm4})
    } \ ,
\end{align}
where $z,r$ denote the $\SU(2)$ and $\U(1)$ fugacities respectively. (See Figure \ref{fig:elliptic-inversion2}.) With only one $[2,2]$ puncture, the elliptic genus
\bea 
    \mathcal{I}^{(0,4),4}_{g = 1; 0,0,1,0} =& \frac{\eta( q)^5 \vartheta_1(v^2) \vartheta_1(v^{-2})}{2 \vartheta_1(v^4)} \int_{\mathrm{JK}} \frac{d s}{2 \pi i s} \frac{\vartheta_1(s^{ \pm 2})}{\vartheta_1(s^{ \pm 1}) \vartheta_1(v^{-2} s^{ \pm 1})} \mathcal{I}^{(0,4),3}_{g = 1;2,0,0,0}(s^{\frac14}/w,s^{-\frac14}/w)\cr 
    = & \ \frac{
      \eta(q)^4\vartheta_1(v^{6})
      \vartheta_1(v^{8})
    }{
      \vartheta_1(v^{2})
      \vartheta_1(v^{4})
      \vartheta_1(v^{4}  w^{\pm4})
      \vartheta_1(v^{2}  w^{\pm4})
    } \ .
\eea

\begin{figure}[ht]
    \centering
\includegraphics[width=0.7\textwidth]{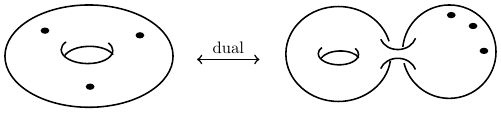}\\
\includegraphics[width=0.8\textwidth]{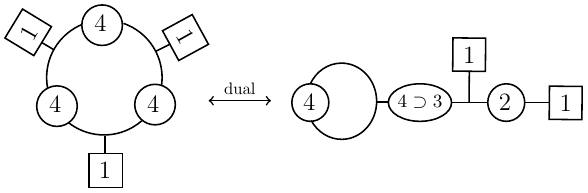} 
    \caption{The S-duality in the class $\cS$ theory of type $A_3$. The right theory involves gauging the $A_3$ trinion theory. }
    \label{fig:SU4-Sduality}
\end{figure}

While the derivation of the elliptic genus for the theory of genus one with a maximal puncture remains unknown, extrapolation from the results in \eqref{SU2-max} and \eqref{SU3-max} allows us to propose the following expression
\begin{equation}
    \mathcal{I}^{(0,4),4}_{g=1; 0,0,0,1}
    = \frac{
        \eta(q)^{12} \vartheta_1(v^{2})\vartheta_1(v^{4})\vartheta_1(v^{6}) \vartheta_1(v^{8})}{\prod_{\substack{A, B = 1}}^4
    \vartheta_1(v^{2}  b_{A} / b_{B})
    } \ .
\end{equation}
The validity of our proposed formula can be tested by examining the S-duality in Figure \ref{fig:SU4-Sduality}. Gauging this theory as in the right figure leads to the theory of genus one with three minimal punctures so that we can compare the result with \eqref{SU4-min}. The detailed computations of the elliptic genera are collected in Appendix \ref{app:04-SU4}.

This allows us to further increase the genus $g$, and finally, the elliptic genus of a general $A_3$ theory is given by
\begin{equation}    \mathcal{I}^{(0,4), 4}_{g; n_1, n_2, n_3, n_4}
    = \bigg(
      \frac{
      \vartheta_1(v^{2})
      \vartheta_1(v^{4})^2
      \vartheta_1(v^{6})^2
      \vartheta_1(v^{8}) 
      }{
      \eta(q)^6
      }
    \bigg)^{g - 1}
    \mathcal{I}^{(0,4), 4}_{1; n_1,0,0,0}
    \mathcal{I}^{(0,4), 4}_{1; 0,n_2,0,0}
    \mathcal{I}^{(0,4), 4}_{1; 0,0,n_3,0}
    \mathcal{I}^{(0,4), 4}_{1; 0,0,0,n_4} \ ,\nonumber
\end{equation}
where
\begin{align}
    \mathcal{I}^{(0,4), 4}_{1; n_1,0,0,0}
    = & \ \prod_{i = 1}^{n_1} \frac{\eta(q)^2\vartheta_1(v^{8})}{
    \vartheta_1(v^{2})
    \vartheta_1(v^{4}  c_i^{\pm4})
    } \\
    \mathcal{I}^{(0,4), 4}_{1; 0,n_2,0,0}
    = & \ \prod_{i = 1}^{n_3} \frac{
      \eta(q)^6 \vartheta_1(v^{6})\vartheta_1(v^{8})
    }{
    \vartheta_1(v^{2})^2
    \vartheta_1(v^{2} z_i^{\pm2} )
    \vartheta_1(v^{3} z_i^{\pm} r^{\pm4}_i)
    } \\
    \mathcal{I}^{(0,4), 4}_{1; 0,0,n_3,0}
    = & \ \prod_{i = 1}^{n_2}
    \frac{\eta(q)^4 \vartheta_1(v^{6}) \vartheta_1(v^{8})}{
    \vartheta_1(v^{2})
    \vartheta_1(v^{4})
    \vartheta_1(v^{2}  w_i^{\pm4})
    \vartheta_1(v^{4}  w_i^{\pm4})
    } \ ,\\
    \mathcal{I}^{(0,4), 4}_{1; 0,0,0,n_4}
    = & \ \prod_{i = 1}^{n_4}
    \frac{
          \eta(q)^{12}\vartheta_1(v^{2}) \vartheta_1(v^{4})\vartheta_1(v^{6}) \vartheta_1(v^{8})}{\prod_{\substack{A, B = 1}}^4
      \vartheta_1(v^{2}  b_{iA} / b_{iB})
      }
  \end{align}

\subsection{Type \texorpdfstring{$A_{N-1}$}{A(N-1)} and TQFT structure}\label{sec:04-AN-1}

From the above results, we can observe a simple TQFT structure in the $\mathcal{N} = (0,4)$ elliptic genus of the class $\mathcal{S}$ theory at genus $g \ge 1$.  This structure suggests that the elliptic genus can be expressed as a straightforward product of contributions from individual punctures and handles. Therefore, the $\mathcal{N} = (0,4)$ elliptic genus corresponding to type $A_{N -1}$ is expected to have the following form:
\begin{equation}\label{04-TQFT}
    \mathcal{I}_{g, n}^{(0,4), N} = (\cH_N)^{g - 1} \prod_{i = 1}^n \mathcal{I}_{\lambda_i}^{(0,4),N}(b_i) \ ,
\end{equation}
where $g$ is the genus of the associated Riemann surface, and $n$ collectively denotes the number of punctures, with their internal data represented by partitions (Young diagrams). The function $\mathcal{I}_{\lambda_i}$ captures the contribution from the $i$-th puncture labelled by a partition $\lambda_i$, and $\cH_N$ encapsulates the contribution originating from a handle. Loosely speaking, this expression resembles the TQFT expression of the 4d $\mathcal{N} = 2$ superconformal index \cite{Gadde:2011ik,Gadde:2011uv}, which involves an infinite sum over representations of $\SU(N)$ schematically as
\begin{equation}
    \mathcal{I}^\text{4d} = \sum_{\mu} H_\mu^{2g - 2 + n} \prod_i \psi_{\mu}^{(\lambda_i)}(b_i) \ .
\end{equation}

We expect the elliptic genus $\mathcal{I}_{g, n}$ to obey a TQFT structure under cutting and gluing. Let us consider the maximal puncture corresponding to the integer partition $[1^N]$, which contributes
\begin{equation}\label{SUN-max}
    \mathcal{I}^{(0,4), N}_{[1^N]}(b) = \frac{\eta(q)^{N^2-N}\prod_{M=1}^N\vartheta_{1}(v^{2M})}{\prod_{A, B}^N\vartheta_{1}(v^{2}b_A^{}/b_B^{})} \ ,
\end{equation}
where $b_A$ denotes the $\SU(N)$ flavor fugacities with constraint $b_1 \cdots b_N = 1$.

\begin{figure}[ht]
    \centering
    \includegraphics[width=\textwidth]{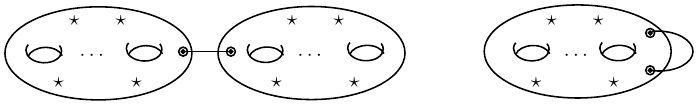}
    \caption{The gluing maximal punctures leads to a new Riemann surface, and (0,4) elliptic genera are consistent with the cut-and-join procedure on Riemann surfaces $C_{g,n}$.}
    \label{fig:04-TQFT}
\end{figure}

Consider two Riemann surfaces, labeled as $C_{g_1,n_1}$ and $C_{g_2, n_2}$, each with a maximal puncture. By $\SU(N)$ gauging, these two maximal punctures can be joined together, which results in a new Riemann surface $C_{g_1+g_2, n_1+n_2-2}$. 
Similarly, if a Riemann surface $C_{g,n}$ possesses more than two maximal punctures, by gauging the diagonal of the $\SU(N)^2$ flavor symmetry originating from these two maximal punctures, we can transform this surface into a new Riemann surface $C_{g+1,n-2}$. These processes can be visualized in Figure \ref{fig:04-TQFT}. For the form of (0,4) elliptic genus \eqref{04-TQFT} to be compatible with these procedures, the handle contribution must be 
\begin{align} \label{SUN-handle}
\cH_N
    = & \ \int_\text{JK} \frac{d\boldsymbol{a}}{2\pi i \boldsymbol{a}}
 \mathcal{I}^{(0,4), N}_{[1^N]}(\boldsymbol{a})
 \mathcal{I}^{(0,4), N}_{[1^N]}(\boldsymbol{a}^{-1})
    \mathcal{I}_\text{vec}^{(0,4)}(a) \cr 
    =&\frac{\prod_{M = 1}^{N} \vartheta_1(v^{2 M})^2}{N!\eta(q)^{N-1}\vartheta_1(v^2)^{N+1}}\int_{\textrm{JK}} \frac{d\boldsymbol{a}}{2\pi i\boldsymbol{a}}\prod_{A\neq B} \frac{\vartheta_1(a_A/a_{B})}{\vartheta_1(v^2 a_A/a_{B})}~,\cr
    = & \frac{\prod_{M = 1}^{N} \vartheta_1(v^{2 M})^2
    }{\eta(q)^{2(N - 1)}\vartheta_1(v^2)\vartheta_1(v^{2 N})}. 
\end{align}
The JK integral is analogous to the one in \eqref{inversion2}. These results \eqref{SUN-max} and \eqref{SUN-handle} reduce to those in the previous examples when $N = 2,3,4$. Furthermore, given that the (0,4) elliptic genus form in \eqref{04-TQFT} receives only local contributions, verifying the following properties is straightforward
\begin{align}
\mathcal{I}^{(0,4),N}_{g_1+g_2, n_1+n_2-2}(b,c)=&\int_{\mathrm{JK}} \frac{d a}{2 \pi i a} \mathcal{I}^{(0,4),N}_{g_1, n_1}(b,a) \mathcal{I}^{(0,4),N}_{g_2, n_2}(c,a^{-1}) \mathcal{I}_{\mathrm{vec}}^{(0,4)}(a)~,\cr 
    \mathcal{I}^{(0,4),N}_{g + 1, n - 2}(b) = &\int_\text{JK} \frac{da}{2\pi i a} 
    \mathcal{I}^{(0,4),N}_{g, n}(b, a, a^{-1}) \mathcal{I}_\text{vec}^{(0,4)}(a)~.
\end{align}

The contribution from other punctures can be derived from the maximal one using nilpotent Higgsing. When transitioning from the maximal one to another type, an operator $\cO$, which is charged under the $\SU(N)$ flavor symmetry, acquires a nilpotent VEV $\langle \cO\rangle$ with Jordan blocks of sizes 1 and $N-1$, which specifies an embedding $\SU(2)\hookrightarrow\SU(N)$ \cite{Chacaltana:2012zy}.
Following \cite{Agarwal:2018ejn}, we propose that this nilpotent Higgsing procedure can be implemented at the level of the elliptic genus as follows. The contribution from a puncture defined by an integer partition $\lambda$ of $N$ is given by
\begin{equation}\label{Higssing}
    \mathcal{I}_\lambda(c) = \lim_{b \to c} \bigg[  \frac{K_\lambda(c)}{K_{[1^N]}(b)} \bigg]_{\Gamma(t^\alpha z) \to \frac{\eta(q)}{\vartheta_1( {v}^{2\alpha}{z})} }\mathcal{I}_{[1^N]}(b) \ .
\end{equation}
Here, $b$ denotes the flavor fugacities associated to the puncture, and the function $K$ is defined using the plethystic exponential \eqref{PE} as
\begin{equation}
    K_\lambda(c) 
    := \operatorname{PE}\bigg[\sum_j \frac{t^{j + 1} - pqt^j}{(1-p)(1 - q)}
    \ch^f_{\mu_j}(c)
    \bigg]  \ .
\end{equation}
The ratio $K_\lambda/K_{[1^N]}$ in \eqref{Higssing} can always be expressed by elliptic Gamma functions, which will be shown at the end of this section.

The replacement $b \to c$ and the $K_\lambda$ should be understood in the following way \cite{Gadde:2013fma}. Recall that the integer partition $\lambda$ captures an embedding of $\SU(2)$ into $\SU(N)$. The adjoint representation of $\SU(N)$ decomposes with respect to this embedding
\begin{equation}
    \mathbf{adj} = \oplus_j \mu_j \otimes \sigma_j \ ,
\end{equation}
where $\sigma_j$ denotes the spin-$j$ representation of the embedded $\SU(2)$, and the $\mu_j$ denotes a representation of the commutant, namely, the flavor symmetry $f$ of the puncture. With the decomposition, one writes down an equality of characters
\begin{equation}\label{adj-decomposition}
    \ch_\mathbf{adj}(b) = \sum_j \ch_{\mu_j}^f(c) \ch_{\sigma_j}^{\SU(2)}(t^{1/2}) \ ,\quad
    \operatorname{ch}_{\sigma_j}^{\SU(2)}(t^{1/2})
    = \sum_{m = -j}^j t^m \ .
\end{equation}
determining the substitution $b \to c$ (up to Weyl transformation)\footnote{To actually obtain the decomposition (\ref{adj-decomposition}) of character, one may start by finding the replacement $b \to c$ associated to the embedding. The process to find $b \to c$ effectively starts by addressing a simpler equation:
\begin{equation}
    \ch_\mathbf{fund}(b) = \sum_j \ch_{\mu'_j}^f(c) \ch_{\sigma_j}^{\SU(2)}(t^{1/2})
\end{equation}
with respect to the decomposition of the $\SU(N)$ fundamental $\mathbf{fund} = \oplus_j \mu'_j \otimes \sigma_j$. Then, the replacement can then be reintroduced back to (\ref{adj-decomposition}) and help determine pairs $\mu, \sigma$ of representations.}.

 For example, the partition $\lambda = [1^N]$ corresponding to the maximal puncture simply means a trivial embedding of $\SU(2)$, and therefore $f = \SU(N)$, $j = 0$ and $\mu_0 = \mathbf{adj}$. The $K$-function then reads
\begin{equation}
    K_{[1^N]}(b) = \operatorname{PE} \bigg[  \frac{t - p q}{(1 - p)(1-q)} \ch_\mathbf{adj}(b) \bigg] \ ,
\end{equation}
where $b = (b_1, b_2, \ldots, b_{N - 1}, b_N)$ denotes the fugacities of the flavor $\SU(N)$.  As another example, when $N = 4$, $\lambda = [2, 2]$, the flavor symmetry is $f = \SU(2)$. The adjoint of $\SU(4)$ decomposes as $\mathbf{adj} = ([j = 1] \oplus [j = 0]) \otimes \sigma_{j = 1} \oplus [j = 1] \otimes \sigma_{j = 0}$ where $[j]$ simply denotes the spin-$j$ representation with respect to $f$. The replacement $b \to c$ reads $(b_1, b_2, b_3, b_4) \to (ct^{1/2}, ct^{-1/2}, c^{-1}t^{1/2}, c^{-1}t^{-1/2})$, and the $K$ function is given by
\begin{equation}
    K_{[2^2]}(c) = \operatorname{PE}\Bigg[
    \frac{(t - p q)}{(1 - q)(1 - p)}(c^2 + \frac{1}{c^2} + 1)
    + \frac{(t^2 - p q t)}{(1 - q)(1 - p)}(c^2 + \frac{1}{c^2} + 2)
    \Bigg] \ .
\end{equation}

Another case of interest is the principal embedding $\lambda = [N]$ corresponding to trivial flavor symmetry. The fugacity takes the principal specialization
\begin{equation}
    (b_1, \cdots, b_N) \to (t^{-\frac{N - 1}{2}}, t^{- \frac{N - 3}{2}}, \cdots, t^{\frac{N - 3}{2}}, t^{- \frac{N - 1}{2}}) \ .
\end{equation}
The adjoint of $\SU(N)$ is decomposed into $\mathbf{adj} = \oplus_{i = 1}^{N - 1}\sigma_{i}$ and the $K$ function is
\begin{equation}
    K_{[N]}(c) =\operatorname{PE}\bigg[
    \sum_{i = 1}^{N - 1} \frac{t^{i + 1} - p q t^i}{(1 - p)(1 - q)}
    \bigg] \ .
\end{equation}
Note that there is no real $c$ dependence since the corresponding flavor symmetry is trivial. We should regard the principal embedding as removing the puncture entirely. Hence,
\begin{equation}
    \lim_{b \to c} \bigg[\frac{K_{[N]}(c)}{K_{[1^N]}(b)}\bigg]_{\Gamma(t^\alpha z) \to \frac{\eta(\tau)}{\vartheta_1(2\alpha \mathsf{v} + \mathsf{z})} } \mathcal{I}_{[1^N]}(b) = 1 \ .
\end{equation}
This illustrates the closure of a puncture.

In this way, the contribution from a puncture with hook-type Young diagram can be read as
\begin{align}
\mathcal{I}_{[N-K,1^K]}=\frac{\prod_{M=N-K+1}^N\vartheta_{1}(v^{2M})}{\eta(q)^K}
\frac{\eta(q)^{2K}}{\prod_{A=1}^K\vartheta_{1}(v^{N-K+1}(r^Nc_{A})^{\pm})}
\frac{\eta(q)^{K^2}}{\prod_{A,B=1}^K\vartheta_{1}(v^2c_{A}/c_{B})}~,
\end{align}
where $r$ and $c_A$ represent the flavor fugacities. In particular, for genus one with a simple puncture, the elliptic genus is given by
\bea 
\mathcal{I}^{(0,4),N}_{g=1,n=1}=&\frac{\eta(q)^2\vartheta_{1}(v^{2N})}{\vartheta_{1}(v^{2}) \vartheta_{1}(v^{N}c^{\pm N})}\cr 
=&\frac{\eta(q)^2}{\vartheta_{1}(v c^{\pm 1})}\prod_{i=2}^{N} \frac{\vartheta_{1}(v^{2i})}{\vartheta_{1}(v^{2(i-1)})}\frac{\vartheta_{1}(v^{i-1} c^{\pm (i-1)})}{\vartheta_{1}(v^{i} c^{\pm i})}~.
\eea 
The first term represents the contribution from a free hypermultiplet, while the subsequent terms account for $\operatorname{Tr}(\Phi\wt\Phi)^i$, $\operatorname{Tr}(\Phi^i)$, and $\operatorname{Tr}(\wt\Phi^i)$ ($i=2,\ldots,N$) where $(\Phi,\wt\Phi)$ is the $\boldsymbol{N}\otimes \overline{\boldsymbol{N}}$ hypermultiplet in the UV quiver theory. Indeed, the Higgs branch of the theory \cite{Witten:1997yu} is 
\begin{align}
\frac{\mathbb{C}^{N}\times \mathbb{C}^{N}}{S_N} = \mathbb{C}^2 \times \frac{\frakt_\mathbb{C} \times \frakt_\mathbb{C}}{S_N}~.
\end{align}
On the left-hand side, the eigenvalues of the mutually commuting $(\Phi, \wt\Phi)$ span $\mathbb{C}^{N}\times \mathbb{C}^{N}$ which receives the action of the Weyl group $S_N$. On the right-hand side, the first factor $\mathbb{C}^2$ is spanned by $\operatorname{Tr}(\Phi)$ and $\operatorname{Tr}(\wt\Phi)$ (the free hypermultiplet), and the second represents the Higgs branch of 4d $\cN=4$ SCFT \cite{Bonetti:2018fqz,Arakawa:2023cki} where $\frakt_\bC$ is the Cartan subalgebras of $\fraksl(N)$. Hence, its quaternionic dimension is $N$, consistent with the right-moving central charge $c_R=6N$, as derived from \eqref{04-cc}.

In all the above ratios of $K$ we will take a replacement $\Gamma(t^\alpha z) \to \frac{\eta(\tau)}{\vartheta_1(2 \alpha \mathsf{v} + \mathsf{z})}$ at the end of computation. This is possible thanks to the fact that $\lim_{b\to c}K_\lambda/K_{[1^N]}$ is always a product of elliptic Gamma functions. It can be seen by explicitly writing out the plethystic exponential
\begin{equation}
    \lim_{b \to c}\frac{K_{\lambda}(c)}{K_{[1^N]}(b)}
    = \operatorname{PE}\bigg[
    \frac{t - pq}{(1-p)(1-q)}\sum_{j} \operatorname{ch}^f_{\mu_j}(c) \Big(t^j - \operatorname{ch}^{\SU(2)}_{\sigma_j}(t^{1/2}) \Big)
    \bigg]\ .
\end{equation}
For each $j$ in the above sum, we write explicitly
\begin{equation}
    (t - pq) (t^j - \operatorname{ch}_{\sigma_j}^{\SU(2)}(t^{1/2}))
    = - (t - pq) (t^{ - j} + t^{-j+1} + \cdots + t^{j - 2} + t^{j - 1}) \ .
\end{equation}
Note that the adjoint representation $\mathbf{adj}$ is real, hence we have
\begin{equation}
    \operatorname{ch}_{\mu_j}^f(c) = \operatorname{ch}_{\mu_j}^f(c^{-1}) \ .
\end{equation}
Therefore,
\begin{align}
     (t - pq) (t^j - \operatorname{ch}_{\sigma_j}^{\SU(2)}(t^{1/2}))\operatorname{ch}_{\mu_j}^f(c)
     = & \ - (t^{- j + 1} + t^{- j + 2} + \cdots + t^{j - 1} + t^j)\operatorname{ch}_{\mu_j}^f(c) \nonumber \\ 
     & + pq (t^{- j + 1} + t^{- j + 2} + \cdots + t^{j - 1} + t^j) \operatorname{ch}_{\mu_j}^f(c)\bigg|_{\substack{t \to t^{-1}\\
     c \to c^{-1}}} \nonumber \ .
\end{align}
Here we see the structure of $x - \frac{pq}{x}$ emerges, and the ratio of $K$ precisely forms a product of elliptic Gamma functions. At the end of the computation, the replacement $\Gamma(t^\alpha z) \to \frac{\eta(\tau)}{\vartheta_1(2 \alpha \mathsf{v} + \mathsf{z})}$ should be performed to derive the contribution from a puncture to the $(0,4)$ elliptic genus.

\bigskip
Our findings give rise to various intriguing questions and potential research directions. Thus, we conclude this section by highlighting a few prospective avenues for future exploration.

\paragraph{Simplicity of forms:} The (0,4) elliptic genus manifests in surprisingly simple forms, primarily as products of theta functions. However, such simplicity is observed only in theories where the genus is greater than zero. The underlying reasons for this remain mysterious.

\paragraph{Non-linear sigma model:} Our investigations point toward an $\mathcal{N}=(0,4)$ non-linear sigma model as an infrared theory. The target space of this model is the moduli space of the hypermultiplet with a nontrivial left-moving bundle. Notably, the form of the (0,4) elliptic genus strongly indicates the existence of an LG dual theory for this class of theories. An immediate challenge is to identify the superpotential of the LG model, which realizes the target space of the non-linear sigma model.

\paragraph{Relation with Schur indices:} Prior works, such as \cite{DelZotto:2016pvm,DelZotto:2017mee,DelZotto:2018tcj,Gu:2019dan}, have brought up the relationship between (0,4) elliptic genera and Schur indices. However, the findings in these works remain observational and lack a foundational understanding. Therefore, a deeper analysis of the (0,4) elliptic genus presented in this paper, in light of these observations, is a promising avenue.

\acknowledgments
The authors express their gratitude to Richard Eager, Guli Lockhart and Eric Sharpe for clarifying the result of \cite{Eager:2019zrc} and offering feedback on our manuscript. Our appreciation also goes to Jaewon Song for his insightful comments on our manuscript, and to Wei Cui, Junkang Huang, Zixiao Huang, Jiaqun Jiang, Marcus Sperling, Jingxiang Wu and Shutong Zhuang for the enlightening discussions. A special thanks is due to Matteo Sacchi for pointing out the issue of (0,2) central charges in the first arXiv version. 
The research of S.N. is supported by National Natural Science Foundation of China No.12050410234 and Shanghai Foreign Expert grant No. 22WZ2502100. The research of Y.P. is supported by the National Natural Science Foundation of China (NSFC) under Grant No. 11905301.

\appendix

\section{Notations and conventions}\label{app:notation}

In this paper, the symbol $q$ is defined as $q := e^{2\pi i \tau}$, where $\tau$ is a complex structure of a two-torus. Throughout the paper, single symbols written in sans-serif type are used to represent chemical potentials. The fugacity $z$ and the chemical potential $\mathsf{z}$ for either gauge or flavor symmetry are related by the equation $z = e^{2\pi i \mathsf{z}}$. Abusing notations, functions with fugacities and chemical potentials will be used interchangeably. For example, the following two notations represent the same theta function 
\begin{align}
  \vartheta_1(\mathsf{z}) = \vartheta_1(z) \ .
\end{align}

The notation $f(a^\pm b^\pm)$ is a shorthand notation used to denote the multiplication of all possible combinations of signs in the arguments. It is defined as follows:
\begin{align}\label{signs}
f(a^\pm b^\pm):=&f(ab)f(a^{-1}b)f(a b^{-1})f(a^{-1}b^{-1})~,\cr
g(\pm\sfa \pm\sfb):=&g(\sfa+\sfb)g(-\sfa+\sfb)g(\sfa-\sfb)g(-\sfa-\sfb)~.
\end{align}

Throughout this paper, we use the following notation for $q$-Pochhammer symbols 
\be 
(z;q):=\prod_{k=0}^{\infty}(1-z q^{k})~.
\ee 

The elliptic gamma function is defined by
\begin{equation}\label{EGF}
\Gamma(z;p, q)=\prod_{m,n=0}^{\infty} \frac{1-p^{m+1} q^{n+1} / z}{1-p^m q^n z}=\operatorname{PE}\bigg[
    \frac{z - \frac{pq}{z}}{(1 - q)(1-p)}
    \bigg]~,
\end{equation}
where PE represents the plethystic exponential
\be \label{PE}
\mathrm{PE}[f(x, y, \cdots)] \equiv \exp \left[\sum_{d=1}^{\infty} \frac{1}{d} f\left(x^d, y^d, \cdots\right)\right]~,
\ee 
which brings the single particle index $f$ to the multi-particle index.
We often use the shorthand notation $\Gamma(z)$ for the elliptic Gamma function.

The Dedekind eta function is
\be \label{eta}
\eta(\tau)=q^{\frac{1}{24}} \prod_{n=1}^{\infty}\left(1-q^n\right)
\ee 
where $q=e^{2 \pi i \tau}$ and $\Im \tau>0$. Often, we also use the notation $\eta(q)$. Its modular properties are
\be 
\eta(\tau+1)=e^{\frac{i \pi}{12}} \eta(\tau), \quad \eta\left(-\frac{1}{\tau}\right)=\sqrt{-i \tau} \eta(\tau)~.
\ee

\subsection{Jacobi theta functions}
The Jacobi theta functions are defined as a Fourier series
$$
\begin{aligned}
& \vartheta_1(\mathsf{z}|\tau):=-i \sum_{r \in \mathbb{Z}+\frac{1}{2}}(-1)^{r-\frac{1}{2}} e^{2 \pi i r \mathsf{z}} q^{\frac{r^2}{2}},
& \vartheta_2(\mathsf{z}|\tau):= & \ \sum_{r \in \mathbb{Z}+\frac{1}{2}} e^{2 \pi i r \mathsf{z}} q^{\frac{r^2}{2}}, \\
& \vartheta_3(\mathsf{z}|\tau):= \sum_{n \in \mathbb{Z}} e^{2 \pi i n \mathsf{z}} q^{\frac{n^2}{2}}, 
& \vartheta_4(\mathsf{z}|\tau):= & \sum_{n \in \mathbb{Z}}(-1)^n e^{2 \pi i n\mathsf{z}} q^{\frac{n^2}{2}} .
\end{aligned}
$$
where $q = e^{2\pi i \tau}$ and $z=e^{2 \pi i\mathsf{z}}$. The Jacobi theta functions can be rewritten in the triple-product form
$$
\begin{aligned}
& \vartheta_1(\mathsf{z}|\tau) = i q^{\frac{1}{8}} z^{-\frac{1}{2}}(q;q)(z;q)(z^{-1}q;q), 
& \vartheta_2(\mathsf{z}|\tau)= & \ q^{\frac{1}{8}}z^{-\frac{1}{2}}(q;q)(-z;q)(- z^{-1}q;q)~, \\
& \vartheta_3(\mathsf{z}|\tau)= (q;q)(-zq^{1/2};q)(- z^{-1}q^{1/2};q)~,
& \vartheta_4(\mathsf{z}|\tau)= & \ (q;q)(zq^{1/2};q)(z^{-1}q^{1/2};q)~.
\end{aligned}
$$
From the Jabobi triple products, we can easily find the relation between $\vartheta_1$ and $\vartheta_4$ as
\begin{equation}\label{theta14}
\vartheta_4(\mathsf{z}|\tau)=-{i} q^{\frac{1}{8}} z^{\frac{1}{2}} \vartheta_1\left(\mathsf{z}+
\frac{\tau}{2}\mid\tau\right)~.
\end{equation}
We will also use the notation $\vartheta_i(z,q)$. In either notation, the $q$ and $\tau$ are often omitted, and we simply write $\vartheta_i(z)$ or $\vartheta_i(\mathsf{z})$.

Let us spell out some properties of the function $\vartheta_1(\mathsf{z}|\tau)$ we use in the main text. Under shifts of $\mathsf{z}$ we have
\begin{equation} \label{shift-property}
\vartheta_1(\mathsf{z}+a+b \tau | \tau)=(-1)^{a+b} e^{-2 \pi i b\mathsf{z}-i \pi b^2 \tau} \vartheta_1(\mathsf{z}|\tau)
\end{equation} 
for $a, b \in \mathbb{Z}$. Furthermore, $\vartheta_1$ is odd with respect to $\mathsf{z}$ while the others are even
$$
\vartheta_1(-\mathsf{z}|\tau)=-\vartheta_1(\mathsf{z}|\tau) , \qquad
\vartheta_{i=2,3,4}(- \mathsf{z} | \tau) = \vartheta_i( \mathsf{z} | \tau) \ .
$$
The function $\vartheta_1(\mathsf{z}|\tau)$ has simple zeros in $\mathsf{z}$ at $\mathsf{z}=\mathbb{Z}+\tau \mathbb{Z}$, and no poles. When computing JK residues, it is notable that the derivative of $\mathsf{z}$ at $0$ relates to $\eta(\tau)$ as follows
$$
\vartheta_1^{\prime}( 0 |\tau )=2 \pi \eta(q)^3 \ .
$$
From this relationship, we deduce a pole at $z=0$ as
\begin{equation}
    \frac{1}{\vartheta_1(\mathsf{z})} = \frac{1}{2\pi \eta(\tau)^3}\frac{1}{z} + \cO(\mathsf{z}) \ ,
\end{equation}
from which one easily extracts residues of ratios of Jacobi theta functions.

Under the modular transformation $\tau \xrightarrow{T} \tau + 1$, $(\mathsf{z},\tau) \xrightarrow{S}(\frac{\mathsf{z}}{\tau}, - \frac{1}{\tau})$, the Jacobi theta function $\vartheta_1$ transforms
$$
\vartheta_1(\mathsf{z}|\tau+1)=e^{\frac{\pi i}{4}} \vartheta_1(\mathsf{z}|\tau), \quad \vartheta_1\left(\frac{\mathsf{z}}{\tau} | -\frac{1}{\tau}\right)=-i \sqrt{-i \tau} e^{\pi i\mathsf{z}^2 / \tau} \vartheta_1(\mathsf{z}|\tau) \ .
$$

\subsection{Eisenstein series}\label{app:Eisenstein}
The twisted Eisenstein series, denoted by $E_k\Big[\substack{\phi\\\theta}\Big]$ with characteristics $\Big[\substack{\phi\\\theta}\Big]$, are defined as a series in $q$,
\[
\begin{aligned}
E_{k \geq 1}\begin{bmatrix}
\phi \\
\theta
\end{bmatrix}:= & -\frac{B_k(\lambda)}{k !} \\
& +\frac{1}{(k-1) !} \sum_{r \geq 0}^\prime \frac{(r+\lambda)^{k-1} \theta^{-1} q^{r+\lambda}}{1-\theta^{-1} q^{r+\lambda}}+\frac{(-1)^k}{(k-1) !} \sum_{r \geq 1} \frac{(r-\lambda)^{k-1} \theta q^{r-\lambda}}{1-\theta q^{r-\lambda}} .
\end{aligned}
\]
Here, $\phi \equiv e^{2 \pi i \lambda}$ determines $0 \leq \lambda<1$. $B_k(x)$ represents the $k$-th Bernoulli polynomial. When $\phi=\theta=1$, the prime in the sum indicates that the $r=0$ term is omitted.

Additionally, we define
\[
E_0\begin{bmatrix}
\phi \\
\theta
\end{bmatrix}=-1 \ .
\]
The standard, or untwisted, Eisenstein series $E_{2 n}$ is obtained from the $\theta, \phi \rightarrow 1$ limit of $E_{2 n}\Big[\substack{\phi\\\theta}\Big]$,
\[
E_{2 n}(\tau)=E_{2 n}\begin{bmatrix}
+1 \\
+1
\end{bmatrix} \ .
\]
In contrary, taking the limit $\theta, \phi \to 1$ for odd $k$ results in 0, with the exception of $E_1\Big[ \substack{\phi \\ \theta} \Big]$, which is singular.

The Eisenstein series with $\phi= \pm 1$ enjoy a useful symmetry property
\[
E_k\begin{bmatrix} 
\pm 1 \\
z^{-1}
\end{bmatrix}=(-1)^k E_k\begin{bmatrix} 
\pm 1 \\
z
\end{bmatrix} .
\]
For instance, under transformations $z \rightarrow q z$ or $z \rightarrow q^{\frac{1}{2}} z$, the twisted Eisenstein series intermix with those of lower weight:
\[
E_n\begin{bmatrix} 
\pm 1 \\
z q^{\frac{k}{2}}
\end{bmatrix}=\sum_{\ell=0}^n\left(\frac{k}{2}\right)^{\ell} \frac{1}{\ell !} E_{n-\ell}\begin{bmatrix}
(-1)^k( \pm 1) \\
z
\end{bmatrix} .
\]
Similarly, for the modular $S$-transformation, an inhomogeneous behavior is observed. For instance,
\begin{align}
    E_n \begin{bmatrix}
    +1 \\ +z
  \end{bmatrix} \xrightarrow{S} &
  \left(\frac{1}{2\pi i}\right)^n\left[\bigg(\sum_{k \ge 0}\frac{1}{k!}(- \log z)^k y^k\bigg)
  \bigg(\sum_{\ell \ge 0}(\log q)^\ell y^\ell E_\ell \begin{bmatrix}
    + 1 \\ z
  \end{bmatrix}\bigg)\right]_n \ ,
\end{align}
where $[ \cdots ]_n$ implies taking the coefficient of $y^n$.

\section{Jeffrey-Kirwan residue integrals}\label{app:JK}
In this appendix, we provide a detailed overview of the Jeffrey-Kirwan (JK) residue computation \cite{jeffrey1995localization,brion1999arrangement,szenes2003toric,Benini:2013xpa} related to the elliptic genus addressed in the main text. Since the elliptic genera receive contributions from both non-degenerate and degenerate poles in general, a thorough review of the JK residue integral definition is beneficial for the paper to be self-contained.

In the context of a rank-$r$ gauge theory, the elliptic genus computed through the JK-residue technique integrates an $r$-form over specific cycles, and is conventionally represented as
\begin{equation}
    \mathcal{I}^\textrm{2d} = \oint_\text{JK} \prod_{i = 1}^r \frac{da_i}{2\pi i a_i} \mathcal{Z}(\boldsymbol{a})
    = \oint_\text{special cycles} \mathcal{Z}(\boldsymbol{a}) d \mathsf{a}_1 \wedge \ldots \wedge d \mathsf{a}_r \ .
\end{equation}
As stated in the main text, $a_i = e^{2\pi i \mathsf{a}_i}$, and similarly for other variables except for $q = e^{2\pi i \tau}$. For our purpose, the integrand $\mathcal{Z}$, as a function of $\mathsf{a}_i$, is separately elliptic in each $\mathsf{a}_i$, namely,
\begin{equation}
    \mathcal{Z}(\ldots, \mathsf{a}_i + \tau, \ldots) = \mathcal{Z}(\ldots, \mathsf{a}_i, \ldots), \quad
    \mathcal{Z}(\ldots, \mathsf{a}_i + 1, \ldots) = \mathcal{Z}(\ldots, \mathsf{a}_i, \ldots) \ .
\end{equation}
More concretely, $\mathcal{Z}$  takes the form of certain ratios of the Jacobi theta functions $\vartheta_1$, and poles come from the zeros of $\vartheta_1$ in the denominator. Each pole is given as a solution to a set of pole equations
\begin{equation}\label{JK-solution}
    \sum_{i = 1}^r Q_{a}^i\mathsf{a}_i + \mathsf{b}_a = m_a + n_a \tau \ , \qquad
     \quad Q_{a}^i\in \mathbb{Z}\ , m_a, n_a \in \mathbb{N} \ , a = 1, 2, \ldots, r \ ,
\end{equation}
coming from some factors $\vartheta_1(Q_{a}^i \mathsf{a}_i + \mathsf{b}_a)^{N_a}$ in the denominator of $\mathcal{Z}$. Note that $m_a, n_a$ only take values in a finite range in $\mathbb{N}$ that will be determined by the charge vectors $Q_{a} = (Q_{a}^1, Q_{a}^2, \ldots, Q_{r}^r)$. A few remarks follow.
\begin{itemize}
    \item Zeros from numerators may arise in certain solutions of the pole equations, reducing the pole's order. If the total order of the pole is below $r$, it is not included in the JK residue.
    \item At some poles $\mathsf{a}_*$, there may be $n > r$ factors of $\vartheta_1^N$'s simultaneously made zero by $\mathsf{a}_*$, associated with $n$ different charge vectors $Q_1, \ldots, Q_n$. This is referred to as a degenerate pole.
    \item A pole associated to precisely $r$ different $\vartheta_1^N$ factors and therefore $r$ different charge vectors $Q_1, \ldots, Q_r$ is referred to as a non-degenerate pole.
    \item The range of $m_a, n_a$ is not unique. We start by rearranging the $Q_{a}$ terms such that $Q^i_{i} \ne 0$. Then $m_a, n_a$ are defined by methods like the Hermite or Smith normal form decomposition of the  (reordered) integral square matrix $(Q)_{ai}:= Q_a^i$. In the Smith decomposition,
    \begin{align}
        U Q V = D, \qquad & U, V \text{ are integral and } |\det U| = |\det V| = 1\cr
        &D \text{ is diagonal.}\nonumber
    \end{align}
    Then we fix the range $m_a, n_a, = 0, 1, \ldots, D_{aa} - 1$. Alternatively, in Hermite decomposition, $U Q = T$ with a unimodular integral $U$, and $T$ is an upper triangular integral matrix. In this case, $m_a, n_a = 0, 1, \ldots, T_{aa} - 1$. Note that although $T_{aa} \ne D_{aa}$ in general, the final result of the JK-residue computation will be the same.
\end{itemize}

For any pole $\mathsf{a}_*$ satisfying \eqref{JK-solution}, we need to compute the corresponding JK residue. We follow the constructive definition of JK residue \cite{Benini:2013xpa}. To begin, one picks a generic reference vector $\upeta \in \mathfrak{h}^*$ of the gauge group. If $\mathsf{a}_*$ is a non-degenerate pole with charge vectors $Q_1, \ldots, Q_r$, its contribution is given by
\begin{equation}
    \mathop{\operatorname{JK-Res}}_{\mathsf{a}_*}(\upeta) \mathcal{Z} 
    = \delta(Q, \upeta) \frac{1}{|\det Q|}
    \operatorname{Res}_{\epsilon_r = 0}\cdots\operatorname{Res}_{\epsilon_1 = 0}
    \mathcal{Z}\bigg|_{Q_a\mathsf{a} + \mathsf{b}_a = m_a + n_a \tau + \epsilon_a} \ ,
\end{equation}
where $\delta(Q, \upeta)$ equals one when $\upeta$ is inside the cone spanned by $Q_1, \ldots, Q_r$, and zero otherwise. The residues are calculated in sequence.

For a degenerate pole, we identify an associated set of charge vectors, $Q_* = \{Q_1, \ldots, Q_n\}$, with $n > r$. From the set $Q_*$, a collection of geometric objects can be defined.
\begin{itemize}
  \item  Given any $r$-sequence of linearly independent charge vectors $ (Q_{a_1}, \ldots, Q_{a_r}) $ from $Q_*$, we can construct a flag, $F$. This flag is essentially a series of nested subspaces of $\mathbb{R}^r$:
     \begin{align}
       \{0\} \subset F_1 \subset \ldots \subset F_r = \mathbb{R}^r, \quad 
       F_{\ell} = \operatorname{span}\{Q_{a_1}, \ldots, Q_{a_\ell}\}  \ .
     \end{align}
  Note that different sequences may give rise to the same flag. When this happens, we only consider one of them.  The sequence $(Q_{a_1}, \ldots, Q_{a_r})$ is often called a basis $\mathcal{B}(F, Q_*)$ of $F$ in $Q_*$. Given an $F$, the basis in $Q_*$ is generally not unique, but we pick an arbitrary one.
  \item From each flag $F$ and its basis $\mathcal{B}(F, Q_*)$, one constructs a sequence of vectors
  \begin{equation}
      \kappa(F, Q_*) := (\kappa_1, \ldots, \kappa_r), \quad
      \kappa_a = \sum_{\substack{Q\in Q_*\\Q \in F_a}}Q \ .
  \end{equation}
  One further defines $\operatorname{sign}F := \operatorname{sign} \det \kappa(F, Q_*)$.
  \item For each $\kappa(F)$, one constructs a closed cone $\mathsf{c}(F, Q_*)$ spanned by $\kappa(F, Q_*)$.
\end{itemize}
With these objects defined, the JK residue of the given degenerate pole $\mathsf{a}_*$ is given by
\begin{equation}
    \mathop{\operatorname{JK-Res}}_{\mathsf{a}_*}(\upeta) \mathcal{Z}
    = \sum_{F} \delta(F, \upeta) \frac{\operatorname{sign}F}{\det \mathcal{B}(F, Q_*)} \mathop{\operatorname{Res}}_{\epsilon_r = 0}\cdots
    \mathop{\operatorname{Res}}_{\epsilon_1 = 0} \mathcal{Z}\bigg|_{\substack{
    Q_{a_1} \mathsf{a} + \mathsf{b}_{a_1} = m_{a_1} + n_{a_1}\tau + \epsilon_1 \\
    \cdots \\
    Q_{a_r} \mathsf{a} + \mathsf{b}_{a_r} = m_{a_r} + n_{a_r}\tau + \epsilon_r
    }} \ ,
\end{equation}
where the sum is over all flags constructed out of $Q_*$ associated to $\mathsf{a}_*$. Again, $\delta(F, \upeta)$ equals one if the closed-cone $\mathsf{c}(F, Q_*)$ contains $\upeta$, and zero otherwise. This definition of $\operatorname{JK-Res}$ naturally extends to non-degenerate poles, where there are precisely $r$ vectors in $Q_*$, and
\begin{equation}
    \kappa(F, Q_*) = \mathcal{B}(F, Q_*), \qquad
    \frac{\operatorname{sign}F}{\det \mathcal{B}(F, Q_*)}
    = \frac{1}{|\det \mathcal{B}(F, Q_*)|} \ .
\end{equation}
The result clearly reduces to the previous definition of JK-residue for the non-degenerate case. Finally, given a generic $\upeta$,
\begin{equation}
    \int_\text{JK}\prod_{i = 1}^r \frac{da_i}{2\pi i a_i}\mathcal{Z}(\boldsymbol{a}) = \sum_{\mathsf{a}_*} \mathop{\operatorname{JK-Res}}_{\mathsf{a}_*}(\upeta) \mathcal{Z}(\boldsymbol{a}) \ .
\end{equation}
Although the structure of poles and the results of individual JK residues often differ drastically when $\upeta$ varies across chambers, the overall result is independent of the choice of $\upeta$.

In the following, we apply the JK-residue prescription to a number of quiver gauge theories discussed in the main text, presenting details of the computations. We will first focus on cases with $\SU(2)$ and $\U(1)$ gauge groups, followed by those with $\SU(N)$ and $\U(1)$ gauge groups.

\section{JK residues of \texorpdfstring{$\mathcal{N} = (0,2)$}{N=(0,2)} elliptic genera}\label{app:02}
In this appendix, we provide detailed computations of JK residue integrals for elliptic genera of 2d $\cN=(0,2)$ quiver gauge theories, complementing the main text. To elucidate the JK residue computations, we use notations based on chemical potentials instead of fugacities here.

\subsection{\texorpdfstring{$g=1, n=2$ }{g=1,n=2}}\label{app:genus1puncture2}

Given genus $g = 1$ and the number of punctures $n = 2$, one can write down different quiver gauge theories with different gauge groups.

\subsubsection{\texorpdfstring{$\SU(2)^2$}{SU(2)SU(2)} gauge theory}

The first theory is an $\SU(2)^2$ gauge theory coupled to two bi-fundamentals,
\begin{center}
    \begin{tabular}{c|c}
        & $\SU(2)_1 \times \SU(2)_2$  \\
        \hline
        $\phi_1$ & $(\boldsymbol{2} , \boldsymbol{2})$ \\
        $\phi_2$ & $(\boldsymbol{2} , \boldsymbol{2})$
    \end{tabular}
\end{center}
The quiver is shown in Figure \ref{fig:SU2-(1,2)-quiver-1}. The elliptic genus of this theory is computed by the JK residue computation of the integral
\begin{equation}
    \mathcal{I}_{1,2}
    = \int_\text{JK}
    \frac{da_1}{2\pi i a_1}
    \frac{da_2}{2\pi i a_2}
    \frac{\eta(\tau)^8}{4}\frac{\prod_{i = 1}^2\vartheta_1(\pm 2 \mathsf{a}_i)}{
    \prod_{i = 1}^2\vartheta_4(\pm \mathsf{a}_1 \pm \mathsf{a}_2 + \mathsf{c}_i)
    } \ .
\end{equation}
Here $\mathsf{c}_{1,2}$ represent the flavor $\U(1)\times \U(1)$ fugacities. The charge covectors are drawn in Figure \ref{fig:SU2-(1,2)-charge-vectors-1}.
\begin{figure}
    \centering    
    \includegraphics[width=0.4\textwidth]{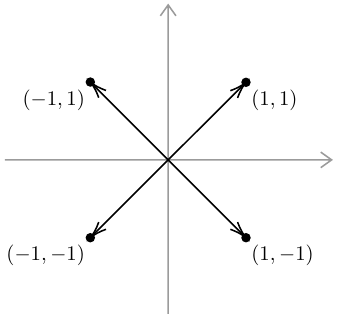}
    \caption{The charge vectors of the genus-one theory as an $\SU(2)^2$ gauge theory.\label{fig:SU2-(1,2)-charge-vectors-1}}
\end{figure}
Various reference vectors, $\upeta$, can be chosen, all of which yield the same result. For example, picking $\upeta = (1, 0)$ picks out one cone in $\mathbb{R}^2$ spanned by the charge vectors $(1,1)$ and $(1,-1)$. The corresponding poles are given by the set of equations
\begin{equation}
    \mathsf{a}_1 - \mathsf{a}_2 + \mathsf{c}_1 + \frac{\tau}{2}= m_1 + n_1 \tau, \quad
    \mathsf{a}_1 + \mathsf{a}_2 + \mathsf{c}_2 + \frac{\tau}{2}= m'_1 + n'_1 \tau \ ,
\end{equation}
and
\begin{equation}
    \mathsf{a}_1 - \mathsf{a}_2 + \mathsf{c}_2 + \frac{\tau}{2}= m_2 + n_2 \tau, \quad
    \mathsf{a}_1 + \mathsf{a}_2 + \mathsf{c}_1 + \frac{\tau}{2}= m'_2 + n'_2 \tau \ ,
\end{equation}
where $m_i, n_i = 0$, and $m_i, n_i = 0, 1$. There are in total 8 poles, all contribute $ - \frac{1}{8}\frac{\eta(\tau)^2}{ \vartheta_1(\mathsf{c}_1)\vartheta_1(\mathsf{c}_2)}$, and therefore,
\begin{equation}
    \mathcal{I}_{1,2} = - \frac{\eta(\tau)^2}{\prod_{i = 1}^2 \vartheta_1(2 \mathsf{c}_i)} \ .
\end{equation}

\subsubsection{First SU(2)\texorpdfstring{$\times$}{x}U(1) gauge theory}\label{app:genus1-puncture2-2}

Additionally, there are two other quiver theories as $\SU(2) \times \U(1)$ gauge theories that correspond to the genus-one Riemann surface with two punctures, with the quiver diagrams on the left in Figure \ref{fig:U1-(1,2)-quiver-1}. The first such gauge theory has an elliptic genus described by
\begin{equation}
    \mathcal{I}_{1,2}' = \int_\text{JK}\frac{da_1}{2\pi i a_1}\frac{da_2}{2\pi i a_2}
    \frac{\eta(\tau)^8}{2} \frac{\vartheta_1(\pm 2\mathsf{a}_1)\vartheta_4(\pm 2\mathsf{a}_2)}{
         \vartheta_4(\mathsf{a}_1 \pm \mathsf{a}_2 \pm \mathsf{b}_1 + \mathsf{d}_1)
         \vartheta_4(-\mathsf{a}_1 \pm \mathsf{a}_2 \pm \mathsf{b}_2 + \mathsf{d}_1)
    } \ . \nonumber
\end{equation}
Here we turn on the $\U(1)$ flavor symmetry that rotates the chiral multiplets from the two $U_2^{(0,2)}$ with the same phase with fugacity $\mathsf{d}_1$. One can also check that the integrand remains separately elliptic with respect to the variables $\mathsf{a}_{1,2}$. The charge vectors are still
\begin{equation}
    (1,1), \quad (-1,1), \quad (1, -1), \quad (-1,-1) \ .
\end{equation}
One can pick any $\upeta$ inside the four quadrants, and the JK residue computation yields the same result,
\begin{equation}\label{SU2-(1,2)-elliptic-genus-1}
    \mathcal{I}'_{1,2} = \frac{2 \eta(\tau)^2 \vartheta_4(2 \mathsf{d}_1)^2}{
    \vartheta_1(2 \mathsf{d}_1 \pm \mathsf{b}_1 \pm \mathsf{b}_2)
    } \ .
\end{equation}

\subsubsection{Second SU(2)\texorpdfstring{$\times$}{x}U(1) gauge theory}\label{app:genus1-puncture2-3}
A distinct $\SU(2) \times \U(1)$ gauge theory is depicted by the quiver diagram on the right side of Figure \ref{fig:U1-(1,2)-quiver-1}. The elliptic genus can be computed as the JK residue of the integrand
\begin{equation}
    \mathcal{Z}''_{1,2} = 
    \frac{\eta(\tau)^{10}\vartheta_4(\pm 2\mathsf{a}_1)\vartheta_1(\pm 2\mathsf{a}_2)}{
    4\vartheta_4(\mathsf{a}_1 + \mathsf{d}_1)^2
    \prod_\pm\vartheta_4(\mathsf{a}_1 \pm 2 \mathsf{a}_2 + \mathsf{d}_1)
    \vartheta_4(- \mathsf{a}_1 \pm \mathsf{b}_1 \pm \mathsf{b}_2 + \mathsf{d}_1) 
    }\ . \nonumber
\end{equation}
From the denominator, we can deduce the charge vectors
\begin{equation}
    (-1, 0), \quad
    (1, -2), \quad
    (1, 0), \quad
    (1, 2) \ .
\end{equation}
See Figure \ref{fig:SU2-(1,2)-charge-vectors-2}.
\begin{figure}
    \centering
    \includegraphics{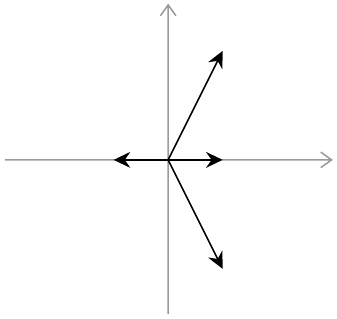}
    \caption{The charge vectors of the JK residues for the second SU(2)$\times$U(1) gauge theory}
    \label{fig:SU2-(1,2)-charge-vectors-2}
\end{figure}
Clearly, there are several choices for $\upeta$. Let us start with $\upeta = (-1,-1)$. In this case, only the cone spanned by $(-1, 0)$ and $(1, -2)$ contribute, corresponding to the poles from the equation (with four different choices of the signs $\pm$)
\begin{align}
    - \mathsf{a}_1 \pm \mathsf{b}_1 \pm \mathsf{b}_2 + \mathsf{d}_1 + \frac{\tau}{2}= & \ m_{1} + n_1 \tau \cr
    \mathsf{a}_1 - 2 \mathsf{a}_2 + \mathsf{d}_1 + \frac{\tau}{2}= & \ m_2 + n_2 \tau \ ,
\end{align}
where $m_1, n_1 = 0$, $m_2, n_2 = 0, 1$. In total, there are $4 \times 4$ different poles. For example, poles having $- \mathsf{a}_1 - \mathsf{b}_1 - \mathsf{b}_2 + \mathsf{d}_1 +\frac{\tau}{2}= 0$ contribute
\begin{equation}
    \frac{\eta(\tau)^4 \vartheta_4(-2 \mathsf{b}_1 - 2 \mathsf{b}_2 + 2 \mathsf{d}_1)^2}{
    \vartheta_1(2 \mathsf{b}_1 + 2 \mathsf{b}_2)
    \vartheta_1(2 \mathsf{b}_1 + 2 \mathsf{b}_2)
    \vartheta_1(- 2 \mathsf{b}_1 - 2 \mathsf{b}_2 + 4 \mathsf{d}_1)
    \prod_{i = 1}^2 \vartheta_1(2 \mathsf{b}_i)} \ . \nonumber
\end{equation}
To summarize, the elliptic genus reads
\begin{align}\label{SU2-(1,2)-elliptic-genus-2}
  \mathcal{I}''_{1,2} = \frac{\eta(\tau)^2}{\prod_{i = 1}^2 \vartheta_1(2\mathsf{b}_i)}
    \sum_{\alpha,\beta = \pm} \frac{\alpha \beta\vartheta_4(2 \alpha \mathsf{b}_1 + 2 \beta \mathsf{b}_2 + 2 \mathsf{d}_1)}{\vartheta_1(2 \alpha \mathsf{b}_1 + 2 \beta \mathsf{b}_2) \vartheta_1(2 \alpha \mathsf{b}_1 + 2 \beta \mathsf{b}_2 + 4 \mathsf{d}_1)} \ .
\end{align}

Alternatively, one can also choose $\upeta = (1,1)$. In this case, the relevant cones are spanned by the charge vectors
\begin{equation}
    (1, 2) \text{ and } (1,0), \qquad
    (1, 2) \text{ and } (1, -2) \ .
\end{equation}
The corresponding poles are from the equations
\begin{equation}\label{SU2-(1,2)-pole-eq-1}
    \mathsf{a}_1 - 2 \mathsf{a}_2 + \mathsf{d}_1 +\frac{\tau}{2} = m_1 + n_1 \tau, 
    \qquad
    \mathsf{a}_1 + 2 \mathsf{a}_2 + \mathsf{d}_1 +\frac{\tau}{2} = m_2 + n_2 \tau \ ,
\end{equation}
with $m_1, n_1 = 0$, $m_2, n_2 = 0,1,2,3$, and the equations
\begin{equation}\label{SU2-(1,2)-pole-eq-2}
    \mathsf{a}_1 + \mathsf{d}_1 +\frac{\tau}{2}= m_1 + n_1 \tau, 
    \qquad
    \mathsf{a}_1 + 2 \mathsf{a}_2 + \mathsf{d}_1 +\frac{\tau}{2}= m_2 + n_2 \tau \ ,
\end{equation}
with $m_1, n_1 = 0$, $m_2, n_2 = 0, 1$. The JK residue computation is, however, more subtle in this setup, due to the presence of degenerate poles
\begin{equation}
    (\mathsf{a}_1, \mathsf{a}_2) = (- \mathsf{d}_1-\frac{\tau}{2}, \frac{m + n \tau}{2} ), \qquad
    m, n = 0, 1 \ .
\end{equation}
Note also that these degenerate poles are precisely the common solutions to (\ref{SU2-(1,2)-pole-eq-1}), (\ref{SU2-(1,2)-pole-eq-2}) (up to shift of $\mathsf{a}_1$ by full periods $1, 1 + \tau, \tau$). At these poles, the factors
\begin{equation}
    \vartheta_4( \mathsf{a}_1 - 2 \mathsf{a}_2 + \mathsf{d}_1), 
    \vartheta_4( \mathsf{a}_1 + 2 \mathsf{a}_2 + \mathsf{d}_1), 
    \vartheta_4( \mathsf{a}_1 + \mathsf{d}_1)^2
\end{equation}
simultaneous vanish. Therefore, there are $12$ non-degenerate poles and $4$ degenerate poles that contribute to the elliptic genus. The former is straightforward to compute. For example,
\begin{equation}
    \text{JK}_{\substack{
    \mathsf{a}_1 - 2 \mathsf{a}_2 + \mathsf{d}_1 +\frac{\tau}{2} = 0\\
    \mathsf{a}_1 + 2 \mathsf{a}_2 + \mathsf{d}_1 +\frac{\tau}{2}= \tau
    }} \mathcal{Z}
    =- \frac{\vartheta_4(2 \mathsf{d}_1)^2}{
    16  \vartheta_4(\pm \mathsf{b}_1 \pm \mathsf{b}_2 + 2 \mathsf{d}_1)
    } \ .
\end{equation}
For the degenerate poles, we follow \cite{Benini:2013xpa}. The relevant charge vectors can be grouped into
\begin{equation}
    Q_* = \{(1, -2), (1, 2), (1, 0)\} \ ,
\end{equation}
and that gives rise to three flags $F_{1,-2}, F_{1,2}, F_{1, 0}$ led by the three vectors in $Q_*$. The corresponding $\kappa(F)$ and other relevant information is collected in the following table.
\begin{center}
    \begin{tabular}{c|c | c | c}
    $F$ & $F_{1, - 2}$ & $F_{1,2}$ & $F_{1,0}$ \\
    \hline
    $\kappa(F)$ & $((1, -2), (3,0))$ & $((1,2), (3,0))$ & $((1,0), (3,0))$  \\
    $\textrm{sign}\det(\kappa(F))$ & $1$ & $-1$ & 0\\
    $\upeta \in \mathsf{c}(F, Q_*)$ & False & True & False
\end{tabular}
\end{center}
From the table, we can compute the contribution from the degenerate poles
\begin{equation}
    = \frac{{\eta(\tau)}^2\vartheta_4(2 \mathsf{d}_1)^2}{4
    \vartheta_1(\pm \mathsf{b}_1 \pm \mathsf{b}_2 + 2 \mathsf{d}_1)
    } \ .
\end{equation}
In the end, the elliptic genus computed using $\upeta = (1,1)$ is

\begin{equation}\label{SU2-(1,2)-elliptic-genus-3}
    \mathcal{I}''_{1,2} = \frac{\eta(\tau)^4}{2}\vartheta_4(2 \mathsf{d}_1)^2
    \sum_{i = 1}^4 (-1)^i \frac{1}{ \vartheta_i(2 \mathsf{d}_1 \pm \mathsf{b}_1 \pm \mathsf{b}_2)} \ .
\end{equation}
Although look different, the elliptic genus (\ref{SU2-(1,2)-elliptic-genus-2}) and (\ref{SU2-(1,2)-elliptic-genus-3}) are actually identical, and are equal to
\begin{equation}
    \mathcal{I}''_{1,2}
    = \eta(\tau)^2\vartheta_4(2 \mathsf{d}_1)^2
    \frac{\vartheta_1(8 \mathsf{d}_1)}{\vartheta_1(4 \mathsf{d}_1)}
    \prod_{i = 1}^2 \frac{\vartheta_1(4 \mathsf{b}_i)}{\vartheta_1(2 \mathsf{b}_i)}
    \times
     \frac{1}{\vartheta_1(4 \mathsf{d}_1 \pm 2 \mathsf{b}_1 \pm 2\mathsf{b}_2)} \ .
\end{equation}
The equivalence of the expressions is checked by the power expansion in $q$.

\subsection{\texorpdfstring{$g = 1, n = 3$ }{g=1,n=3}}

Corresponding to the genus-one Riemann surfaces $C_{1, 3}$ with three punctures, there are several different 2d $\mathcal{N} = (0,2)$ theories.

\subsubsection{\texorpdfstring{$\SU(2)^3$}{SU(2)cube} gauge theory}
Let us consider first the $\SU(2) \times \SU(2) \times \SU(2)$ gauge theory drawn in Figure \ref{fig:SU2-(1,3)-quiver-1}. The elliptic genus is the JK residue of the integrand
\begin{equation}
    \mathcal{Z}_{1,3}
    = - \frac{\eta(\tau)^{12}
    \vartheta_1(\pm 2 \mathsf{a}_1)
    \vartheta_1(\pm 2 \mathsf{a}_2)
    \vartheta_1(\pm 2 \mathsf{a}_3)}{
    \prod_{A < B}
    \vartheta_4(\pm \mathsf{a}_A \pm \mathsf{a}_B + \mathsf{c}_1)
    }\ .
\end{equation}
We can choose $\upeta = (1, 1 + \frac{1}{1000}, 1 + \frac{1}{2000})$. There is no degenerate pole, and the elliptic genus is given by
\begin{equation}
    \mathcal{I}_{1,3} = \frac{\eta(\tau)^3}{\prod_{i = 1}^3 \vartheta_1( 2 \mathsf{c}_i)} \ .
\end{equation}

\subsubsection{First \texorpdfstring{$\SU(2)^2\times\U(1)$}{SU(2)SU(2)U(1)} theory}\label{app:SU2-g1p3-2}

Let us now consider an $\SU(2) \times \SU(2) \times \U(1)$ gauge theory illustrated in the left of Figure \ref{fig:U1-(1,3)-quiver-1}, whose elliptic genus is the JK residue of the integrand
\begin{equation}
    \mathcal{Z}'_{1,3}= 2 \frac{\eta(\tau)^{12}
      \vartheta_4(\pm 2 \mathsf{a}_1)
      \vartheta_1(\pm 2 \mathsf{a}_2)
      \vartheta_1(\pm 2 \mathsf{a}_3)
    }{4
      \vartheta_4(\pm \mathsf{a}_2 \pm \mathsf{a}_3 + \mathsf{c}_1)
      \vartheta_4(\mathsf{a}_1 \pm \mathsf{a}_2\pm \mathsf{b}_2 + \mathsf{d}_1)
      \vartheta_4(- \mathsf{a}_1 \pm \mathsf{a}_3 \pm \mathsf{b}_3 + \mathsf{d}_1)
    } \ .
\end{equation}
An arbitrary reference vector can be chosen, for example, $\upeta = (\frac{999}{1000}, \frac{1999}{2000}, \frac{2999}{3000})$ which leads to $40$ non-degenerate poles. The JK residue is straightforward, yielding,
\begin{align}
    \frac{\mathcal{I}'_{1,3}}{\eta(\tau)^3}
    = & \ -\frac{1}{2\vartheta_1(2 \mathsf{c}_1)\prod_{i = 2}^3\vartheta_1(2 \mathsf{b}_i)} \sum_{\alpha, \beta = \pm} \frac{ \alpha \beta
        \vartheta_1(\alpha \mathsf{b}_2 + \beta \mathsf{b}_3 + \mathsf{c}_1)^2
    }{
    \vartheta_4(\alpha \mathsf{b}_2 +\beta \mathsf{b}_3 + \mathsf{c}_1 \pm 2 \mathsf{d}_1)
    } \nonumber \\
    & \ + \frac{
      \vartheta_4(2 \mathsf{d}_1)^2
    }{
      2 \vartheta_1(2 \mathsf{b}_2)\vartheta_1(2 \mathsf{c}_1)
    }
    \sum_{\alpha=\pm} \frac{
        \alpha \vartheta_1(2 \alpha \mathsf{b}_2 + 2 \mathsf{c}_1)
    }{
        \vartheta_4(\alpha \mathsf{b}_2 + \mathsf{c}_1 \pm \mathsf{b}_3 
        \pm 2 \mathsf{d}_1)
    }\\
    & \ + \frac{
      \vartheta_4(2 \mathsf{d}_1)^2
    }{
        \vartheta_1(2 \mathsf{b}_3)\vartheta_1(4 \mathsf{d}_1)
    }
    \sum_{\alpha=\pm} \frac{
        \alpha \vartheta_1(2 \alpha \mathsf{b}_3 + 4 \mathsf{d}_1)
    }{
        \vartheta_4(\alpha \mathsf{b}_3 + 2\mathsf{d}_1 \pm \mathsf{b}_2 
        \pm  \mathsf{c}_1)
    } \ . \nonumber
\end{align}
It is straightforward to check that this complicated expression is actually equal to
\begin{equation}
    \frac{\mathcal{I}'_{1,3}}{\eta(\tau)^3}
    = 2 \frac{\vartheta_4(2 \mathsf{d}_1)^2}{\vartheta_1(4 \mathsf{d}_1)}
    \frac{\vartheta_1(2 \mathsf{c}_1 + 4 \mathsf{d}_1)}{\vartheta_1(2 \mathsf{c}_1)}
    \frac{1}{
    \vartheta_4(\mathsf{c}_1 + 2 \mathsf{d}_1 \pm \mathsf{b}_2 \pm \mathsf{b}_3)
    } \ ,
\end{equation}
which signals an LG description of the quiver gauge theory.

\subsubsection{Second \texorpdfstring{$\SU(2)^2\times\U(1)$}{SU(2)SU(2)U(1)} theory}
Another quiver gauge theory with the gauge group $\SU(2) \times \SU(2) \times \U(1)$ is depicted to the right in Figure \ref{fig:U1-(1,3)-quiver-1}. Its elliptic genus is determined by the JK residue of the integrand
\begin{equation}
    \mathcal{Z} = \mathcal{I}_{U_2}^{(0,2)}(\mathsf{c}_1, \mathsf{b}_1, \mathsf{a}_3)
    \mathcal{I}_{U_2}^{(0,2)}(\mathsf{a}_2  + \mathsf{d}_1, \mathsf{b}_2, -\mathsf{a}_3)
    \mathcal{I}_{U_2}^{(0,2)}(-\mathsf{a}_2 + \mathsf{d}_1, \mathsf{a}_1, - \mathsf{a}_1) \vartheta_4(\pm \mathsf{a}_2)
    \prod_{i=\{1,3\}} \mathcal{I}_{\textrm{vec}}^{(0,2)}(\mathsf{a}_i) \ . \nonumber
\end{equation}
The six charge vectors are given by
\begin{equation}
    (\pm 1, 0,0), \quad (\pm 1,0,1), \quad (0, \pm 2, -1) \ .
\end{equation}
There are various choices for $\upeta$. For example, let us begin with
\begin{equation}
    \upeta = ( \frac{999}{1000},\frac{1999}{2000}, \frac{2999}{3000}).
\end{equation}
With this choice, there are $64$ non-degenerate poles, giving the elliptic genus
\begin{align}
    & \ \frac{\mathcal{I}_{0,3}''}{\eta(\tau)^3} \nonumber \\
    = & \ \frac{1}{\vartheta_1(2 \mathsf{c}_1)\prod_{i = 1}^2 \vartheta_1(2 \mathsf{b}_i)}
    \sum_{\alpha \beta = \pm}
    \frac{\alpha \beta \vartheta_4(2\alpha \mathsf{b}_1 + 2 \beta \mathsf{b}_2 + 2 \mathsf{c}_1 + 2 \mathsf{d}_1)^2}{
    \vartheta_1(2 \alpha \mathsf{b}_1 + 2 \beta \mathsf{b}_2 + 2 \mathsf{c}_1)
    \vartheta_1(2 \alpha \mathsf{b}_1 + 2 \beta \mathsf{b}_2
    + 2 \mathsf{c}_1 + 4 \mathsf{d}_1)
    } \nonumber\\
    & \ + \frac{2\vartheta_4(2 \mathsf{d}_1)^2}{\vartheta_1(4 \mathsf{d}_1)}
    \sum_{i = 1}^4 \frac{(-1)^i}{2\vartheta_i(\pm \mathsf{b}_1 \pm \mathsf{b}_2 + \mathsf{c}_1)} \ .
\end{align}
For instance, another intriguing choice for $\upeta$ is 
\begin{equation}
    \upeta = (- \frac{999}{1000}, \frac{1}{1000},1 ) \ .
\end{equation}
With this choice of $\upeta$, there are $56$ non-degenerate poles and $8$ degenerate poles. The latter poles are given by
\begin{align}
    & \ \mathsf{a}_1 = - \frac{m + n \tau}{2}, \quad
    \mathsf{a}_2 = \mathsf{d}_1 + \frac{\tau}{2} , \quad 
    \mathsf{a}_3 = + \mathsf{b}_2 - 2 \mathsf{d}_1 - \tau,  \quad m, n = 0, 1 \cr
    \text{or, } \qquad
    & \ \mathsf{a}_1 = - \frac{m + n \tau}{2},
    \quad
    \mathsf{a}_2 = \mathsf{d}_1 + \frac{\tau}{2}, \quad
    \mathsf{a}_3 = - \mathsf{b}_2 - 2 \mathsf{d}_1 - \tau  , \quad m, n = 0, 1\ .
\end{align}
The elliptic genus reads
{\small\begin{align}
  & \ 4 \frac{\vartheta_1(2 \mathsf{b}_2) \vartheta_1(4 \mathsf{d}_1)}{\eta(\tau)^3 \vartheta_4(2 \mathsf{d}_1)^2}\mathcal{I}''_{0,3} \nonumber \\
  = & \ \sum_{i = 1}^4 \sum_{\alpha = \pm}
  \frac{(-1)^i \alpha \vartheta_1(2 \alpha \mathsf{b}_2 + 4 \mathsf{d}_1)}{
  \vartheta_i(\pm \mathsf{b}_1 + \alpha \mathsf{b}_2 \pm \mathsf{c}_1 + 2 \mathsf{d}_1)
  }\\
  & \ + \sum_{\alpha, \beta, \gamma = \pm} \frac{2 \alpha \beta\vartheta_1(4 \mathsf{d}_1)}{
  \vartheta_1(2 \mathsf{b}_1)\vartheta_1(2 \mathsf{c}_1)
  \vartheta_4(2 \mathsf{d}_1)^2
  } \frac{ \vartheta_4(2 \alpha \mathsf{b}_1 + 2 \beta \mathsf{b}_2 + 2 \mathsf{c}_1 + 2 \gamma \mathsf{d}_1)^2}{
  \vartheta_1(2(\alpha \mathsf{b}_1 + \beta \mathsf{b}_2 + \mathsf{c}_1 + 2 \gamma \mathsf{d}_1))
  \vartheta_1(2( \alpha \mathsf{b}_1 + \beta \mathsf{b}_2 + \mathsf{c}_1))
  } \ . \nonumber
\end{align}}
Although the two expressions for $\mathcal{I}''_{1,3}$ look different, they are both equal to the simple ratio
\begin{equation}
    \frac{\mathcal{I}_{1,3}''}{\eta(\tau)^3}
    = \frac{\vartheta_4(2 \mathsf{d}_1)^2 \vartheta_1(4 \mathsf{c}_1 + 8 \mathsf{d}_1)}{
    \vartheta_1(4 \mathsf{d}_1) \vartheta_1(2 \mathsf{c}_1)}
    \frac{1}{\vartheta_1(\pm 2 \mathsf{b}_1 \pm 2\mathsf{b}_2 + 2 \mathsf{c}_1 + 4 \mathsf{d}_1)}
    \prod_{i = 1}^2\frac{\vartheta_1(4 \mathsf{b}_i)}{\vartheta_1(2 \mathsf{b}_i)}\ ,
\end{equation}
suggesting an LG description of the theory.

\subsection{\texorpdfstring{$g = 2, n = 0$ }{g=2,n=0}}

Associated to the genus-two Riemann surface with no puncture, there are two possible quiver gauge theories with $\SU(2)^2 \times \U(1)$ gauge groups. (See Figure \ref{fig:SU2-genus2}.)

\subsubsection{First \texorpdfstring{$\SU(2)^2\times\U(1)$}{SU(2)SU(2)U(1)} theory}

The theory on the left of Figure \ref{fig:SU2-genus2} has an elliptic genus given by the JK-residue of the integrand
\begin{equation}
    \mathcal{Z}_{2,0} = 2 \mathcal{I}_{U_2}^{(0,2)}(\mathsf{a}_3 + \mathsf{d}_1, \mathsf{a}_1, \mathsf{a}_2)
    \mathcal{I}_{U_2}^{(0,2)}(-\mathsf{a}_3 + \mathsf{d}_1, -\mathsf{a}_1, -\mathsf{a}_2)
    \vartheta_4(\pm 2\mathsf{a}_3)
    \prod_{i = 1}^2\mathcal{I}_{\textrm{vec}}^{(0,2)}(\mathsf{a}_i) \ .
\end{equation}
The eight charge vectors are given by $(\pm 1, \pm 1, \pm 1)$. There are many essentially equivalent choices of $\upeta$. For example, we pick
\begin{equation}
    \upeta = (1, \frac{999}{1000}, \frac{4999}{5000}) \ .
\end{equation}
With this choice of $\upeta$, there are $32$ non-degenerate poles. All of them contribute identically to the total JK residue. Finally, the elliptic genus reads
\begin{equation}
    \mathcal{I}_{2,0} = - \frac{2\vartheta_4(2 \mathsf{d}_1)^2}{\eta(\tau)\vartheta_1(4 \mathsf{d}_1)} \ .
\end{equation}

\subsubsection{Second \texorpdfstring{$\SU(2)^2\times\U(1)$}{SU(2)SU(2)U(1)} theory}

The theory on the right of Figure \ref{fig:SU2-genus2} has an elliptic genus integrand
\begin{equation}
    \mathcal{Z}'_{2,0} = 2\mathcal{I}_{U_2}^{(0,2)}(\mathsf{a}_3 + \mathsf{d}_1, \mathsf{a}_1, - \mathsf{a}_1)
    \mathcal{I}_{U_2}^{(0,2)}(-\mathsf{a}_3 + \mathsf{d}_1, -\mathsf{a}_2, -\mathsf{a}_2)
    \vartheta_4(\pm 2 \mathsf{a}_3)
    \prod_{i = 1}^2 \mathcal{I}_{\textrm{vec}}^{(0,2)}(\mathsf{a}_i)
    \ .
\end{equation}
The charge vectors are given by
\begin{equation}
    (\pm 2, 0, 1), \quad
    (0, \pm 2, -1), \quad
    (0, 0, \pm 1) \ .
\end{equation}
Let us consider a choice of $\upeta$
\begin{equation}
    \upeta = (\frac{1001}{1000}, \frac{501}{500}, \frac{1003}{1000}) \ .
\end{equation}
With this choice, there are $48$ non-degenerate poles and $16$ degenerate poles. The latter takes the form
\begin{equation}
    \mathsf{a}_1 = \frac{m_1 + n_1 \tau}{2}, \quad
    \mathsf{a}_2 = - \mathsf{d}_1 -\frac{\tau}{2}+ \frac{m_2 + n_2 \tau}{2}, \quad
    \mathsf{a}_3 = - \mathsf{d}_1 -\frac{\tau}{2}\ .
\end{equation}
It turns out that all $64$ poles share identical contributions to the elliptic genus. In the end, we have
\begin{equation}
    \mathcal{I}'_{2,0} = \frac{2\vartheta_4(2 \mathsf{d}_1)^2}{\eta(\tau)\vartheta_1(4 \mathsf{d}_1)} \ .
\end{equation}
Apparently, up to a sign,
\begin{equation}
    \mathcal{I}_{2,0} = \mathcal{I}'_{2,0} \ .
\end{equation}

\subsection{Genus two with \texorpdfstring{$n$}{n} punctures}

Let us briefly summarize the computation for genus-two theories with $n$ punctures. Given $n$, there are essentially only two $\SU(2)^{n + 2} \times\U(1)$ quiver gauge theories one can consider: if the $\U(1)$ node in the quiver diagram is removed/ungauged, one frame continues to have a connected quiver diagram, while the other frame is cut into two disconnected pieces. 

The first frame is simpler. The integrand reads
\begin{equation}
    \mathcal{Z}_{2,n} = \vartheta_4(\pm 2\mathsf{a}_{n + 3})\prod_{j=1}^{n+2}\mathcal{I}_\text {vec}^{(0,2)}(\mathsf{a}_j)
    \cdot 
    \prod_{i = 1}^{n + 2} \mathcal{I}_{U_2}^{(0,2)}(\mathsf{c}_i, -\mathsf{a}_{i - 1}, \mathsf{a}_i)\Bigg|_{\substack{
    \mathsf{c}_{n+1} = \mathsf{a}_{n + 3} + \mathsf{d}_1\\
    \mathsf{c}_{n+2} =  - \mathsf{a}_{n + 3} + \mathsf{d}_1 \\
    \mathsf{a}_0 = \mathsf{a}_{n + 2}\\
    }} \ .
\end{equation}
One can pick a simple and generic $\upeta = (\upeta_1, \ldots, \upeta_{n + 3})$, for example, 
\begin{equation}
    \upeta_i = 1 - \frac{1}{1000000 i} \ .
\end{equation}
With this choice, there are only $2^{2g + n + 1}$ non-degenerate poles, which lead to a simple elliptic genus
\begin{equation}
    \mathcal{I}_{2,n} = 2 (-1)^{n + 1} \frac{\vartheta_4(2 \mathsf{d}_1)^2}{\eta(\tau)\vartheta_1(4 \mathsf{d}_1)} \prod_{i = 1}^n\frac{
    \eta(\tau)
    }{\vartheta_1(2 \mathsf{c}_i)} \ .
\end{equation}

The second frame has an integrand
\begin{equation}
    \mathcal{Z}_{2,n}'
    = \vartheta_4(\pm 2\mathsf{a}_{n + 3})\prod_{j=1}^{n+2}\mathcal{I}_\text {vec}^{(0,2)}(\mathsf{a}_j)
    \cdot 
    \mathcal{I}_{U_2}^{(0,2)}( - \mathsf{a}_{n + 3} + \mathsf{d}, \mathsf{a}_{n + 2}, - \mathsf{a}_{n + 2})
    \prod_{i = 1}^{n + 1} \mathcal{I}_{U_2}^{(0,2)}(\mathsf{c}_i, -\mathsf{a}_{i - 1}, \mathsf{a}_i)\Bigg|_{\substack{
    \mathsf{c}_{n + 1} = \mathsf{a}_{n + 3} + \mathsf{d}_1\\
    \mathsf{a}_0 = \mathsf{a}_{n + 1}\\
    }} \ .
\end{equation}
There are different $\upeta$ to choose from. For example, 
\begin{equation}
    \upeta = (\upeta_1, \ldots), \qquad \upeta_i = 1 - \frac{1}{1000 \times i} \ .
\end{equation}
With this choice, only non-degenerate poles contribute, yielding
\begin{equation}
    \mathcal{I}_{2, n}' = 2 (-1)^{n} \frac{\vartheta_4(2 \mathsf{d}_1)^2}{\eta(\tau)\vartheta_1(4 \mathsf{d}_1)} \prod_{i = 1}^n\frac{\eta(\tau)}{\vartheta_1(2 \mathsf{c}_i)} \ .
\end{equation}
While other choices of $\upeta$ may lead to degenerate poles, the final result of the elliptic genus remains independent of $\upeta$. Up to a sign, both the $\SU(2)^{n + 3} \times \U(1)$ quiver gauge theory share the same elliptic genus,
\begin{equation}
    \mathcal{I}_{2, n} = \mathcal{I}'_{2,n} \ .
\end{equation}

\subsection{SU(3) theory for \texorpdfstring{$g=1,n=1$}{g=1,n=1} }\label{app:SU3-g1p1}

The elliptic genus for the $\mathcal{N} = (0,2)$ $\SU(3)$ theory coupled to an adjoint chiral multiplet is given by the integral
\begin{equation}
  \mathcal{I}_{1,1}^{(0,2),3}=  \int_\text{JK}\frac{da_1}{2\pi i a_1}\frac{da_2}{2\pi i a_2}
    \mathcal{I}_{U_{3}}^{(0,2)}(\mathsf{c}, \mathsf{a}, -\mathsf{a}) \mathcal{I}_\text{vec}^{(0,2)}(\mathsf{a}) \ ,
\end{equation}
where
\begin{align}
    \mathcal{I}_{U_{3}}^{(0,2)}(\mathsf{c}, \mathsf{a}, \mathsf{b})
    = & \ \prod_{A, B = 1}^3 \frac{\eta(\tau)}{\vartheta_4(\mathsf{c} + \mathsf{a}_A + \mathsf{b}_B)} \bigg | _{\substack{
    \mathsf{a}_3 = - \mathsf{a}_1 - \mathsf{a}_2\\
    \mathsf{b}_3 = - \mathsf{b}_1 - \mathsf{b}_2
    }} \\
    \mathcal{I}_\text{vec}^{(0,2)}(\mathsf{a}) = & \ \frac{\eta(\tau)^3}{3!} \prod_{\substack{A, B = 1\\
    A\ne B}}^3 \frac{\vartheta_1(\mathsf{a}_A - \mathsf{a}_B)}{\eta(\tau)} \ .
\end{align}
Subsequent discussions involving $\SU(N)$ gauge group will continue to use these basic building blocks. One can pick some generic reference vector $\upeta$, for example,
\begin{equation}
    \upeta = (1, 0), \qquad \text{or} \qquad (0,1) \ ,
\end{equation}
there are always $18$ non-degenerate poles, yielding
\begin{equation}
  \mathcal{I}_{1,1}^{(0,2),3}=\frac{\eta(\tau)}{\vartheta_4(3 \mathsf{c})} \ .
\end{equation}

\subsection{\texorpdfstring{$g = 1, n = 2$}{g=1,n=2} with SU(3) and/or U(1) gauge group}\label{app:SU3-g1p2}

For the genus-one Riemann surface with two punctures, several $A_2$ type $\mathcal{N} = (0,2)$ theories can be defined.

\subsubsection{\texorpdfstring{$\SU(3)^2$}{SU(3)SU(3)} gauge theory}

The simplest theory is an $\SU(3)^3$ theory, with integrand
\begin{equation}
    \mathcal{Z} = \mathcal{I}_{U_3}^{(0,2)}(\mathsf{c}_1, \mathsf{a}_1, \mathsf{a}_2)
    \mathcal{I}_{U_3}^{(0,2)}(\mathsf{c}_2, - \mathsf{a}_1, - \mathsf{a}_2)
    \prod_{i = 1}^2 \mathcal{I}_{\textrm{vec}}^{(0,2)}(\mathsf{a}_i) \ .
\end{equation}
One can pick, for example,
\begin{equation}
    \upeta = (1, \frac{1}{1000}, - \frac{1}{19}, \frac{16}{17}) \ .
\end{equation}
There are $108$ non-degenerate poles, giving
\begin{equation}
    \mathcal{I}_{1,2}^{(0,2),3}= \frac{\eta(\tau)^2}{\vartheta_4(3 \mathsf{c}_1) \vartheta_4(3 \mathsf{c}_2)} \ .
\end{equation}

\subsubsection{SU(3)\texorpdfstring{$\times$}{x}U(1) gauge theory}\label{app:SU3U1-genus1-puncture2}

The integrand of the elliptic genus for the $\SU(3) \times\U(1)$ theory, depicted on the left of Figure \ref{fig:SU3U1-(1,2)-quiver-1}, is
\begin{equation}
    \mathcal{Z} =
    \mathcal{I}_{U_3}^{(0,2)}(\mathsf{a}_2 + \mathsf{d}_1, \mathsf{a}_1, \mathsf{b})
    \mathcal{I}_{U_3}^{(0,2)}(- \mathsf{a}_2 + \mathsf{d}_1, -\mathsf{a}_1, \mathsf{b}')
    \mathcal{I}_{\textrm{vec}}^{(0,2)}(\mathsf{a}_1)\vartheta_4(\pm 3 \mathsf{a}_2) \ ,
\end{equation}
where $\vartheta_4(\pm 3 \mathsf{a}_2)$ accounts for the contribution from two Fermi multiplets with $\pm 3$ $\U(1)$ gauge charges. With a simple choice of
\begin{equation}
    \upeta = (\frac{999}{1000}, \frac{1999}{2000}, \frac{2999}{3000}) \ ,
\end{equation}
there are $162$ non-degenerate poles, giving a fairly complicated expression as the elliptic genus,
\begin{align}
    \mathcal{I}'_{1,2}
    = & \ -\eta(\tau)^7
    \sum_{A, A' =1}^3 \frac{\vartheta_1(- \mathsf{b}_A - \mathsf{b}'_{A'} + \mathsf{d}_1)^2}{
    \prod_{B\ne A}\vartheta_1(\mathsf{b}_B - \mathsf{b}_A)
    \prod_{B'\ne A'}\vartheta_1(\mathsf{b}'_{B'} - \mathsf{b}'_{A'})
    } \nonumber\\
    & \ \qquad \qquad \quad\times \frac{1}{
    \vartheta_1(\mathsf{b}_A + \mathsf{b}'_{A'} + 2\mathsf{d}_1)
    \prod_{B \ne A, B'\ne A'}\vartheta_1(\mathsf{b}_B + \mathsf{b}'_{B'} + 2 \mathsf{d}_1)
    } \ .
\end{align}
Although the expression is complicated, it can be reformulated as a simpler ratio
\begin{equation}
    \mathcal{I}'_{1,2} =  \frac{3 \eta(\tau)^7  \vartheta_1(3 \mathsf{d}_1)^2}{
    \prod_{A, B = 1}^3 \vartheta_1(\mathsf{b}_A + \mathsf{b}'_B + 2 \mathsf{d}_1)
    } \ .
\end{equation}

\section{JK residues of \texorpdfstring{$\mathcal{N} = (0,4)$}{N=(0,4)} elliptic genera}\label{app:04}

In this Appendix, we present the detailed computation of the 
$\mathcal{N} = (0,4)$ elliptic genus. Recall that the basic building blocks of the elliptic genus for the Lagrangian theories of type $A_{N-1}$, as illustrated in Figure \ref{fig:04-block}, are given by \cite{Putrov:2015jpa}
\begin{align}
    \mathcal{I}_{0,3}^{(0,4),N}(\mathsf{c}, \mathsf{a}, \mathsf{b}) = & \ \prod_{A,B = 1}^N \frac{\eta(\tau)}{
      \vartheta_1(\mathsf{v} \pm (\mathsf{c} + \mathsf{a}_A + \mathsf{b}_B))
    }  \ ,\\
    \mathcal{I}_{\textrm{vec}}^{(0,4),N}= & \ \frac{1}{N!} \big(\vartheta_1(2 \mathsf{v} )\eta(\tau) \big)^{N - 1}
    \prod_{\substack{A, B=1 \\ A\ne B}}^N \frac{\vartheta_1(2 \mathsf{v} + \mathsf{a}_A - \mathsf{a}_B)
    \vartheta_1(\mathsf{a}_A - \mathsf{a}_B)}{\eta(\tau)^2} \ .
\end{align}
where all the $\SU(N)$ chemical potentials satisfy the traceless condition such as $\sum_{A = 1}^N\mathsf{a}_A = 0$. The chemical potentials $\mathsf{a}, \mathsf{b}, \mathsf{c}$ make manifest the flavor symmetry $\SU(N) \times \SU(N) \times \U(1) \subset\SU(2N)\times \U(1)$. Note that our convention is slightly different from that in \cite{Putrov:2015jpa} by some constant factor.

\subsection{Type \texorpdfstring{$A_{N-1}$}{AN-1} theories with minimal punctures at genus-one}

We begin with the theory of type $A_2$ at genus-one with one minimal puncture associated to an $\U(1)$ flavor symmetry. The elliptic genus can be computed by the JK-residue
\begin{align}
  \mathcal{I}^{(0,4),3}_{1,1,0}(\mathsf{c})
  = \int_\text{JK} \prod_{A=1}^2\frac{da_A}{2\pi i a_A}
  \mathcal{I}^{(0,4),3}_{0,3}(\mathsf{c}, \mathsf{a}, - \mathsf{a}) \mathcal{I}_{\textrm{vec}}^{(0,4),3}(\mathsf{a}) \ .
\end{align}
The integration simplifies when choosing a suitable reference vector as $\upeta = (1, 1-\frac{1}{1000})$, resulting in only $54$ non-degenerate poles. The result of the JK-residue computation is
\begin{align}
  \mathcal{I}^{(0,4),3}_{1,1,0}(\mathsf{c})
  = \eta(\tau)^2
  \Bigg[ & \ 
  - \frac{
    \vartheta_1(4 \mathsf{v} - 2 \mathsf{c})\vartheta_1(3 \mathsf{v}- \mathsf{c})}{
    \vartheta_1( \mathsf{v} - 3 \mathsf{c})
    \vartheta_1(3 \mathsf{v} - 3 \mathsf{c})
    \vartheta_1(2 \mathsf{c})
    \vartheta_1(\mathsf{v} + \mathsf{c})
  } \cr 
  & \ + \frac{
    \vartheta_1(3 \mathsf{v} - \mathsf{c})
    \vartheta_1(3 \mathsf{v} + \mathsf{c})
  }{
    \vartheta_1(\mathsf{v} - 3 \mathsf{c})
    \vartheta_1(\mathsf{v} - \mathsf{c})
    \vartheta_1(\mathsf{v} + \mathsf{c})
    \vartheta_1(\mathsf{v} + 3 \mathsf{c})
  } \cr 
  & \ + \frac{
    \vartheta_1(3 \mathsf{v} + \mathsf{c})
    \vartheta_1(4 \mathsf{v} + 2 \mathsf{c})
  }{
    \vartheta_1(\mathsf{v} - \mathsf{c})
    \vartheta_1(2 \mathsf{c})
    \vartheta_1(\mathsf{v} + 3 \mathsf{c})
    \vartheta_1(3 \mathsf{v} + 3 \mathsf{c})
  } \quad
  \Bigg] \ .
\end{align}
The sum of three terms can be reorganized into a simple ratio,
\begin{align}
  \mathcal{I}^{(0,4),3}_{1,1,0}(\mathsf{c})
  = & \ \frac{\eta(\tau)^2\vartheta_1(6 \mathsf{v})}{
  \vartheta_1(2 \mathsf{v})
  \vartheta_1(3 \mathsf{v} + 3 \mathsf{c})
  \vartheta_1(3 \mathsf{v} - 3 \mathsf{c})
  } \nonumber\\
  = & \ q^{-1/6}(1 + 2v^2 + (c^3 +c^{-3})v^3 + 3 v^4\\
  & \ + (- v^{-6} - 2 v^{-4} - v^{-2} + 6 + 2(c^3 + c^{-3})v + 15 v^2 + \ldots) q + \ldots) \nonumber
\end{align}

The same computation can be done for $\SU(4)$, just more tedious. There are both non-degenerate and degenerate poles, and in the end, the elliptic genus reads
\begin{align}
  \mathcal{I}^{(0,4),4}_{1,1,0}(\mathsf{c}) = \eta (\tau)^2
  \Bigg[ & \ 
    -\frac{\vartheta_1(5 \mathsf{v}-3 \mathsf{c}) \vartheta_1(4 \mathsf{v}-2 \mathsf{c}) \vartheta_1(3 \mathsf{v}-\mathsf{c})}{\vartheta_1(2 \mathsf{c}) \vartheta_1(2 \mathsf{v}-4 \mathsf{c}) \vartheta_1(4 \mathsf{v}-4 \mathsf{c}) \vartheta_1(\mathsf{v}-3 \mathsf{c}) \vartheta_1(\mathsf{v}+\mathsf{c})} \nonumber\\
    & \ -\frac{\vartheta_1(4 \mathsf{v}-2 \mathsf{c}) \vartheta_1(3 \mathsf{v}-\mathsf{c}) \vartheta_1(3 \mathsf{v}+\mathsf{c})}{\vartheta_1(2 \mathsf{c}) \vartheta_1(4 \mathsf{c}) \vartheta_1(2 \mathsf{v}-4 \mathsf{c}) \vartheta_1(\mathsf{v}-\mathsf{c}) \vartheta_1(\mathsf{v}+\mathsf{c})}\\
    & \ +\frac{\vartheta_1(2 \mathsf{v}) \vartheta_1(4 \mathsf{v}) \vartheta_1(3 \mathsf{v}-\mathsf{c}) \vartheta_1(3 \mathsf{v}+\mathsf{c})}{(\vartheta_1(2 \mathsf{c}))^2 \vartheta_1(\mathsf{v}-3 \mathsf{c}) \vartheta_1(2 \mathsf{v}-2 \mathsf{c}) \vartheta_1(2 \mathsf{v}+2 \mathsf{c}) \vartheta_1(\mathsf{v}+3 \mathsf{c})} \nonumber\\
    & \ -\frac{\vartheta_1(3 \mathsf{v}-\mathsf{c}) \vartheta_1(3 \mathsf{v}+\mathsf{c}) \vartheta_1(4 \mathsf{v}+2 \mathsf{c})}{\vartheta_1(2 \mathsf{c}) \vartheta_1(4 \mathsf{c}) \vartheta_1(\mathsf{v}-\mathsf{c}) \vartheta_1(\mathsf{v}+\mathsf{c}) \vartheta_1(2 \mathsf{v}+4 \mathsf{c})}\nonumber\\
    & \ +\frac{\vartheta_1(3 \mathsf{v}+\mathsf{c}) \vartheta_1(4 \mathsf{v}+2 \mathsf{c}) \vartheta_1(5 \mathsf{v}+3 \mathsf{c})}{\vartheta_1(2 \mathsf{c}) \vartheta_1(\mathsf{v}-\mathsf{c}) \vartheta_1(\mathsf{v}+3 \mathsf{c}) \vartheta_1(2 \mathsf{v}+4 \mathsf{c}) \vartheta_1(4 \mathsf{v}+4 \mathsf{c})}
  \Bigg]\nonumber \ .
\end{align}
Similar to the $\SU(3)$ case, the expression can be recast into a simple ratio,
\begin{align}
  \mathcal{I}^{(0,4),N}_{1,1,0}(\mathsf{c})
  = \frac{
    \eta(\tau)^2\vartheta_1(2N \mathsf{v})
  }{
    \vartheta_1(2 \mathsf{v})
    \vartheta_1(N \mathsf{v} \pm N \mathsf{c})
  } \ , \quad N = 4 \ .
\end{align}
Here, the expression on the right is expected to hold for all genus-one $A_{N - 1}$ theories with one minimal puncture. Furthermore, through an even more intricate computation, we derive the elliptic genus for a circular $\SU(3)^2$ quiver with $\U(1)^2_{x_1, x_2}$ flavor symmetry, or equivalently, a theory of type $A_2$ at genus-one with two minimal punctures. The expression from the direct JK-residue computation is too complicated to detail here; however, it can be reorganized into a simple form
\begin{align}
  \mathcal{I}^{(0,4), 2}_{g = 1, n_1 = 2}(\mathsf{c}_1, \mathsf{c}_2)
  = \frac{\eta(\tau)^4\vartheta(6 \mathsf{v})^2}{
  \vartheta_1(2 \mathsf{v})^2 
  \prod_{i = 1}^2 \vartheta(3 \mathsf{v} \pm 3 \mathsf{c}_i)
  } \ .
\end{align}
Here $n$ denotes the number of minimal punctures. Extrapolating from the above results, it is natural to conjecture that for all the $\SU(N)$-type theory at genus-one with $n_{1}$ minimal punctures, the elliptic genus can be written as
\begin{align}
  \mathcal{I}^{(0,4), N}_{g=1, n_1}(\mathsf{c}_1, \cdots, \mathsf{c}_{n_1})
  = \prod_{i = 1}^{n_{U(1)}}
  \frac{\eta(\tau)^2\vartheta_1(2N \mathsf{v})}{
    \vartheta_1(2 \mathsf{v})
    \vartheta_1(N \mathsf{v} \pm N \mathsf{c}_i)
  } \ .
\end{align}
We will apply this conjecture to later computations.

\subsection{Type \texorpdfstring{$A_{2}$}{A2} theories at genus \texorpdfstring{$g \ge 1$}{g>0}}\label{app:04-SU3}

The elliptic genus of $\mathcal{N} = (0,4)$ $T_3$ theory was computed in \cite{Putrov:2015jpa} using an elliptic inversion formula. In more detail, the elliptic genus of the $\mathcal{N} = (0,4)$ $T_3$ theory and the $\SU(3)$ SQCD with six fundamental flavors is related by the Argyres-Seiberg duality,
\begin{equation}
    \mathcal{I}^{(0,4), N = 3}_{\text{SQCD}}(\mathsf{a},\mathsf{b},\mathsf{r},\mathsf{s})
    = \int_\text{JK} \frac{dz}{2\pi i z}
    \frac{\eta(\tau)}{\vartheta_1(\mathsf{v} \pm \mathsf{s} \pm \mathsf{z})}
    \mathcal{I}_{\textrm{vec}}^{(0,4),2}(\mathsf{z})
    \mathcal{I}_{T^3}(\mathsf{a}, \mathsf{b}, \mathsf{c}) \ .
\end{equation}
Within the integral, we gauge an $\SU(2) \subset \SU(3)_c$, leading to $\mathsf{c}_1 = \mathsf{r} + \mathsf{z}, \mathsf{c}_2 = \mathsf{r} - \mathsf{z}, \mathsf{c}_3 = -2 \mathsf{r}$. The integral can be inverted to compute the elliptic genus $\mathcal{I}_{T_3}$. Explicitly, it is given by a simple JK-residue computation
\begin{align}
  \mathcal{I}_{T_3}(\mathsf{a}, \mathsf{b}, \mathsf{c})
  = \frac{\eta(\tau)^5}{2 \vartheta_1(2 \mathsf{v} \pm 2 \mathsf{z})} \int_\text{JK}\frac{ds}{2\pi i s}
    \frac{\vartheta_1(\pm 2 \mathsf{s}) \vartheta_1(\pm 2 \mathsf{v})}{
    \vartheta_1(- \mathsf{v} \pm \mathsf{s} \pm \mathsf{z})
    } \mathcal{I}^{(0,4), N = 3}_{\text{SQCD}}(\mathsf{a}, \mathsf{b}, \mathsf{r}, \mathsf{s}) \ . \nonumber
\end{align}
Note that the coefficient in front of the integral has been adjusted according to our convention. On the right, $\mathsf{a}, \mathsf{b}, \mathsf{r}, \mathsf{s}$ represent the $\SU(3)_a \times \SU(3)_b \times\U(1)_x \times\U(1)_y$ flavor chemical potentials, where
\begin{align}
  \mathsf{x} = \frac{\mathsf{s}}{3} + \mathsf{r}, \qquad \sfy = \frac{\mathsf{s}}{3} - \mathsf{r}\  .
\end{align}
After the integral, $\mathsf{z}, \mathsf{r}$ combine into $\SU(3)$ chemical potentials
\begin{align}
  \mathsf{c}_1 = \mathsf{r} + \mathsf{z}, \qquad
  \mathsf{c}_2 = \mathsf{r} - \mathsf{z}, \qquad
  \mathsf{c}_3 = - 2 \mathsf{r} \ .
\end{align}
The $\sfa,\sfb,\sfc$ denote the $\SU(3)^3 \subset E_6$ chemical potentials for the 
$\mathcal{N} = (0,4)$ $E_6$ theory.

From the elliptic inversion formula for $\mathcal{I}_{T_3}$, it is evident that the diagonal of the two $\SU(3)$ flavor subgroups can be gauged, which yields the elliptic genus of the $\SU(3)$-type genus-one theory with one maximal puncture associated to an $\SU(3)_c$ flavor symmetry. Now the situation is much better: the right-hand side involves $\mathcal{I}^{(0,4),3}_{g = 1, n_1 = 2}$ which is shown to be a simple ratio of elliptic theta functions, where $n_1$ denotes the number of minimal punctures. (See Figure \ref{fig:elliptic-inversion1}.) When performing the JK-residue computation, we encounter only non-degenerate poles, and we obtain a simple result for the (0,4) elliptic genus with $n_3 = 1$ maximal puncture,
\begin{align}
  \mathcal{I}^{(0,4),3}_{g = 1, n_1 = 0, n_3 = 1}(\mathsf{c})
  =    \frac{\eta(\tau)^6\vartheta_1(2 \mathsf{v})\vartheta_1(4 \mathsf{v}) \vartheta_1(6 \mathsf{v})}{
    \prod_{\substack{A, B = 1}}^3 \vartheta_1(2 \mathsf{v} +\mathsf{b}_{iA} - \mathsf{b}_{iB})
  }\ .
\end{align}

Instead of directly gauging the diagonal of the two existing $\SU(3)$ groups, one can alternatively gauge the diagonal of the $\SU(3)^2$ flavor subgroup of $\mathcal{I}_{T_3}$ and a $\SU(3)$ linear quiver. This approach can be used to generate a genus-one theory with additional minimal punctures. In other words,
\begin{align}
  \mathcal{I}^{(0,4),3}_{g = 1, n_1, n_3 = 1}
  = \text{elliptic inversion of \ } \mathcal{I}^{(0,4),3}_{g = 1, n_1 + 2, n_3 = 0} \ .
\end{align}
Effectively, the elliptic inversion formula represents the fusion of any two minimal punctures into a single maximal one. Therefore, the inversion, or fusing two minimal punctures, can be performed successively, yielding more maximal punctures. Since the elliptic genus on the right are all simple ratios, the JK residue can be easily carried out, and at each step of the fusion, the outcome continues to be a simple ratio. In conclusion, we obtain the following general result for the theory of type $A_2$ for any $g \ge 1$ that has $n_1$ minimal and $n_3$ maximal punctures
\begin{align}
  \mathcal{I}^{(0,4), 3}_{g, n_1, n_3}
  = \bigg(
  \frac{\vartheta_1(2\mathsf{v}) \vartheta_1(4 \mathsf{v})^2 \vartheta_1(6 \mathsf{v})}{
    \eta(\tau)^4
  }
  \bigg)^{g - 1}
 \mathcal{I}^{(0,4), 3}_{g = 1, n_1, 0}
  \mathcal{I}^{(0,4), 3}_{g = 1, 0, n_3} \ ,
\end{align}
where with $N = 3$,
\begin{align}
  \mathcal{I}^{(0,4), 3}_{g = 1, n_1, 0}
  = & \ \prod_{i = 1}^{n_1}
  \frac{\eta(\tau)^2\vartheta_1(2N \mathsf{v})}{
    \vartheta_1(2 \mathsf{v})
    \vartheta_1(N \mathsf{v} \pm N \mathsf{c}_i)
  }  \ , \\
  \mathcal{I}^{(0,4), 3}_{g = 1, 0, n_3}
  = & \ \prod_{i = 1}^{n_{3}}   \frac{\eta(\tau)^6\vartheta_1(2 \mathsf{v})\vartheta_1(4 \mathsf{v}) \vartheta_1(6 \mathsf{v})}{
    \prod_{\substack{A, B = 1}}^3 \vartheta_1(2 \mathsf{v} +\mathsf{b}_{iA} - \mathsf{b}_{iB})
  }\ .
\end{align}

\subsection{Type \texorpdfstring{$A_{3}$}{A3} theories at genus \texorpdfstring{$g \ge 1$}{g>=1}}\label{app:04-SU4}

Let us now consider theories of type $A_3$ of class $\mathcal{S}$. There are four punctures (and the corresponding flavor symmetry) to be considered, minimal ($\U(1)$), $[2,1^2]$ ($\SU(2)\times \U(1)$), $[2^2]$ ($\SU(2)$), and maximal ($\SU(4)$). To begin, we have
\begin{equation}
    \mathcal{I}^{(0,4), N = 4}_{g = 1; n_1, 0, 0, 0}
    = \prod_{i = 1}^{n_1}
  \frac{\eta(\tau)^2\vartheta_1(2N \mathsf{v})}{
    \vartheta_1(2 \mathsf{v})
    \vartheta_1(N \mathsf{v} \pm N \mathsf{c}_i)
  }\ , \qquad
  N = 4 \ ,
\end{equation}
where the subscript $n_1, n_2, n_3, n_4$ denotes the number of punctures associated with the respective flavor symmetries: $\U(1)$, $\SU(2)\times \U(1)$, $\SU(2)$, and $\SU(4)$.

In the context of 4d $\mathcal{N} = 2$ SCFTs, $R_{0,N}$ are non-Lagrangian theories with $\SU(2N) \times \SU(2)$ flavor symmetry, and they arise from a strong coupling limit of the $\mathcal{N} = 2$ $\SU(N)$ theory with $2N$ fundamental hypermultiplets. Concretely, as a theory of $A_3$-type class $\mathcal{S}$, $R_{0,4}$ corresponds to the three-punctured sphere with two maximal-puncture and a $[2, 1^2]$ puncture, where the manifest flavor subgroup is $\SU(4)^2 \times \U(1) \times \SU(2)$.

In 2d, the  $\mathcal{N} = (0,4)$  elliptic genus of $R_{0,4}$ theory was computed in \cite{Agarwal:2018ejn} using an elliptic inversion. The Argyres-Seiberg-like duality implies an integral equality between the elliptic genus of $\SU(4)$ SQCD with eight fundamental flavors and that of the $R_{0,4}$ theory,
\begin{equation}
    \mathcal{I}^{(0,4), N = 4}_{\text{SQCD}}(\mathsf{a}, \mathsf{b}, \mathsf{r}, \mathsf{s})
    = \int_\text{JK} \frac{dz}{2\pi i z}
    \frac{\eta(\tau)}{\vartheta_1(\mathsf{v} \pm \sfs \pm \mathsf{z})}
    \mathcal{I}_{\textrm{vec}}^{(0,4),2}(\mathsf{z})
    \mathcal{I}^{(0,4)}_{R_{0,4}}(\mathsf{a}, \mathsf{b}, \mathsf{r}, \mathsf{z}) \ ,
\end{equation}
which can be inverted to give
\begin{align}
  \mathcal{I}^{(0,4)}_{R_{0,4}}(\mathsf{a}, \mathsf{b}, \mathsf{r}, \mathsf{z})
  = \frac{\eta(\tau)^5}{2 \vartheta_1(2 \mathsf{v} \pm 2 \mathsf{z})}
  \int_\text{JK}\frac{ds}{2\pi i s}
  \frac{\vartheta_1(\pm 2 \mathsf{s}) \vartheta_1(- 2 \mathsf{v})}{
  \vartheta_1(- \mathsf{v} \pm \mathsf{s} \pm \mathsf{z})
  }
  \mathcal{I}^{(0,4), N = 4}_{\text{SQCD}}(\mathsf{a}, \mathsf{b}, \mathsf{r}, \mathsf{s}) \ . \nonumber
\end{align}
Here, $\mathsf{a},\mathsf{b},\mathsf{r},\mathsf{z}$ denote the chemical potentials of $\SU(4) \times \SU(4) \times\U(1) \times \SU(2) \subset \SU(8) \times \SU(2)$ flavor symmetry of the $R_{0,4}$ theory. On the right, $\mathsf{a},\mathsf{b},\mathsf{r},\mathsf{s}$ are the chemical potentials of the $\U(8) = \SU(8)_{\mathsf{a},\mathsf{b},\mathsf{r}} \times \U(1)_\mathsf{s}$ flavor symmetry of the $\SU(4)$ gauge theory with $8$ fundamental hypermultiplets. Chemical potentials $\mathsf{r}, \mathsf{s}$ are related to the standard $\mathsf{x}, \mathsf{y}$ (associated to the two minimal punctures of the $\SU(4)$ SQCD) by
\begin{align}
  \mathsf{x} = \frac{\mathsf{s}}{4} + \mathsf{r}, \qquad \mathsf{y} = \frac{\mathsf{s}}{4} - \mathsf{r} \ .
\end{align}
One can start gauging in the theory of $4^2$ free hypermultiplets or handles to the punctures associated to $\mathsf{a}, \mathsf{b}$ on both sides of the above equation, so that the $\mathcal{I}^{(0,4), N = 4}_{\text{SQCD}}$ on the right becomes $\mathcal{I}^{(0,4), N = 4}_{g, n_1 + 2, 0, 0, n_4}$, while on the left, it becomes $\mathcal{I}^{(0,4), 4}_{g; n_1, 0, 1, n_4}$. Effectively, the elliptic inversion merges two minimal punctures into a $[2, 1^2]$ puncture. (See Figure \ref{fig:elliptic-inversion2}.) In the simplest case, when $g = 1$, $n_1 = 0$, $n_4 = 0$, 
\begin{equation}
    \mathcal{I}^{(0,4),4}_{g = 1; 0,1,0,0}
    = \frac{ \eta(\tau)^6 \vartheta_1(6 \mathsf{v}) \vartheta_1(8 \mathsf{v})}{
        \vartheta_1(2 \mathsf{v})^2
        \vartheta_1(2 \mathsf{v} \pm 2 \mathsf{z})
        \vartheta_1(3\mathsf{v} \pm \mathsf{z} \pm 4 \mathsf{r})
    } \ .
\end{equation}

The $[2,1,1]$ puncture can be further Higgsed to a $[2,2]$ puncture with $\SU(2)$ flavor symmetry. The class $\mathcal{S}$ theory corresponding to a three-punctured sphere with two maximal and one $[2,2]$ puncture has an enhanced $E_7$ flavor symmetry. The elliptic genus of the corresponding 2d $\mathcal{N} = (0,4)$ theory can also be computed using an elliptic inversion formula \cite{Agarwal:2018ejn}. Following the same approach as above, we perform additional gluing/gauging operations on the two maximal punctures, and we obtain the contribution from one $[2,2]$ puncture,
\begin{align}
    \mathcal{I}^{(0,4),4}_{g = 1; 0,0,1,0}(\mathsf{w})
    = & \ \int_\text{JK}\frac{ds}{2\pi i s} \frac{\eta(\tau)^5 \vartheta_1(\pm 2 \mathsf{v})}{2 \vartheta_1(4 \mathsf{v})} \frac{\vartheta_1(\pm 2 \mathsf{s})}{
        \vartheta_1(\pm \mathsf{s})
        \vartheta_1(-2\mathsf{v} \pm \mathsf{s})
    } \mathcal{I}^{(0,4),4}_{1; 2, 0,0,0}(\mathsf{c}_1, \mathsf{c}_2) \nonumber \\
    = & \ \frac{
      \eta(\tau)^6\vartheta_1(6 \mathsf{v})
      \vartheta_1(8 \mathsf{v})
    }{
      \vartheta_1(2 \mathsf{v})
      \vartheta_1(4 \mathsf{v})
      \vartheta_1(4 \mathsf{v} \pm 4 \mathsf{w})
      \vartheta_1(2 \mathsf{v} \pm 4 \mathsf{w})
    } \ .
\end{align}
Again, $\mathsf{w}$ represents the $\SU(2)$ flavor chemical potential. As before, the two $\U(1)$ chemical potentials on the right are related to $\mathsf{w}$ by $\mathsf{c}_1 = \frac{\mathsf{s}}{4} + \mathsf{w}$ and $\mathsf{c}_2 = \frac{\mathsf{s}}{4} - \mathsf{w}$.

The contribution from the maximal puncture can be obtained by considering the generalized S-duality as shown in Figure \ref{fig:SU4-Sduality}, which requires
\begin{align}
    \mathcal{I}^{(0,4),4}_{1; 3, 0,0,0}(\mathsf{x}_{1,2,3})
    = \int_\text{JK} & \ \frac{db_1}{2\pi i b_1} \prod_{A = 1}^2 \frac{da_A}{2\pi i a_A}
    \mathcal{I}^{(0,4),4}_{1; 0,0,0,1}(\mathsf{c})\bigg|_{\mathsf{c}_{A = 1,2,3} = \mathsf{a}_A + \mathsf{r}} \nonumber \\
    & \ \times \mathcal{I}_{\textrm{vec}}^{(0,4),3}(\mathsf{a})
    \prod_{A = 1}^3\prod_{i = 1}^2 \frac{\eta(\tau)^2}{\vartheta_1(\mathsf{v} \pm (- \mathsf{a}_A + \mathsf{b}_i + \mathsf{x}'_1))}\\
    & \ \times \mathcal{I}_{\textrm{vec}}^{(0,4),2}(\mathsf{b})
    \prod_{i = 1}^2 \frac{\eta(\tau)^2}{\vartheta_1(\mathsf{v} \pm ( - \mathsf{b}_i + \mathsf{x}'_2))} \ . \nonumber
\end{align}
In the above $a_A$ and $b_i$ are $\SU(3)$ and $\SU(2)$ fugacities with $a_3 = 1/(a_1 a_2)$ and $b_1 = 1/b_2$. The $\U(1)$ fugacities on both sides are identified by
\begin{equation}
    x'_1 = \frac{x_1^{4/3} x_3^{2/3}}{x_2^{2/3}}, \qquad
    x'_2 = x_2^2 x_3^2, \qquad
    r = \Big(\frac{x_3}{x_1 x_2} \Big)^{1/3} \ .
\end{equation}
Explicitly, the elliptic genus of the $\mathcal{N} = (0,4)$ genus-one theory with one maximal puncture reads
\begin{equation}
    \mathcal{I}^{(0,4),4}_{1; 0,0,0,1}
    = \frac{
        \eta(\tau)^{12} \vartheta_1(2 \mathsf{v})\vartheta_1(4 \mathsf{v})\vartheta_1(6 \mathsf{v}) \vartheta_1(8 \mathsf{v})}{
        \prod_{A, B = 1 }^4
    \vartheta_1(2 \mathsf{v} + \mathsf{c}_{A} - \mathsf{c}_{B})
    } \ .
\end{equation}

Finally, by gauging the maximal punctures, we determine the elliptic genus for genus $g \ge 1$ with arbitrary punctures
\begin{equation}
    \mathcal{I}^{(0,4), 4}_{g; n_1, n_2, n_3, n_4}
    = \bigg(
      \frac{
      \vartheta_1(2\mathsf{v})
      \vartheta_1(4\mathsf{v})^2
      \vartheta_1(6\mathsf{v})^2
      \vartheta_1(8\mathsf{v})^2
      }{
      \eta(\tau)^6
      }
    \bigg)^{g - 1}
    \mathcal{I}^{(0,4), 4}_{1; n_1,0,0,0}
    \mathcal{I}^{(0,4), 4}_{1; 0,n_2,0,0}
    \mathcal{I}^{(0,4), 4}_{1; 0,0,n_3,0}
    \mathcal{I}^{(0,4), 4}_{1; 0,0,0,n_4} \ ,\nonumber
\end{equation}
where
\begin{align}
  \mathcal{I}^{(0,4), 4}_{1; n_1,0,0,0}
  = & \ \prod_{i = 1}^{n_1} \frac{\eta(\tau)^2\vartheta_1(8 \mathsf{v})}{
  \vartheta_1(2 \mathsf{v})
  \vartheta_1(4 \mathsf{v} \pm 4 \mathsf{x}_i)
  } \\
  \mathcal{I}^{(0,4), 4}_{1; 0,n_2,0,0}
  = & \ \prod_{i = 1}^{n_3} \frac{
    \eta(\tau)^6 \vartheta_1(6 \mathsf{v})\vartheta_1(8 \mathsf{v})
  }{
  \vartheta_1(2 \mathsf{v})^2
  \vartheta_1(2 \mathsf{v} \pm 2 \mathsf{z}_i)
  \vartheta_1(3 \mathsf{v} \pm \mathsf{z}_i \pm 4\mathsf{r}_i)
  } \\
  \mathcal{I}^{(0,4), 4}_{1; 0,0,n_3,0}
  = & \ \prod_{i = 1}^{n_2}
  \frac{\eta(\tau)^6 \vartheta_1(6\mathsf{v}) \vartheta_1(8 \mathsf{v})}{
  \vartheta_1(2 \mathsf{v})
  \vartheta_1(4 \mathsf{v})
  \vartheta_1(2 \mathsf{v} \pm 4 \mathsf{w}_i)
  \vartheta_1(4 \mathsf{v} \pm 4 \mathsf{w}_i)
  } \ ,\\
  \mathcal{I}^{(0,4), 4}_{1; 0,0,0,n_4}
  = & \ \prod_{i = 1}^{n_4}
  \frac{
        \eta(\tau)^{12} \vartheta_1(2 \mathsf{v})\vartheta_1(4 \mathsf{v})\vartheta_1(6 \mathsf{v}) \vartheta_1(8 \mathsf{v})}{
        \prod_{A, B = 1 }^4
    \vartheta_1(2 \mathsf{v} + \mathsf{b}_{iA} - \mathsf{b}_{iB})
    }
\end{align}

\bibliographystyle{JHEP}
\bibliography{references}

\end{document}